\documentclass[iop,numberedappendix]{emulateapj}
\usepackage[colorlinks=true,linkcolor=blue,anchorcolor=blue,citecolor=blue,urlcolor=blue]{hyperref}
\usepackage{color}
\hypersetup{pdfauthor={Keiichi Umetsu}, pdftitle={CLASH: Full Lensing Analysis of MACS1206}}

\usepackage{natbib}
 
\newcommand{\simgt}{\lower.5ex\hbox{$\; \buildrel > \over \sim \;$}}
\newcommand{\simlt}{\lower.5ex\hbox{$\; \buildrel < \over \sim \;$}}

\def\btheta{\mbox{\boldmath $\theta$}} 

\def\bkappa{\mbox{\boldmath $\kappa$}}
\def\bSigma{\mbox{\boldmath $\Sigma$}}

\def\bd{\mbox{\boldmath $d$}}
\def\bs{\mbox{\boldmath $s$}}

\lefthead{Umetsu et al.}

\righthead{Bayesian Analysis of Weak Lensing Distortion and
Magnification Measurements}  

\begin{document}

\title{
Cluster Mass Profiles from a Bayesian Analysis of Weak Lensing
Distortion and Magnification Measurements: Applications to Subaru Data
\altaffilmark{1}}

\author{
Keiichi Umetsu\altaffilmark{2},
Tom Broadhurst\altaffilmark{3,4},
Adi Zitrin\altaffilmark{5},
Elinor Medezinski\altaffilmark{6},
Li-Yen Hsu\altaffilmark{7}
} 

\altaffiltext{1}
 {Based in part on data collected at the Subaru Telescope,
  which is operated by the National Astronomical Society of Japan.}
\altaffiltext{2}
 {Institute of Astronomy and Astrophysics, Academia Sinica,
  P.~O. Box 23-141, Taipei 10617, Taiwan.}
\altaffiltext{3}
 {Theoretical physics, University of the Basque Country, Bilbao 48080,
 Spain.}
\altaffiltext{4}
{Ikerbasque, Basque Foundation for Science, Alameda Urquijo, 36-5 Plaza
 Bizkaia 48011, Bilbao, Spain.} 
\altaffiltext{5}
  {School of Physics and Astronomy, Tel Aviv University, Tel Aviv 69978,
 Israel.}
\altaffiltext{6}
{Johns Hopkins University, 3400 North Charles Street, Baltimore, MD
 21218, USA.} 
\altaffiltext{7}
 {Leung center for Cosmology and Particle Astrophysics, National Taiwan
  University, Taipei 10617, Taiwan.}


\begin{abstract}
We directly construct model-independent mass profiles of galaxy clusters
from combined weak-lensing distortion and magnification measurements within
a Bayesian statistical framework,
which allows for a full parameter-space extraction of the underlying
signal.    
This method applies to the full range of radius outside the Einstein
radius, and recovers the absolute mass normalization.
We apply our method to deep Subaru imaging of five high-mass
($>10^{15}M_\odot$) clusters, A1689, A1703, A370, Cl0024+17, and
 RXJ1347-11,  to  obtain accurate profiles to beyond the 
virial radius ($r_{\rm vir}$).  For each cluster the lens distortion and
 magnification data are shown to be consistent with each other, and the
 total signal-to-noise ratio of the combined measurements ranges from 13
 to 24 per cluster.  
We form a model-independent mass profile from stacking the clusters,
which is detected at 
$37\sigma$ out to $R\approx 1.7r_{\rm vir}$. 
The projected logarithmic slope 
$\gamma_{\rm 2D}(R)\equiv d\ln{\Sigma}/d\ln{R}$ steepens from
 $\gamma_{\rm 2D}=-1.01\pm 0.09$ at $R\approx 0.1r_{\rm vir}$ to 
 $\gamma_{\rm 2D}=-1.92\pm 0.51$ at $R\approx 0.9r_{\rm vir}$.
We also derive for each cluster inner strong-lensing based mass profiles
 from deep Advanced Camera for Surveys observations with the {\it Hubble
 Space Telescope}, which we show overlap well with the outer
 Subaru-based profiles and together are well described by a generalized
 form of the Navarro-Frenk-White profile, except for the ongoing merger
 RXJ1347-11, with modest variations in the central cusp slope
 ($-d\ln{\rho}/d\ln{r}\simlt 0.9$).  
The improvement here from adding the magnification measurements is
significant, $\sim 30\%$ in terms of cluster mass profile measurements,
 compared with the lensing distortion signal.
\end{abstract}
 

\section{Introduction} 
\label{sec:intro}

Galaxy clusters provide an independent means of examining any viable
model of cosmic structure formation through the growth of structure and
by the form of their equilibrium mass profiles, complementing cosmic
microwave background and galaxy clustering observations.
A consistent framework of structure formation
requires that most of the matter in the Universe is in the hitherto
unknown form of dark matter, of an unknown nature, and that most of the
energy filling the Universe today is in the form of a mysterious
``dark energy'', 
characterized by a negative pressure. This model actually requires
that the expansion rate of the Universe has recently changed sign and is 
currently accelerating.

Clusters play a direct role in testing cosmological
models, providing several independent checks of any viable cosmology,
including the current consensus $\Lambda$ cold dark matter ($\Lambda$CDM)
model. 
A spectacular example has been recently provided from detailed lensing
and X-ray observations of the ``Bullet Cluster''
\citep[aka, IE0657-56;][]{2004ApJ...606..819M,Clowe+2006_Bullet},
which is a  consequence of a high-speed collision
between two cluster components with a mass ratio 
of the order of $6:1$ \citep{Mastropietro+Burkert2008},
displaying a prominent bow shock preceding a cool 
bullet lying between the two clusters, 
implying these clusters passed through each other recently 
\citep{2002ApJ...567L..27M,2004ApJ...606..819M}. 
Here the Bullet system reveals lensing mass contours that follow the
bimodal distribution of cluster members, demonstrating that the bulk of
the dark matter is relatively collisionless as galaxies
\citep{Clowe+2006_Bullet}, as also shown by a comprehensive analysis of
galaxy and dark-matter dynamics for A1689
\citep{Lemze+2011_DM}. 
Other cases of merging systems show that in general displacement of the
hot gas relative to the dark matter is related to interaction
\citep{2005ApJ...618...46J,Okabe+Umetsu2008}. 
For dynamically-relaxed clusters, the form of the equilibrium mass
profile reflects closely the distribution of dark matter
\citep[see][]{Mead+2010_AGN}  which, unlike galaxies, does not suffer
from halo compression by adiabatic contraction of cooled gas.
The majority of baryons in clusters are in the form of hot, diffuse
X-ray emitting gas, and represents only a minor fraction of the total
lensing mass near the centers of clusters
\citep{2008MNRAS.386.1092L,Umetsu+2009}.

The predicted Navarro-Frenk-White profile
\citep[hereafter, NFW;][]{1996ApJ...462..563N,1997ApJ...490..493N} 
derived from simulations based on collisionless, cold (non-relativistic)
dark matter has a continuously-declining logarithmic gradient
$\gamma_{\rm 3D}(r)=d\ln{\rho}/d\ln{r}$
towards the center of mass, much shallower than the isothermal case
($\gamma_{\rm 3D}=-2$) within the characteristic scale radius, $r_s$
($\simlt 300\,{\rm kpc}\,h^{-1}$ for cluster-sized halos).
A useful index of the degree of concentration, $c_{\rm vir}$, compares
the virial radius, $r_{\rm vir}$,  to $r_s$ of the NFW profile, $c_{\rm
vir}\equiv r_{\rm vir}/r_s$.  
This has been confirmed thoroughly with higher resolution
simulations 
\citep{1998MNRAS.300..146G,Fukushige+Makino1997,Okamoto+Habe1999,
Power+2003,Navarro+2004,2007MNRAS.381.1450N}, 
with some intrinsic variation related to the individual assembly
history of a cluster \citep{Jing+Suto2000,Tasitsiomi+2004}.
Gravitational lensing observations are underway to provide reliable and
representative cluster mass profiles to test this since the first careful
measurements showed that the NFW profile provides a good fit to the
entire mass profile when weak and strong lensing are combined
\citep{2003A&A...403...11G,BTU+05,BUM+08,Umetsu+2010_CL0024}. 
Other well studied clusters with similarly high quality data are also in 
good agreement providing strong support for the CDM scenario
\citep[e.g.,][]{Okabe+2010_WL}.

Interestingly these studies reveal that although the dark matter is
consistent with being cold, the predicted profile concentration of the
standard $\Lambda$CDM model falls short of some lensing results 
\citep[e.g.,][]{BTU+05,BUM+08,Oguri+2009_Subaru}.
This observed tendency for higher proportion of mass to lie at smaller
radius in projection is also indicated by the generally large Einstein
radii determined from strong lensing of well studied clusters
\citep{Broadhurst+Barkana2008,Zitrin+2011_MACS} finding a substantial
discrepancy with the predictions despite careful accounting for
potential selection biases inherent to lensing
\citep{2007ApJ...654..714H,Meneghetti+2010_MARENOSTRUM}.
These observations could suggest either substantial mass projected along
the line of sight, perhaps in part due to halo triaxiality
\citep{2005ApJ...632..841O}, or a large overconcentration of
mass; the latter could imply modification within the context of the CDM
family of models.

The abundance of massive clusters is very sensitive to the amplitude
of the initial mass power spectrum \citep{Sheth+Mo+Tormen2001}
representing the most massive objects to have collapsed under their
own gravity, and confirmed by $N$-body simulations of Hubble volumes
\citep{Evrard+2002}.
Such calculations predict for example that the
single most massive cluster to be found in the universe is expected to
be with $M_{\rm vir}=4\times10^{15}M_{\odot}$ out to $z=0.4$
\citep[see Figure 5 of][]{Broadhurst+Barkana2008}, 
similar to the most massive known clusters detected
locally \citep{BUM+08}.\footnote{A370 at $z=0.375$ is currently the most
massive known cluster measured reliably by lensing,
$M_{\rm vir}=(2.9\pm 0.3)\times10^{15}M_{\odot}$.}
At higher redshifts this comparison becomes more sensitive to the
cosmological model, with an order of magnitude decline in the
abundance of $10^{15}M_{\odot}$ clusters at $z>0.8$ compared to the
present \citep{Evrard+2002}.
Hence, the existence of such massive clusters like XMMUJ2235-25 at
$z=1.45$ \citep{Jee+2009}, from lensing work, begins to motivate
alternative ideas such as departures from Gaussian initial density
fluctuation spectrum, or higher levels of dark energy in the past
\citep{Sadeh+Rephaeli2008}, although some non-Gaussian models can be
ruled out by using the cosmic X-ray background measurements
\citep{Lemze+2009_CXRB}. 

The main attraction of gravitational lensing in the cluster regime
\citep[e.g.,][]{1999PThPS.133....1H,1999PThPS.133...53U,2001PhR...340..291B,
Umetsu2010_Fermi} 
is the model-free determination of mass profiles allowed over a wide
range of radius when the complementary effects of strong and weak
lensing are combined  
\citep{BTU+05,UB2008,Merten+2009,Umetsu+2010_CL0024,Meneghetti+2010a,Zitrin+2010_A1703}.
In practice, the quality of data required 
challenges with few facilities, which are able to generate data of
sufficient precision to provide a significant detection of the weak
lensing signal on an individual cluster basis.

In this paper we aim to pursue in greater depth the utility of massive
clusters for defining highest-precision mass profiles by combining all
lensing information available in the cluster regime.
In particular, we shall make a full use of the magnification
information, afforded by measuring spatial variations in the
surface number density of faint background galaxies
\citep{1995ApJ...438...49B,1998ApJ...501..539T,2005PhRvL..95x1302Z,UB2008,Rozo+Schmidt2010,vanWaerbeke+2010},    
which we show here is readily detectable in the high-quality images we
have obtained for this purpose. 
In our earlier work, particularly on A1689 and Cl0024+17, the
magnification information determined from the background counts was
used only as a consistency check of distortion measurements 
\citep{BUM+08,Umetsu+2010_CL0024}.
Recently, we have shown in \citet{UB2008}
how to overcome the intrinsic clustering of background galaxies,
which otherwise perturbs locally the magnification signal, and
how to combine the two independent weak-lensing data sets 
to improve the quality of two-dimensional mass reconstruction
using  regularized maximum-likelihood techniques.
Here we further explore new statistical methods designed to
obtain an optimal combination of the complementary lens distortion and
magnification effects.
Our aim here is to develop and apply techniques to a sample of
five well-studied, high-mass clusters with
$M\simgt 10^{15}M_\odot$, A1689, A1703, A370, Cl0024+17, and RXJ1347-11, 
for examining the underlying mass profiles extracted from the combined 
weak-lensing data sets.  
We then add to this detailed strong-lensing information  for the inner 
$\simlt 200\,$kpc region of these clusters, for which we have identified
many new sets of multiple images from Advanced Camera for Surveys (ACS)
observations \citep{2005ApJ...621...53B,Zitrin+2009_CL0024,Zitrin+2010_A1703}
with the {\it Hubble Space Telescope} ({\it HST}),
to derive improved inner mass profiles for a full
determination of the entire mass profiles of the five well-studied
clusters.

The paper is organized as follows. We briefly summarize in
\S~\ref{sec:basis} the basis of cluster weak gravitational lensing. 
In \S~\ref{sec:method} we present our comprehensive lensing method in a
Bayesian framework 
for a direct reconstruction of the projected cluster mass profile from
combined weak-lensing shape distortion and magnification bias
measurements.  In \S~\ref{sec:results} we apply our method to Subaru
weak-lensing  observations of five massive clusters to derive projected
mass profiles to beyond the cluster virial radius;
we also combine our new weak-lensing mass profiles with inner
strong-lensing based information from {\it HST}/ACS observations to make
a full determination of the entire cluster mass profiles.
Finally, summary and discussions are given in \S~\ref{sec:discussion}.

Throughout this paper, e use the AB magnitude system, and
adopt a concordance $\Lambda$CDM cosmology with $\Omega_{m}=0.3$, 
$\Omega_{\Lambda}=0.7$, and $h\equiv H_0/(100\, {\rm km\, s^{-1}\,
Mpc^{-1}})=0.7$. 
Errors represent a confidence level of $68.3\%$ ($1\sigma$)
unless otherwise stated.

\section{Basis of Cluster Weak Lensing}
\label{sec:basis}

The deformation of the image for a background source can be described by
the 
Jacobian matrix $\cal{A}_{\alpha\beta}$
($\alpha,\beta=1,2$) of the lens mapping.\footnote{Throughout the paper 
we assume in our weak lensing analysis that the angular size of
background galaxy images is sufficiently small 
compared to the scale over which the underlying lensing fields vary, so
that the higher-order weak lensing effects, such as {\it flexion}, can
be safely neglected; see, e.g.,
\cite{2005ApJ...619..741G,HOLICs1,HOLICs2}.} 
The real, symmetric Jacobian ${\cal A}_{\alpha\beta}$ 
can be decomposed as
${\cal A}_{\alpha\beta} = (1-\kappa)\delta_{\alpha\beta}
 -\Gamma_{\alpha\beta}$,
where 
$\delta_{\alpha\beta}$ is Kronecker's delta, 
$\kappa$ is the lensing convergence, 
and 
$\Gamma_{\alpha\beta}$ is the trace-free, symmetric shear matrix,
\begin{eqnarray}
\label{eq:jacob} 
\Gamma_{\alpha\beta}&=&
\left( 
\begin{array}{cc} 
+{\gamma}_1   & {\gamma}_2 \\
 {\gamma}_2  & -{\gamma}_1 
\end{array} 
\right),
\end{eqnarray}
with $\gamma_{\alpha}$ being the components of 
spin-2
complex gravitational
shear $\gamma:=\gamma_1+i\gamma_2$.
In the strict weak lensing limit where $\kappa,|\gamma|\ll 1$, 
$\Gamma_{\alpha\beta}$ induces a quadrupole anisotropy
of the  
background image, which can be observed from ellipticities 
of background galaxy images~\citep{1995ApJ...449..460K}.
The local area distortion due to gravitational lensing, or
magnification, 
is given by the inverse Jacobian determinant,
\begin{equation}
\label{eq:mu}
\mu = 
\frac{1}{{\rm det}{\cal A}}
=
\frac{1}{(1-\kappa)^2-|\gamma|^2},
\end{equation}
where we assume subcritical lensing, i.e., 
${\rm det}{\cal A}(\btheta)>0$.  The lens magnification $\mu$ can be
measured from characteristic variations in the number density of
background galaxies~\citep[][see also
\S~\ref{subsec:magbias}]{1995ApJ...438...49B}.

The lensing convergence $\kappa$ is a weighted projection of the matter
density contrast along the line of sight
\citep[e.g.,][]{2000ApJ...530..547J}.
For gravitational lensing in the cluster regime
\citep[e.g.,][]{Umetsu2010_Fermi}, $\kappa$ is expressed as
$\kappa(\btheta)=\Sigma_{\rm crit}^{-1}\Sigma(\btheta)$,
namely the projected mass density $\Sigma(\btheta)$
in units of the critical surface mass density for gravitational
lensing, defined as
\begin{eqnarray} 
\label{eq:sigmacrit}
\Sigma_{\rm crit} = \frac{c^2}{4\pi G D_d} \langle\beta\rangle^{-1};
\ \ \ \beta(z_s) \equiv {\rm max}\left[
0,\frac{D_{ds}(z_s)}{D_s(z_s)}\right],
\end{eqnarray}
where $D_s$, $D_d$, and $D_{ds}$ are the proper angular diameter
distances from the observer to the source, from the observer to the
deflecting  lens, and from the lens to the source, respectively, and 
$\langle\beta\rangle=\langle D_{ds}/D_s\rangle$ is the mean distance
ratio averaged over the 
population of source galaxies in the cluster field.

In general, the observable quantity for quadrupole weak lensing
is not the gravitational shear $\gamma$ but the complex {\it reduced}
shear (see \S~\ref{subsec:gt}), 
\begin{equation}
\label{eq:redshear}
g(\btheta)=\frac{\gamma(\btheta)}{1-\kappa(\btheta)}
\end{equation}
in the subcritical regime where ${\rm det}{\cal A}>0$
(or $1/g^*$ in the negative parity region with ${\rm det}{\cal A}<0$). 
The reduced shear $g$ is invariant under the following
global linear transformation:
\begin{equation}
\label{eq:invtrans}
\kappa(\btheta) \to \lambda \kappa(\btheta) + 1-\lambda, \ \ \ 
\gamma(\btheta) \to \lambda \gamma(\btheta)
\end{equation}
with an arbitrary scalar constant $\lambda\ne 0$ 
\citep{1995A&A...294..411S}.
This transformation is equivalent to scaling 
the Jacobian matrix ${\cal A}(\btheta)$ with $\lambda$, 
$\cal {A}(\btheta) \to \lambda {\cal
A}(\btheta)$, and hence leaves the critical curves ${\rm det}{\cal
A}(\btheta)=0$ invariant.  
Furthermore, the curve $\kappa(\btheta)=1$, on which the gravitational
distortions disappear, is left invariant under the transformation
(\ref{eq:invtrans}).

This mass-sheet degeneracy can be unambiguously broken
by measuring the magnification effects (see \S~\ref{subsec:magbias}),
because the magnification $\mu$ transforms under the invariance
transformation 
(\ref{eq:invtrans}) as 
\begin{equation}
\mu(\btheta) \to \lambda^2 \mu(\btheta).
\end{equation}
Alternatively, the constant $\lambda$ can be determined such that
the mean $\kappa$ averaged over the outermost cluster region
vanishes, if a sufficiently wide sky coverage is available.\footnote{Or,
one may constrain the constant $\lambda$ such that the enclosed mass
within a certain aperture is consistent with cluster mass
estimates from some other observations
\citep[e.g.,][]{Umetsu+Futamase1999}.}  

\section{Cluster Weak Lensing Methodology}
\label{sec:method}

In this section we develop a Bayesian method to reconstruct the
projected cluster mass profile $\Sigma(\theta)$ from observable lens
distortion and magnification profiles,
without assuming particular functional forms for the mass
distribution, i.e., in a model-independent fashion. 
Although the methodology here is presented for the analysis of
individual clusters, it can be readily generalized for a statistical  
analysis using stacked lensing profiles of a sample of clusters.
In \S~\ref{subsec:stack}, alternatively, we provide a method to stack
reconstructed projected mass profiles of individual clusters to obtain
an ensemble-averaged profile.

\subsection{Lens Distortion Profile}
\label{subsec:gt} 

The observable quadrupole distortion of an object due to gravitational
 lensing is 
 described by the spin-2 reduced shear, $g=g_1+i g_2$ (equation
 [\ref{eq:redshear}]), which is coordinate dependent. For a given 
 cluster center
 on the sky, one can form coordinate-independent
 quantities, the tangential distortion $g_+$ and the $45^\circ$ rotated
 component, from linear combinations of the distortion coefficients
 as
$g_+ = -(g_1 \cos 2\phi + g_2\sin 2\phi)$ and
$g_{\times} = -(g_2 \cos 2\phi - g_1\sin 2\phi)$,
with $\phi$ being the position angle of an object with respect to
the cluster center.
In the strict weak-lensing limit, the
azimuthally-averaged tangential distortion profile $ g_+\approx
\gamma_+$  
satisfies the following identity \citep[e.g.,][]{2001PhR...340..291B}: 
$\gamma_+ (\theta) = \bar{\kappa}(<\theta)-\kappa(\theta)$,
where 
$\kappa(\theta)$ is the azimuthal average of $\kappa(\btheta)$ at
radius $\theta$,
and $\bar{\kappa}(<\theta)$ is the mean convergence 
interior to radius $\theta$.
With the assumption of quasi-circular symmetry in the
projected mass distribution \citep[see][]{UB2008},  the tangential
distortion is expressed as 
\begin{equation}
\label{eq:gt2kappa_nlin}
g_+(\theta) =
\frac{\bar{\kappa}(<\theta)-\kappa(\theta)}
{1-\kappa(\theta)}
\end{equation}
in the nonlinear but subcritical
(${\rm det}{\cal A}(\btheta)>0$) regime.\footnote{
In general, a wide spread of the redshift distribution of background
galaxies, in conjunction with the single-plane approximation,
may lead to an overestimate of the gravitational shear in
the nonlinear regime \citep{2000ApJ...532...88H}. 
Thanks to the deep Subaru photometry, we found that this bias in the
observed reduced shear is approximately $\Delta g/g\sim 0.02\kappa$ to
the first order of $\kappa$.  See \S~3.4 of \citet{Umetsu+2010_CL0024}
for details.
}
In the absence of higher order effects, weak
lensing only induces curl-free tangential distortions, 
while the azimuthal averaged $\times$ component
is expected to vanish. In practice, the presence of $\times$ modes can
be used to check for systematic errors.

From shape measurements of background galaxies,
we calculate the weighted average of 
$g_{+}$ in a set of $N$ radial bands ($i=1,2,...,N$) as 
\begin{equation}
\label{eq:gt}
g_+(\overline\theta_i) = 
\left(\displaystyle\sum_{k\in i} w_{(k)}\right)^{-1}
\left(\displaystyle\sum_{k\in i} w_{(k)}\, g_{+(k)}\right),
\end{equation}
where 
$\overline\theta_i$ is the center of the $i$th radial band
of $[\theta_i,\theta_{i+1}]$, 
the index $k$ runs over all of the objects located within the $i$th
annulus, 
$g_{+(k)}$ is the tangential distortion of the $k$th object,
and $w_{(k)}$ is the statistical weight 
\citep[see][]{UB2008,Umetsu+2009,Umetsu+2010_CL0024}
for the $k$th object, given by
\begin{equation}
w_{(k)} =\frac{1}{\sigma_{g(k)}^2+\alpha_g^2}
\end{equation}
with $\sigma_{g(k)}^2$ being the variance for the shear estimate of
the $k$th galaxy and
$\alpha_g^2$ being the softening constant variance
\citep[e.g.,][]{2003ApJ...597...98H}. 
In our analysis, we choose $\alpha_g=0.4$, which is
a typical value of the mean rms $\sigma_g$ over the background sample 
\cite[e.g.,][]{2003ApJ...597...98H,UB2008,Umetsu+2009,Umetsu+2010_CL0024}.  
We use the continuous limit of the
area-weighted band center for $\overline\theta_i$
(see equation [\ref{eq:medianr}]).
We perform a bootstrap error analysis to assess the uncertainty
$\sigma_{+}(\theta)$ in the tangential distortion profile $g_+(\theta)$.

\subsection{Magnification Bias Profile}
\label{subsec:magbias}

Lensing magnification, $\mu(\btheta)$, influences the observed surface
density of background sources, expanding the area of sky, and
enhancing the observed flux of background sources 
\citep{1995ApJ...438...49B}.
The former effect reduces
the effective observing area in the source plane, decreasing
the number of background sources per solid angle; on the other
hand, the latter effect amplifies the flux of background sources,
increasing the number of sources above the limiting flux.
The net effect is known as magnification bias, and depends on the
intrinsic slope of the luminosity function of background sources.

The number counts for a given magnitude cutoff $m_{\rm cut}$,
approximated locally as a power-law cut with slope $s=d\log_{10}
N(<m)/dm$ ($s>0$), are modified in the presence of lensing as 
\begin{equation}
\label{eq:magbias}
N(<m_{\rm cut})\approx
 N_{0}(<m_{\rm cut})\mu^{2.5s-1}
\end{equation}
\citep{1995ApJ...438...49B},
where $N_{0}(<m_{\rm cut})$ is the unlensed counts, and $\mu$ is the
magnification, $\mu=(1-\kappa)^{-2}(1-|g|^2)^{-1}$.  In the strict weak 
lensing limit, the magnification bias is $\delta N/N_0\approx
(5s-2)\kappa$.

For the number counts to measure magnification, we use a sample of
{\it red} background galaxies, for which the intrinsic count slope $s$
at faint magnitudes is relatively flat, $s\sim 0.1$, so that a net
count depletion results \citep{BTU+05,UB2008,Umetsu+2010_CL0024}.
On the other hand, the faint blue background population tends to have a
steeper intrinsic count slope close to the lensing invariant slope
($s=0.4$). 
The count-in-cell statistic $N(\btheta; <m_{\rm cut})$
is measured from a flux-limited sample of
red background galaxies on a regular grid of equal-area cells, 
each with a solid angle of $\Delta\Omega=(\Delta\theta)^2$.
Note that a practical difficulty of the
magnification bias measurement is contamination due to the intrinsic
clustering of background galaxies, 
which locally can be larger than the lensing-induced
signal in a given cell. In order to obtain a clean measure of the
lensing signal, such intrinsic clustering needs to be downweighted
\citep[e.g.,][]{1995ApJ...438...49B,2005PhRvL..95x1302Z}. 

For a mass profile analysis,
we calculate the mean number density 
$n_\mu(\overline\theta_i)=dN(\overline\theta_i)/d\Omega$ of  
the red background sample as a function of radius from the cluster
center, by azimuthally averaging $N(\btheta)$, using the same 
radial bins ($i=1,2,...,N$) as done for the distortion measurement.
The lens magnification bias is expressed in terms of the number density of
background galaxies 
as $n_\mu(\theta)=n_0\mu(\theta)^{2.5s-1}$ with $n_0$ being
the unlensed mean surface number density of background galaxies.
The normalization and slope parameters ($n_0,s$) can be
estimated from the source counts in cluster outskirts using wide-field
imaging data (\S \ref{subsec:back}). 
In practice, we adopt the following prescription:
\begin{itemize}
\item A positive tail of $>3\sigma$ cells is excluded in each annulus to
      remove inherent small scale clustering of the background
      \citep{BUM+08}. 
\item Each grid cell is weighted by the fraction of its area lying
      within the respective annular bins
      \citep{2002MNRAS.335.1037M,UB2008}. 
\item The uncertainty $\sigma_\mu(\theta)$ in $n_\mu(\theta)$ includes not
      only the Poisson 
      contribution but also the variance due to variations of the counts
      along the azimuthal direction, i.e., contributions from the
      intrinsic clustering of background galaxies \citep{Umetsu+2010_CL0024}.
\item The cell size $\Delta\theta$ 
can be as large as the typical radial band width for a mass profile
analysis, which can cause an additional variance due to Poisson and
sampling errors.
We thus average over a set of radial profiles obtained using 
different girds offset with respect to each other by half a grid spacing
in each direction.
\item The masking effect due to bright cluster galaxies, bright
foreground objects, and saturated pixels is properly taken into
      account and corrected for \citep{UB2008}.  In our analysis, we use
      Method B of Appendix \ref{appendix:mask} developed in this work.
\end{itemize}

\subsection{Bayesian Mass Profile Reconstruction}
\label{subsec:bayesian}

The relation between distortion and convergence is nonlocal,
and the convergence derived from distortion data alone
suffers from a mass-sheet degeneracy (\S~\ref{sec:basis}). However, by
combining the distortion and magnification measurements the
convergence can be obtained unambiguously with the correct
mass normalization.
Here we aim to derive a discrete convergence profile 
from observable lens distortion and 
magnification profiles (see \S~\ref{subsec:gt} and
\S~\ref{subsec:magbias}) within a Bayesian statistical framework, allowing
for a full parameter-space extraction of model and calibration parameters.
A proper Bayesian statistical analysis is of particular importance to
explore the entire parameter space and 
investigate the parameter degeneracies, arising in part from the
mass-sheet degeneracy.   

In this framework, we sample
from the posterior probability density function (PDF) of the underlying
signal  
$\bs$ given the data $\bd$, $P(\bs|\bd)$. 
Expectation 
values of any statistic of the signal $\bs$ shall converge to the
expectation values of the a posteriori marginalized PDF,
$P(\bs|\bd)$.  In our problem,
the signal $\bs$ is a vector containing 
the discrete convergence profile,  
$\kappa_i\equiv \kappa(\overline{\theta}_i)$ $(i=1,2,..,N)$, and 
the average convergence within the inner radial boundary $\theta_{\rm 
min}\equiv \theta_1$ of the weak
lensing data,  $\overline{\kappa}_{\rm min}\equiv
\overline{\kappa}(<\theta_{\rm min})$, so that $\bs
=\{\overline{\kappa}_{\rm min},\kappa_1,\kappa_2,...,\kappa_N\}$, being
specified by $(N+1)$ parameters.
The Bayes' theorem states that
\begin{equation}
P(\bs|\bd) \propto P(\bs) P(\bd|\bs),
\end{equation}
where ${\cal L}(\bs)\equiv P(\bd|\bs)$ is the likelihood of the data
given the model ($\bs$), and $P(\bs)$ is the prior probability
distribution for the model parameters.

\subsubsection{Weak Lensing Likelihood Function}

We combine complementary and independent weak lensing information of
tangential distortion and magnification bias to constrain the
underlying cluster mass distribution $\bs=\{\overline{\kappa}_{\rm
min},\kappa_i\}_{i=1}^{N}$.  The total likelihood function ${\cal L}$
of the model $\bs$ for
combined weak lensing observations is given as a product of the
two separate likelihoods, ${\cal L}={\cal L}_g{\cal L}_\mu$,
where ${\cal L}_g$ and ${\cal L}_\mu$ are the likelihood functions for
distortion and magnification, respectively.
The log-likelihood for the tangential distortion is given as
\begin{equation}
-\ln{{\cal L}_{g}} =
 \frac{1}{2}\sum_{i=1}^{N}
 \frac{\left[g_{+,i}-\hat{g}_{+,i}(\bs)\right]^2}{\sigma_{+.i}^2}
 +{\rm const}.,
\end{equation}
where $\hat{g}_{+,i}(\bs)$ is the theoretical prediction for the
observed distortion $g_{+,i}$, and 
the errors $\sigma_{+,i}$ ($i=1,2,...,N$) due primarily to the
variance of the intrinsic source ellipticity distribution
can be conservatively
estimated from the data using Bootstrap techniques (\S~\ref{subsec:gt}).
Similarly, 
the log-likelihood function for the magnification bias is given as
\begin{equation}
-\ln{{\cal L}_\mu} =  \frac{1}{2}\sum_{i=1}^{N}
 \frac{\left[n_{\mu,i}-\hat{n}_{\mu,i}(\bs)\right]^2}{\sigma_{\mu.i}^2}
 +{\rm const}.,
\end{equation}
where $\hat{n}_{\mu,i}(\bs)$ is the theoretical prediction for the
observed counts $n_{\mu,i}$, and 
the errors $\sigma_{\mu,i}$ include both contributions from 
Poisson errors in the counts, $\sigma_{{\rm Poisson},i}$, 
and contamination due to intrinsic
clustering of red background galaxies, $\sigma_{{\rm clust},i}$,
as discussed in
\S~\ref{subsec:magbias}:
\begin{equation}
\sigma_{\mu,i}^2 = \sigma^2_{{\rm Poisson},i} + \sigma^2_{{\rm clust},i}.
\end{equation}
The lensing observables $g_{+,i}$ and $n_{\mu,i}$ ($i=1,2,...,N$) can be
readily expressed 
by the given model parameters $\bs$, as shown in Appendices
\ref{appendix:avkappa} and \ref{appendix:lprof}.

\subsubsection{Prior Information}
\label{subsubsec:prior}


For each parameter of the model $\bs$, we consider a simple flat prior with a
lower bound of $\bs=0$, that is,
\begin{eqnarray}
\overline{\kappa}_{\rm min}&>&0, \\
\kappa_i &>&0 \ \ \ (i=1,2,...,N).
\end{eqnarray}
Additionally, we account for the calibration uncertainty in the
observational parameters, i.e., the normalization and slope parameters
$(n_0,s)$ of
the background counts and the relative lensing depth $\omega$
due to population-to-population variations between the background
samples used for the magnification and distortion measurements (see
Appendix \ref{appendix:lprof}).



\section{Applications: Subaru Observations of Five Strong-Lensing Clusters}
\label{sec:results}



\begin{deluxetable}{cccccc}
\centering
\tablecolumns{8}
\tablecaption{
 \label{tab:sample}
The Cluster Sample: Redshift and Subaru Filter Information
} 
\tablewidth{0pt} 
\tablehead{ 
 \multicolumn{1}{c}{Cluster} &
 \multicolumn{1}{c}{Redshift} &
 \multicolumn{1}{c}{Filters} &
 \multicolumn{1}{c}{Detection band} &
 \multicolumn{1}{c}{Seeing\tablenotemark{a}} &
 \multicolumn{1}{c}{Refs.}
\\
 \colhead{} &
 \colhead{} &
 \colhead{} &
 \colhead{} &
 \colhead{(arcsec)} &
 \colhead{}
}
\startdata  
A1689 & 0.183 & $i'V$& $i'$ & 0.82& 1\\
A1703 & 0.281 & $g'r'i'$& $r'$ & 0.78 & 2,3\\
A370  & 0.375 & $B_{\rm j}R_{\rm c}z'$& $R_{\rm c}$ & 0.60 & 2,3\\
Cl0024+17 & 0.395 & $B_{\rm j}R_{\rm c}z'$& $R_{\rm c}$ & 0.80 & 4\\
RXJ1347-11 & 0.451 & $VR_{\rm c}z'$& $R_{\rm c}$ & 0.76 & 2,3\\
\enddata
\tablenotetext{a}{Seeing FWHM in units of arcsec in the final co-added
 detection image.}
 \tablecomments{
For observational details, see the references.
Refs.~[2] and [3] adopted a cluster redshift of $z=0.258$
 for A1703. }
\tablerefs{ 
 [1] \cite{UB2008};
 [2] \cite{BUM+08};
 [3] \cite{Medezinski+2010}; 
 [4] \cite{Umetsu+2010_CL0024}.
 }
\end{deluxetable}


\begin{deluxetable}{ccc}
\tablecolumns{3}
\tablecaption{
 \label{tab:rein}
Einstein Radius Information
} 
\tablewidth{0pt} 
\tablehead{ 
 \multicolumn{1}{c}{Cluster} &
 \multicolumn{1}{c}{Einstein radius, $\theta_{\rm ein}$} &
 \multicolumn{1}{c}{Refs.}
\\
 \colhead{} &
 \colhead{(arcsec)} &
 \colhead{}
}
\startdata  
A1689 & $53\pm 3\arcsec$ $(z_s=3.04)$\tablenotemark{a}& 1,2 \\
A1703 & $31\pm 3\arcsec$ $(z_s=2.627)$& 3, 4, 5 \\ 
A370  & $37\pm 3\arcsec$ $(z_s=2)$ & 6, 7\\
Cl0024+17  & $30\pm 3\arcsec$ $(z_s=1.675)$ & 8\\
RXJ1347-11 & $35\pm 2\arcsec$ $(z_s=2.2)$& 7, 9, 10\\
\enddata
\tablenotetext{a}{This Einstein radius constraint translates into
 $\theta_{\rm ein}\simeq 47\arcsec$ at $z_s=2$ (cf. $\theta_{\rm
 ein}=47.0\pm 1.2\arcsec$ at $z_s=2$ by Ref.~[2]).}
\tablerefs{ 
 [1] \cite{2005ApJ...621...53B};
 [2] \cite{Coe+2010};
 [3] \cite{Limousin+2008_A1703};
 [4] \cite{Richard+2009_A1703};
 [5] \cite{Zitrin+2010_A1703};
 [6] \cite{Richard+2010_A370}; 
 [7] Zitrin et al. (2011), in preparation;
 [8] \cite{Zitrin+2009_CL0024};
 [9] \cite{Halkola+2008_RXJ1347};
 [10] \cite{Bradac+2008_RXJ1347}.
 }
\end{deluxetable}

 
\begin{deluxetable*}{c|ccc|ccc|c} 
\tabletypesize{\scriptsize}
\tablecolumns{10}
\tablecaption{ 
 \label{tab:back}
Background galaxy samples
} 
\tablewidth{0pt} 
\tablehead{ 
 \multicolumn{1}{c|}{Cluster} &
 \multicolumn{3}{c|}{Distortion analysis (full background)} &
 \multicolumn{3}{c|}{Magnification analysis (red background)} &
 \multicolumn{1}{c}{$\omega$\tablenotemark{d}} 
\\
 \colhead{} &
 \multicolumn{1}{|c}{$\overline{n}$\tablenotemark{a}} &
 \multicolumn{1}{c}{$\overline{z}_{s,\beta}$\tablenotemark{b}} &
 \multicolumn{1}{c}{$\langle \beta\rangle$\tablenotemark{c}} &
 \multicolumn{1}{|c}{$\overline{n}$} &
 \multicolumn{1}{c}{$\overline{z}_{s,\beta}$} &
 \multicolumn{1}{c|}{$\langle\beta\rangle$} &
 \multicolumn{1}{c}{} 
\\
 \colhead{} &
 \multicolumn{1}{|c}{(${\rm arcmin}^{-2}$)} &
 \multicolumn{1}{c}{} &
 \multicolumn{1}{c}{} &
 \multicolumn{1}{|c}{(${\rm arcmin}^{-2}$)} &
 \multicolumn{1}{c}{} &
 \multicolumn{1}{c|}{} &
 \multicolumn{1}{c}{}
}
\startdata 
 A1689\tablenotemark{a}  & 
 8.8&
 $0.71\pm 0.12$ &
 $0.69$ &
 12.0 &
  $0.71\pm 0.12$  &
 $0.69$ &
 $1.00$
\\
 A1703  & 
 10.0&
 $1.10\pm 0.18$ &
 $0.68$ &
 6.9 &
 $0.93\pm 0.12$ &
 $0.64$ &
 $0.95$
\\
 A370\tablenotemark{b}  & 
 16.7&
 $1.11\pm 0.12$ &
 $0.58$ &
 21.6 &
 $1.05\pm 0.11$ &
  $0.57$ &
 $0.98$
\\
 Cl0024+17  & 
17.2&
 $1.29\pm 0.15$ &
 $0.61$ &
 17.8 &
 $1.14 \pm 0.09$ &
 $0.56$ &
 $0.92$
\\
 RXJ1347-11  & 
6.4&
 $0.98\pm 0.06$ &
 $0.47$ &
 7.7 &
 $1.04\pm 0.07$ &
 $0.49$ &
 $1.05$\\ 
\enddata
\tablenotetext{a}{Mean surface number density of background galaxies.}
\tablenotetext{b}{Effective mean source redshift of the background
 sample corresponding to the
 mean depth $\langle\beta \rangle$, defined such that
 $\beta(\overline{z}_{s,\beta})= \langle\beta \rangle.$} 
\tablenotetext{c}{Distance ratio averaged over the redshift
 distribution of the background sample, taken from
 \citet[][A1689]{UB2008}, \citet[][A1703, A370, RXJ1347-11]{Medezinski+2010} and \citet[][Cl0024+17]{Umetsu+2010_CL0024}.}
\tablenotetext{d}{Relative lensing depth of the red background sample,
 $\omega=\langle\beta({\rm red}) \rangle/\langle\beta({\rm full})\rangle$,
 with respect to the full background sample.}
\tablecomments{
We assume $5\%$ uncertainty in our estimates of the relative lensing
 depth $\omega$.
}
\end{deluxetable*}


\begin{deluxetable}{cccc}
\tablecolumns{4}
\tablecaption{
 \label{tab:magbias}
Normalization and slope of red background counts
} 
\tablewidth{0pt} 
\tablehead{ 
 \multicolumn{1}{c}{Cluster} &
 \multicolumn{1}{c}{$m_{\rm cut}${a}} &
 \multicolumn{1}{c}{Normalization, $n_0$\tablenotemark{b}} &
 \multicolumn{1}{c}{Slope, $s$\tablenotemark{c}}
\\
 \colhead{} &
 \colhead{(AB mag)} &
 \colhead{(${\rm arcmin}^{-2}$)} &
 \colhead{}
}
\startdata  
A1689      & $i'=25.5$ & $11.5\pm 0.3$  & $0.124 \pm 0.060$\\
A1703      & $i'=26$   & $ 8.2\pm 0.2$  & $0.099 \pm 0.070$\\
A370       & $z'=26$   & $22.5\pm 0.4$  & $0.042 \pm 0.070$\\
Cl0024+17  & $z'=25.5$ & $18.3 \pm 0.3$ & $0.121 \pm 0.040$\\
RXJ1347-11 & $z'=26$   & $ 7.0 \pm 0.3$ & $0.025 \pm 0.070$\\
\enddata
\tablenotetext{a}{Fainter magnitude cut of the red background sample,
 $m_{\rm cut}$.}
\tablenotetext{b}{Normalization of unlensed red background counts,
 $n_0$.}
\tablenotetext{c}{Slope of the unlensed red background counts at $m_{\rm
 cut}$, $s=d\log_{10}N_0(<m_{\rm cut})/dm$.}
\end{deluxetable}


\begin{figure*}[!htb] 
 \begin{center}
$
\begin{array}{c@{\hspace{.1in}}c@{\hspace{.1in}}c@{\hspace{.1in}}c}
 \includegraphics[width=45mm,angle=0]{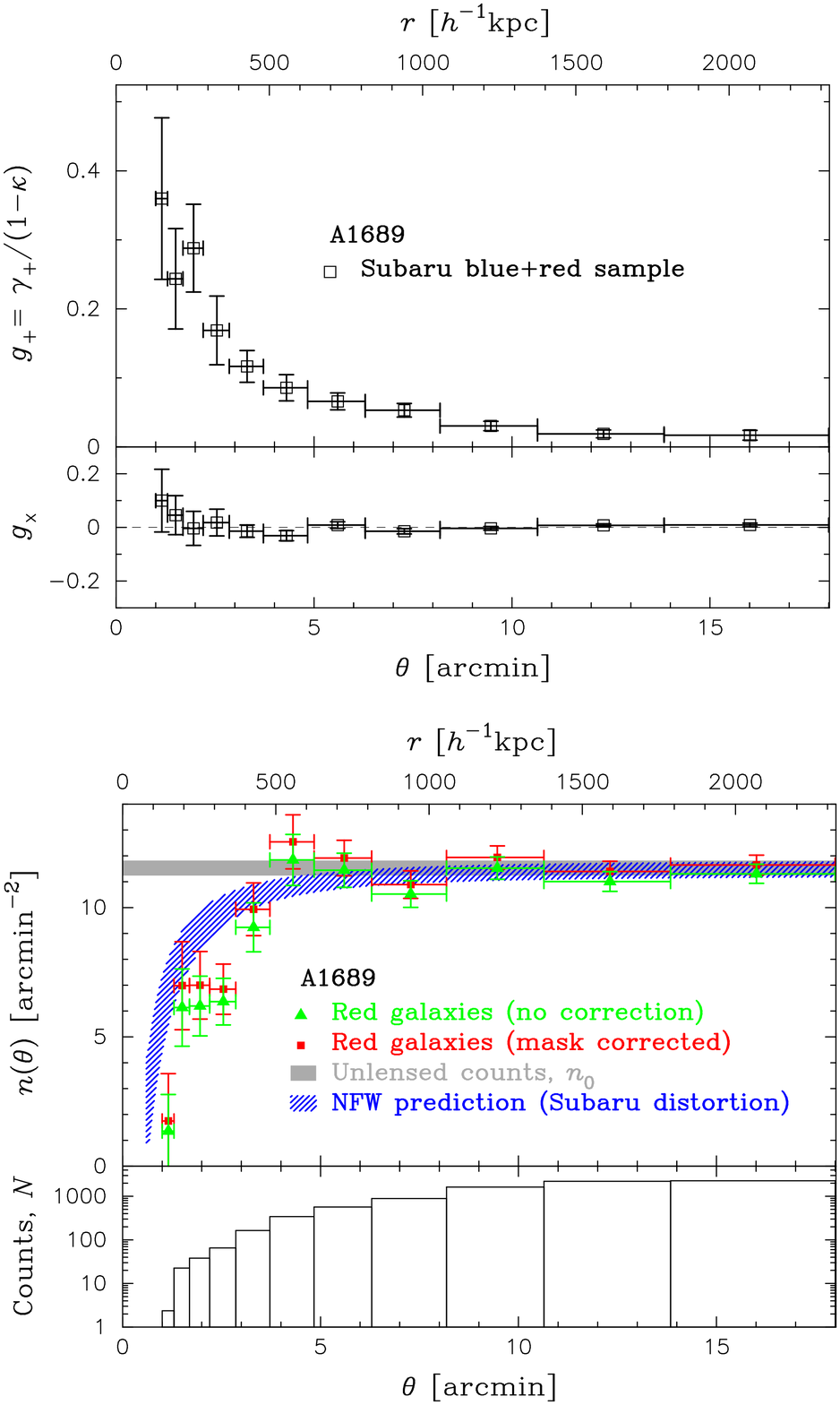} &
 \includegraphics[width=45mm,angle=0]{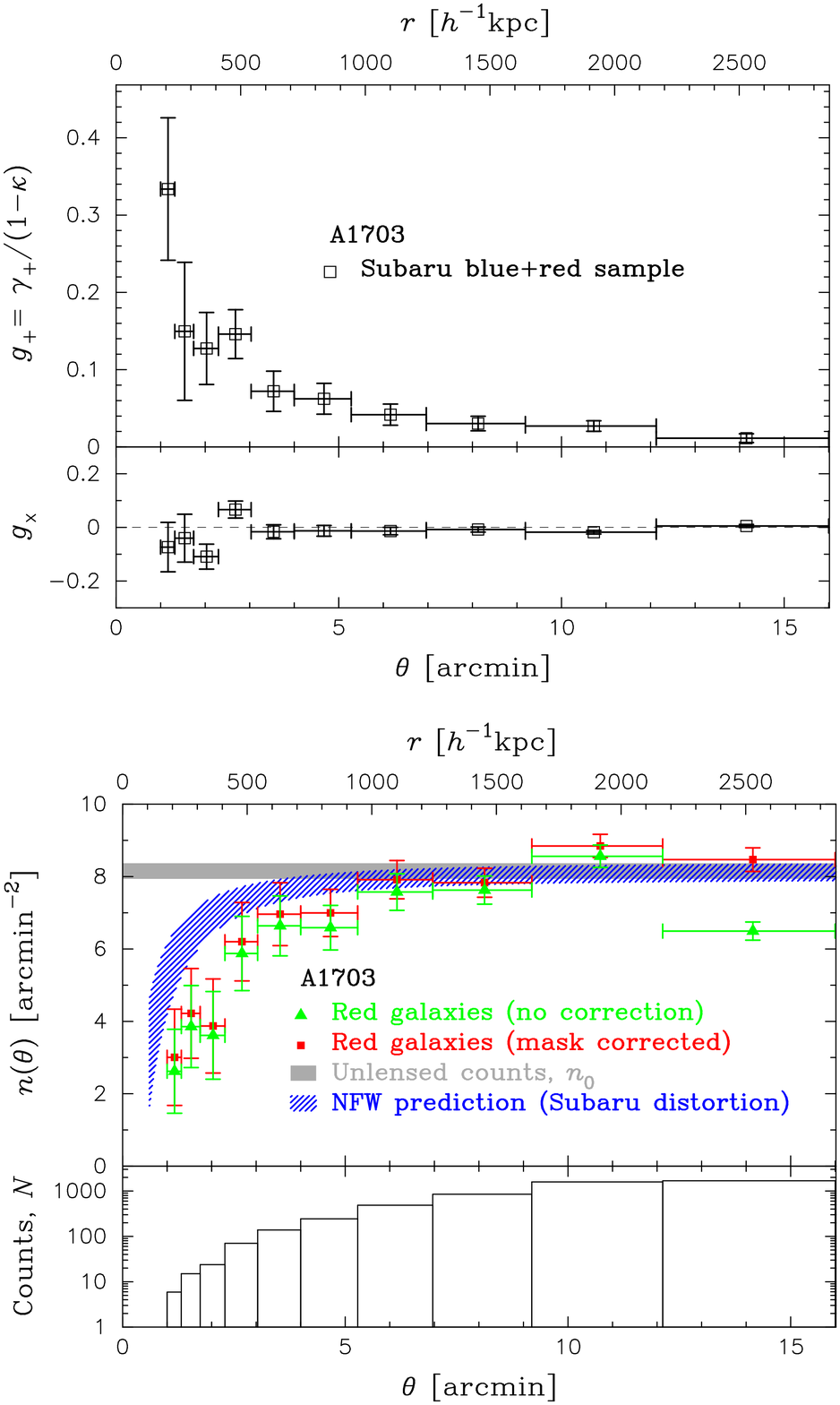} &
 \includegraphics[width=45mm,angle=0]{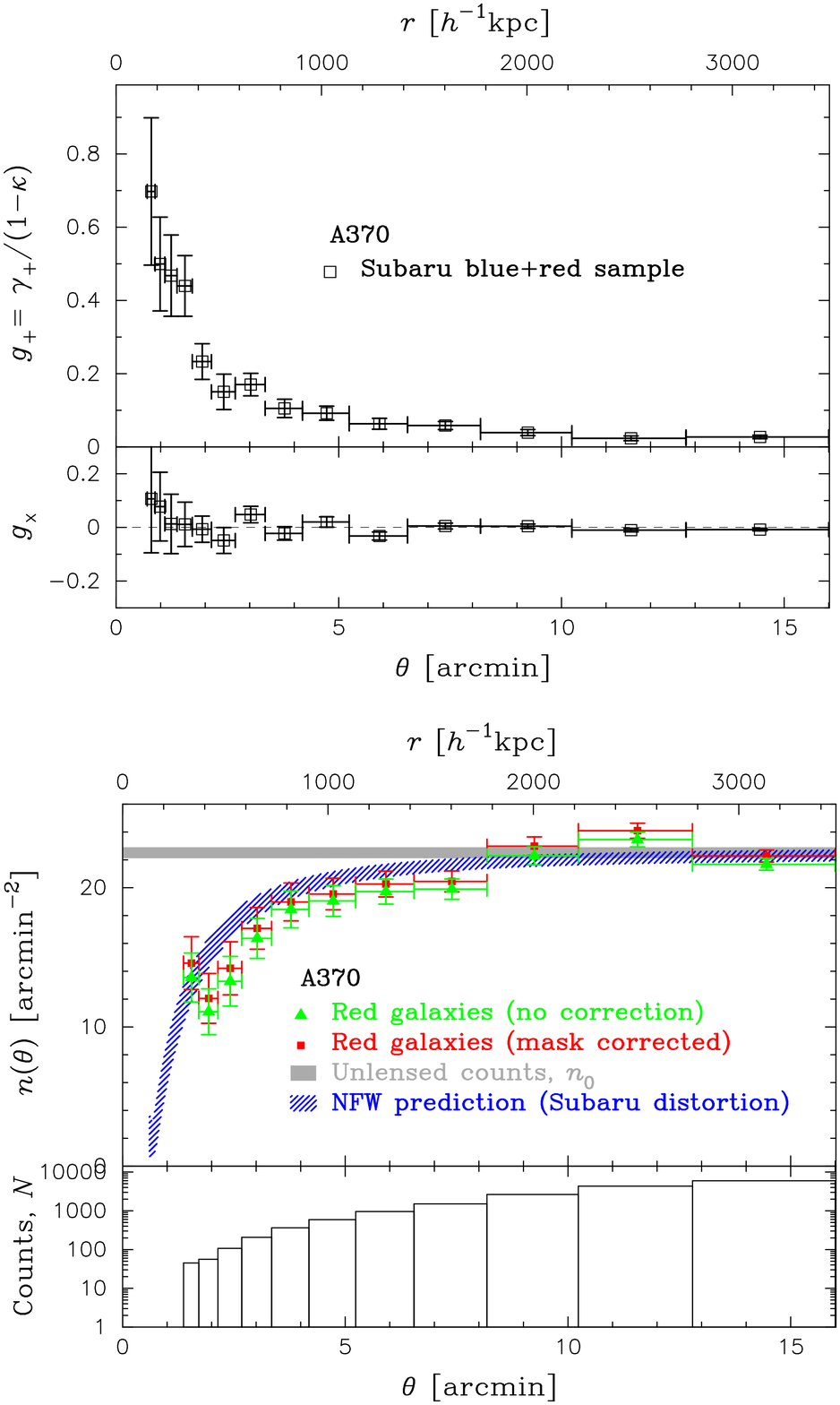} 
\end{array}
$
$
\begin{array}{c@{\hspace{.1in}}c@{\hspace{.1in}}c}
 \includegraphics[width=45mm,angle=0]{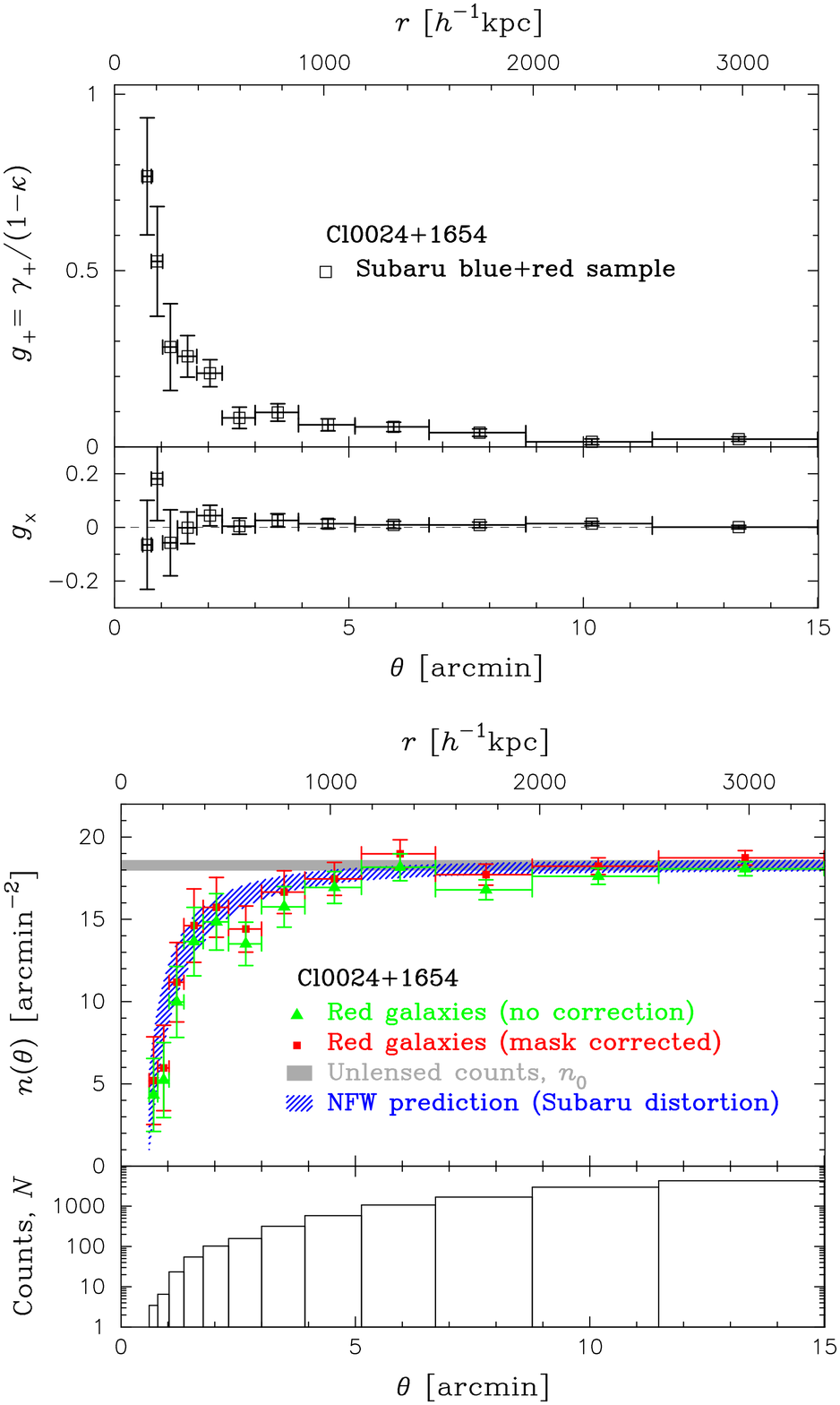} &
 \includegraphics[width=45mm,angle=0]{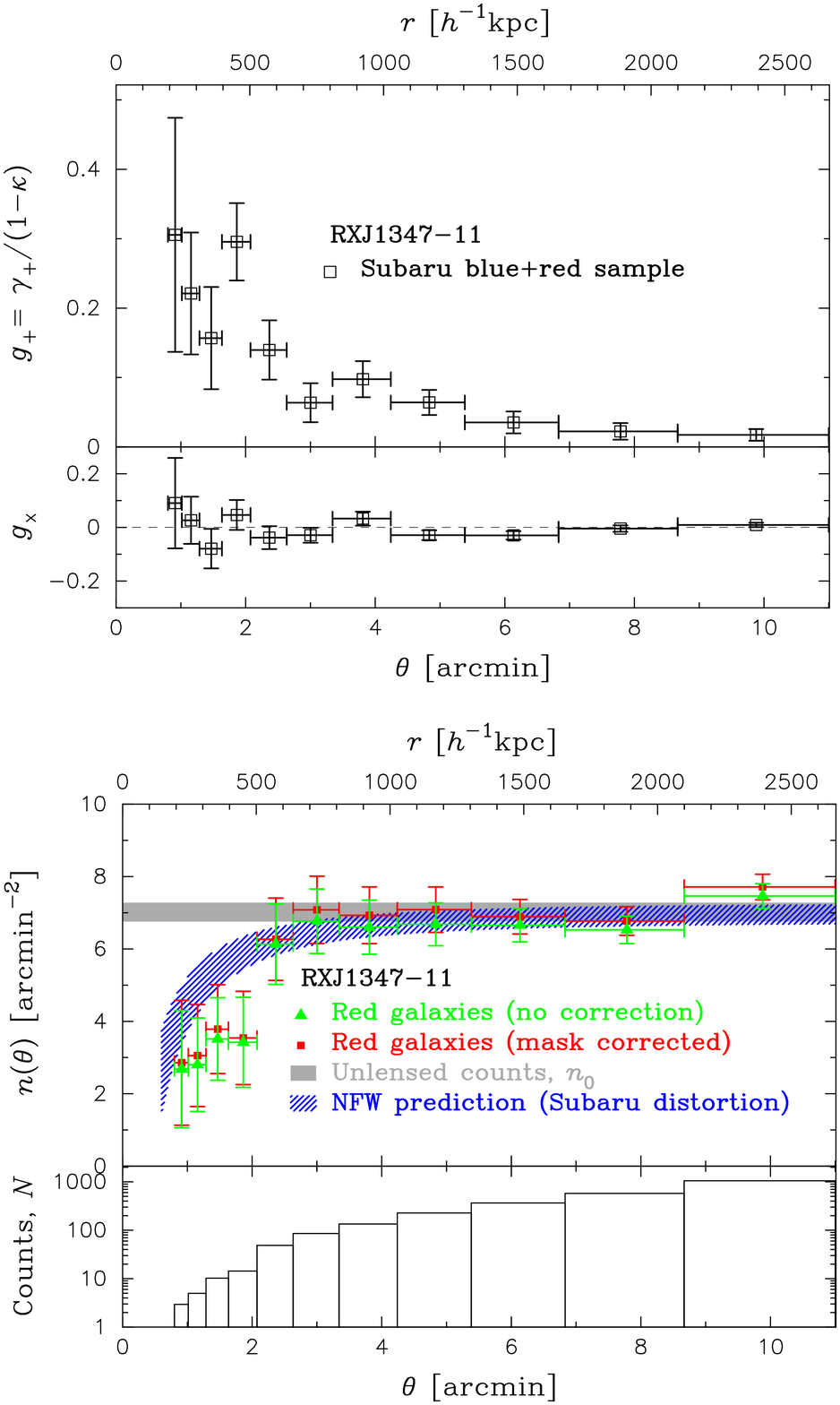} 
\end{array}
$
 \end{center}
\caption{
Radial profiles of the averaged spin-2 distortion and magnification bias
 measurements based on background galaxies registered in deep Subaru
 images, shown separately for the five clusters, A1689 (top-left), A1703
 (top-middle), A370 (top-right), Cl0024+17 (bottom-left), and RXJ1347-11
 (bottom-right). 
For each cluster, the top panels show the radial profiles (open squares
 and error bars) of the 
tangential distortion (upper panel) and the $45^\circ$ 
rotated ($\times$) component (lower panel) based on the spin-2 shape
measurements of the full background galaxy sample (Table
 \ref{tab:back}).  
The bottom panels show the count profiles of red background galaxies.
The squares and triangles (upper panel) show the
 respective results with and without the mask 
correction due to bright foreground objects and cluster
 members. 
The blue hatched area
represents the 68.3\% confidence bounds for the predicted count
depletion curve from an NFW model constrained by our Subaru distortion
analysis, demonstrating clear consistency between these two independent
lensing observables.
The gray horizontal bar represents the constraints on the unlensed count
 normalization $n_0$.
The histogram in the lower panel shows the observed counts of the red
 background galaxies in each annular bin.
\label{fig:data}
} 
\end{figure*} 


\begin{figure*}[!htb] 
 \begin{center}
$
\begin{array}
  {c@{\hspace{.1in}}c@{\hspace{.1in}}c@{\hspace{.1in}}c@{\hspace{.1in}}c@{\hspace{.1in}}c@{\hspace{.1in}}c}
 \includegraphics[width=20mm,angle=270]{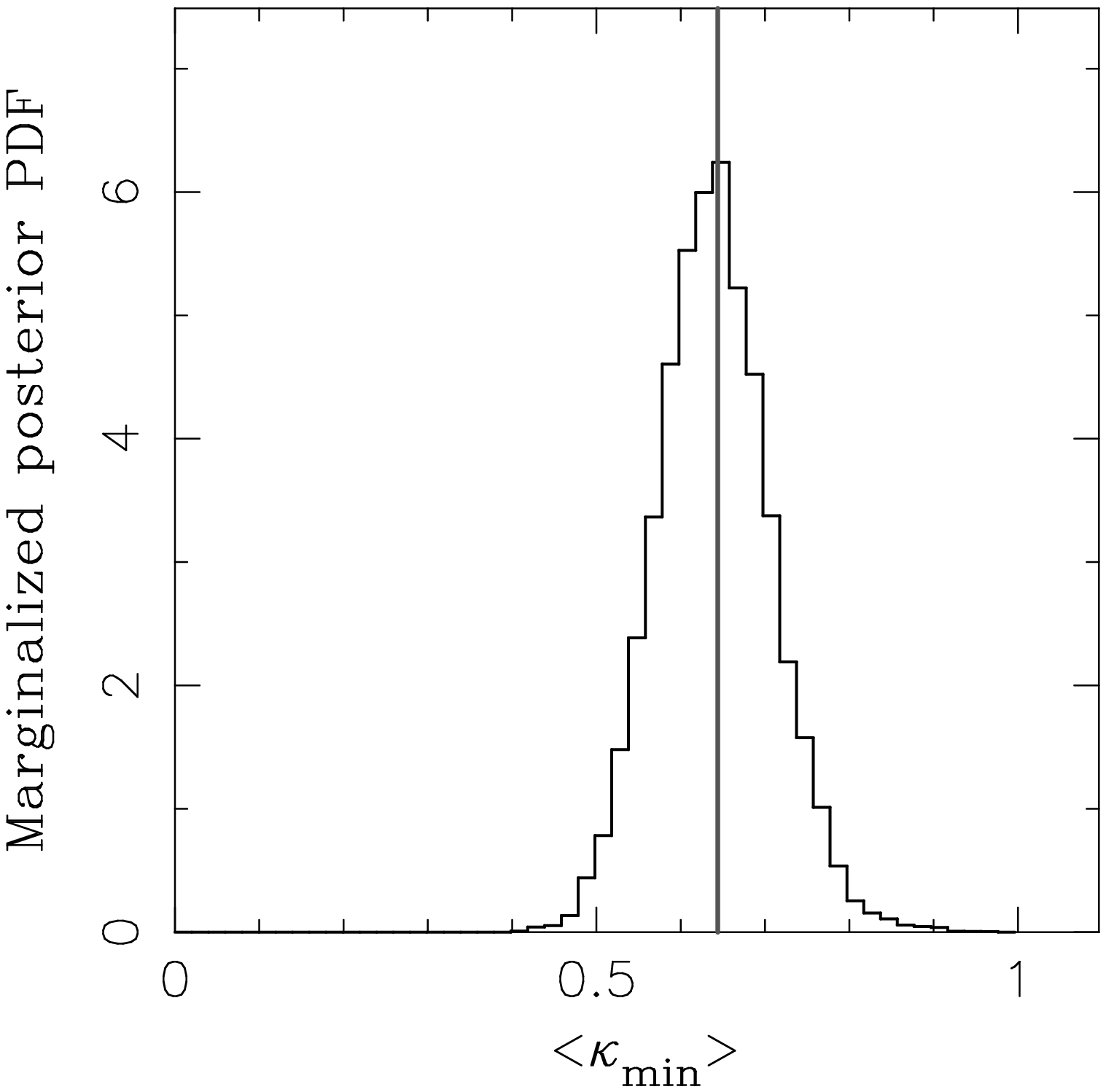} &
 \includegraphics[width=20mm,angle=270]{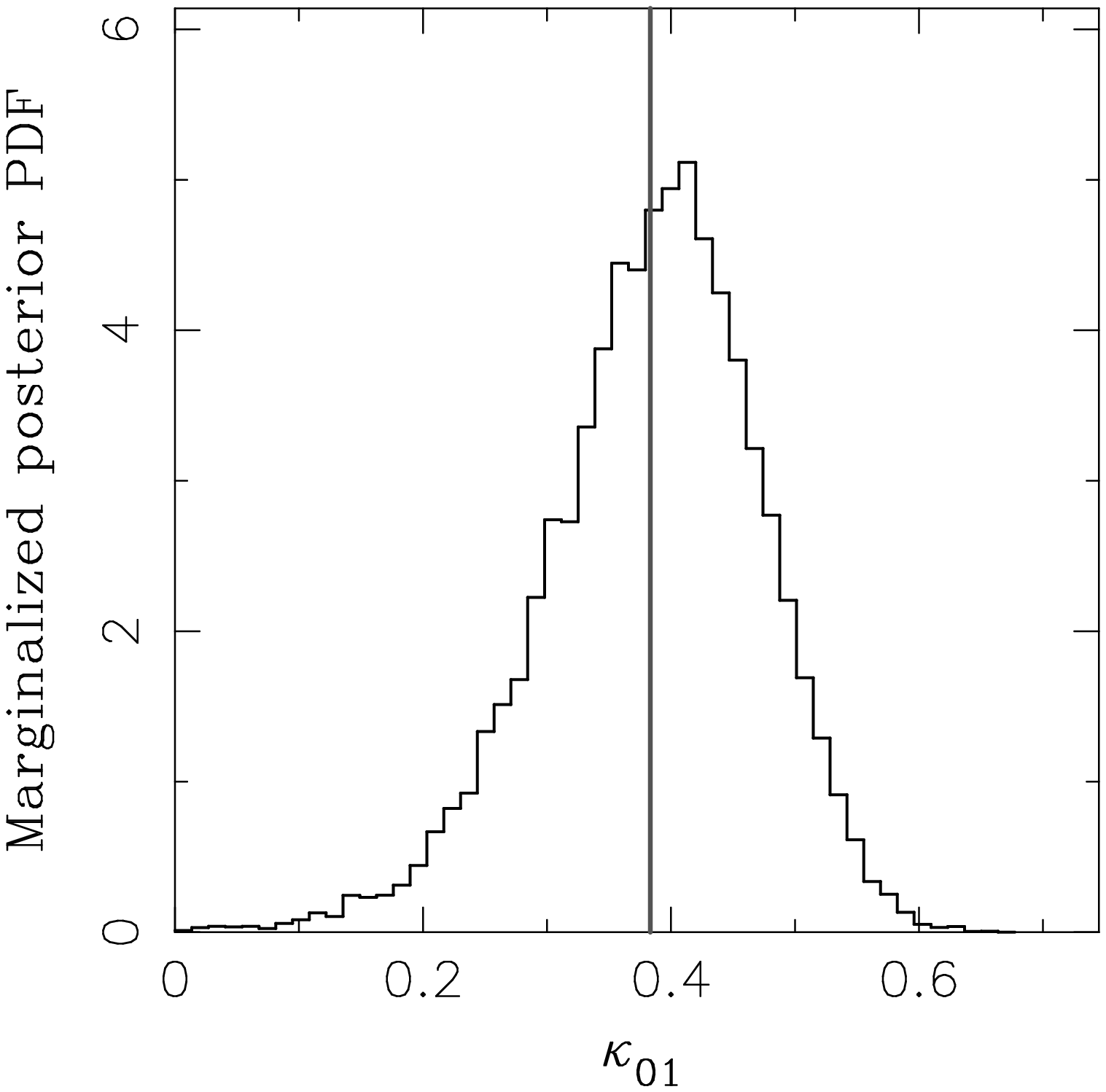} &
 \includegraphics[width=20mm,angle=270]{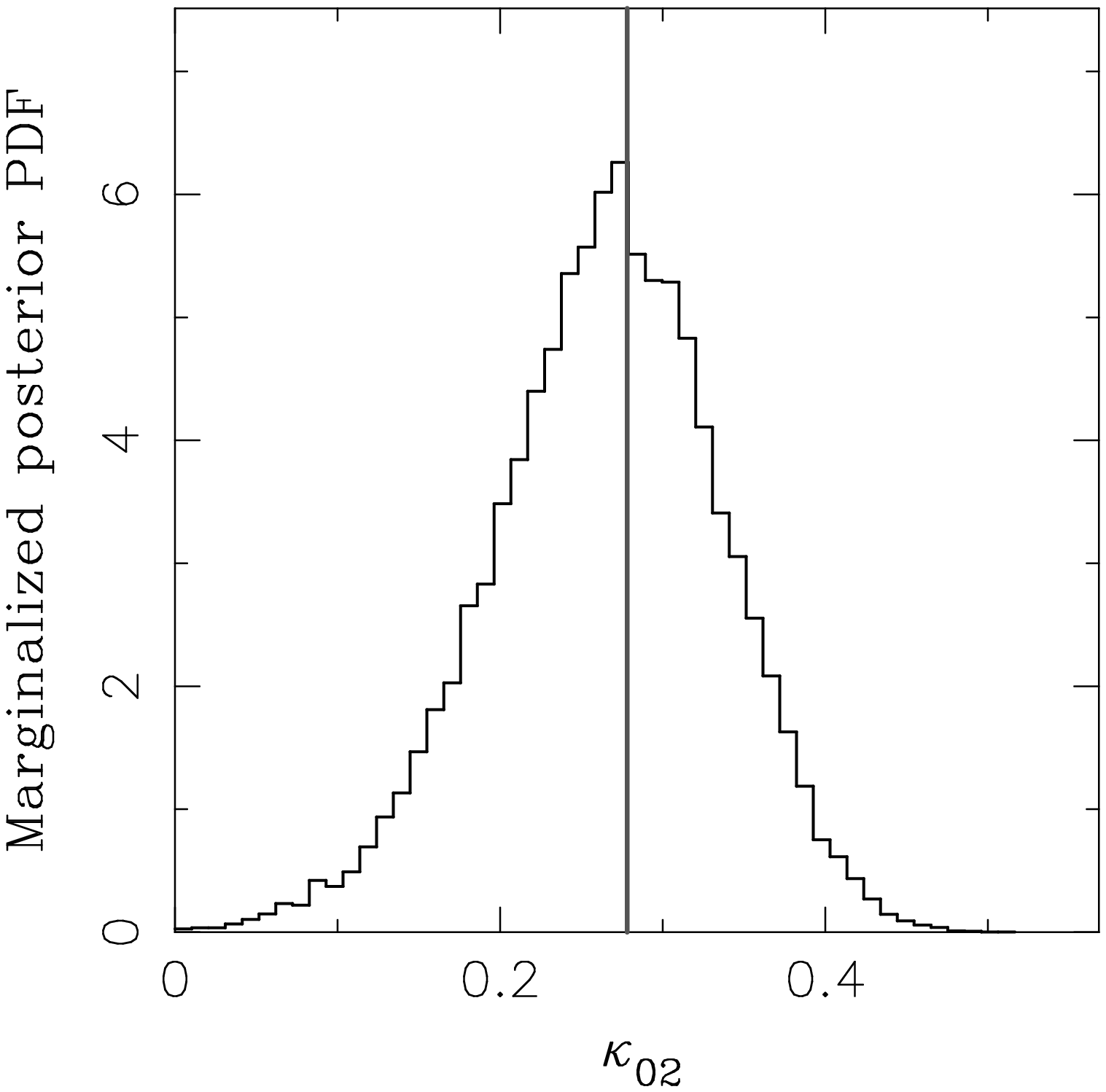} &
 \includegraphics[width=20mm,angle=270]{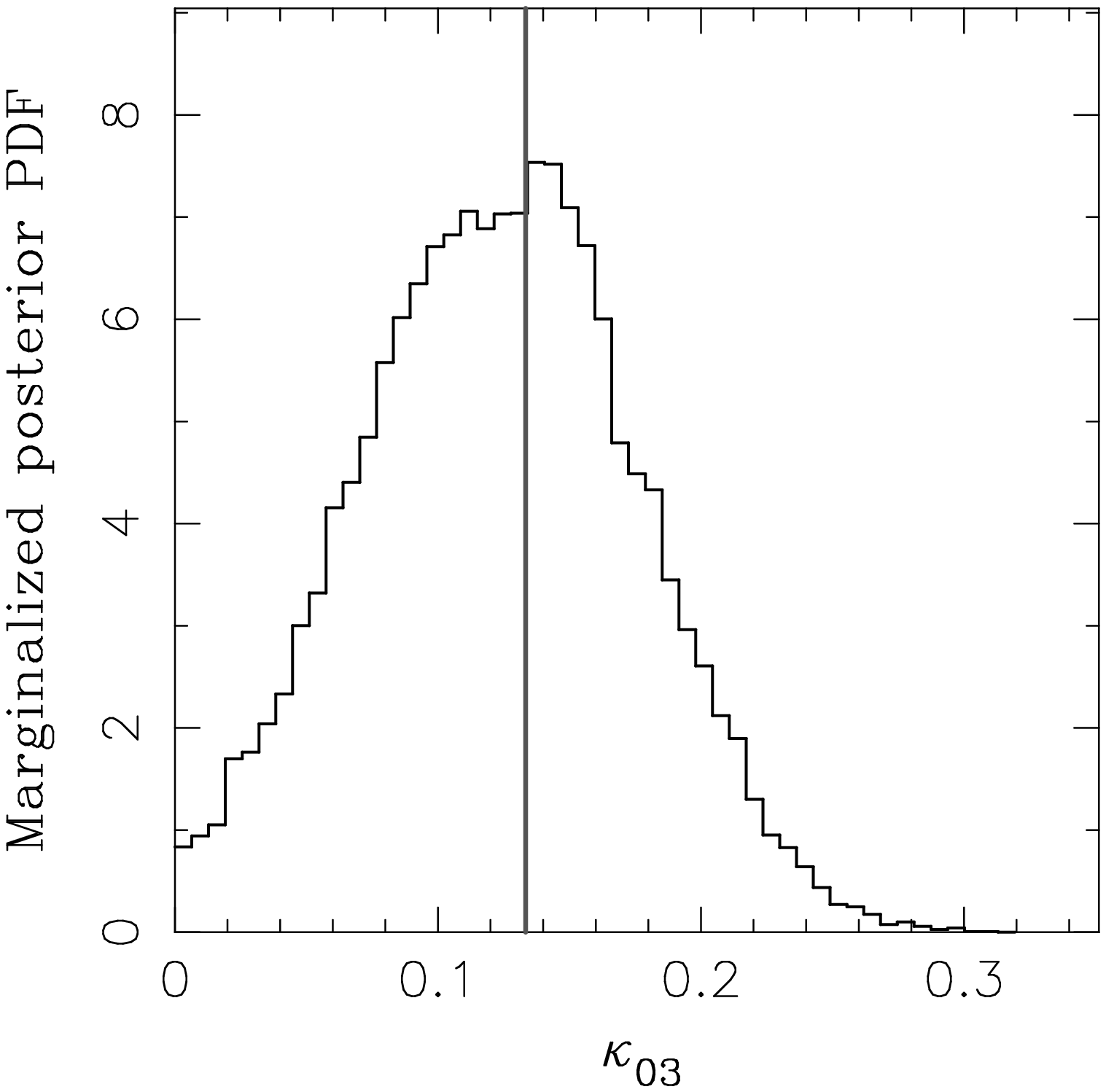} &
 \includegraphics[width=20mm,angle=270]{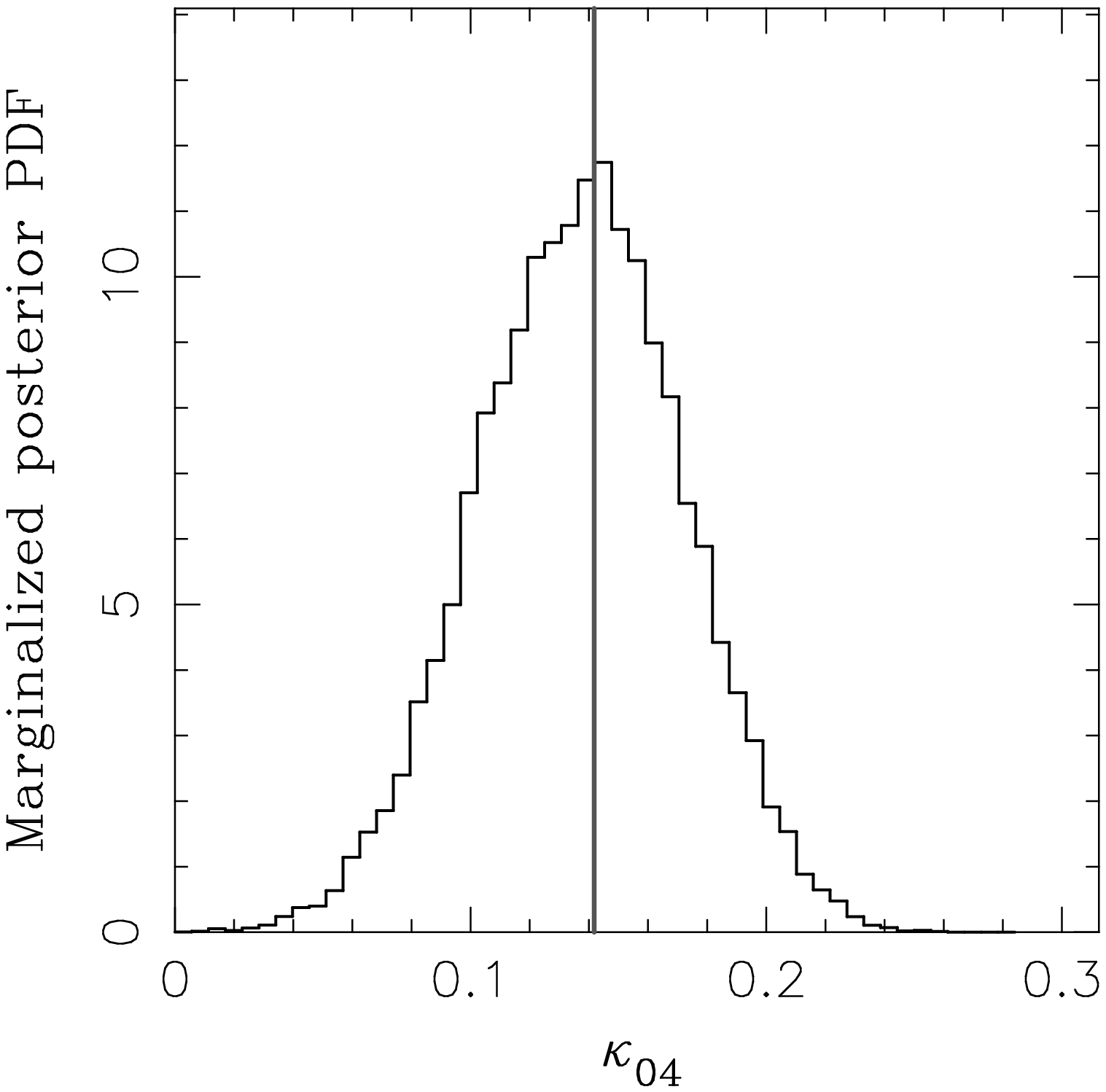} &
 \includegraphics[width=20mm,angle=270]{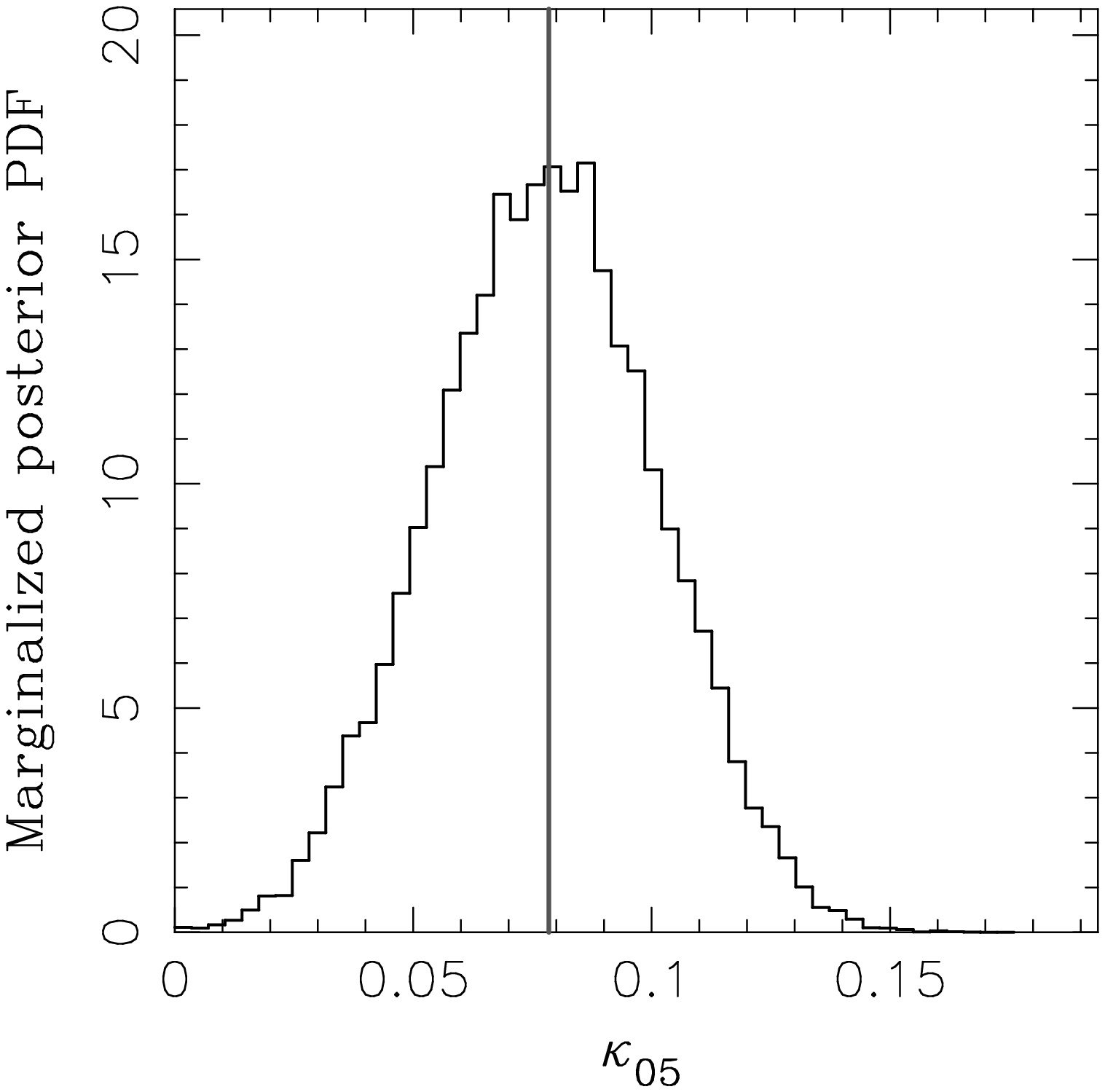}\\
\end{array}
$
$
\begin{array}
  {c@{\hspace{.1in}}c@{\hspace{.1in}}c@{\hspace{.1in}}c@{\hspace{.1in}}c@{\hspace{.1in}}c@{\hspace{.1in}}c}
 \includegraphics[width=20mm,angle=270]{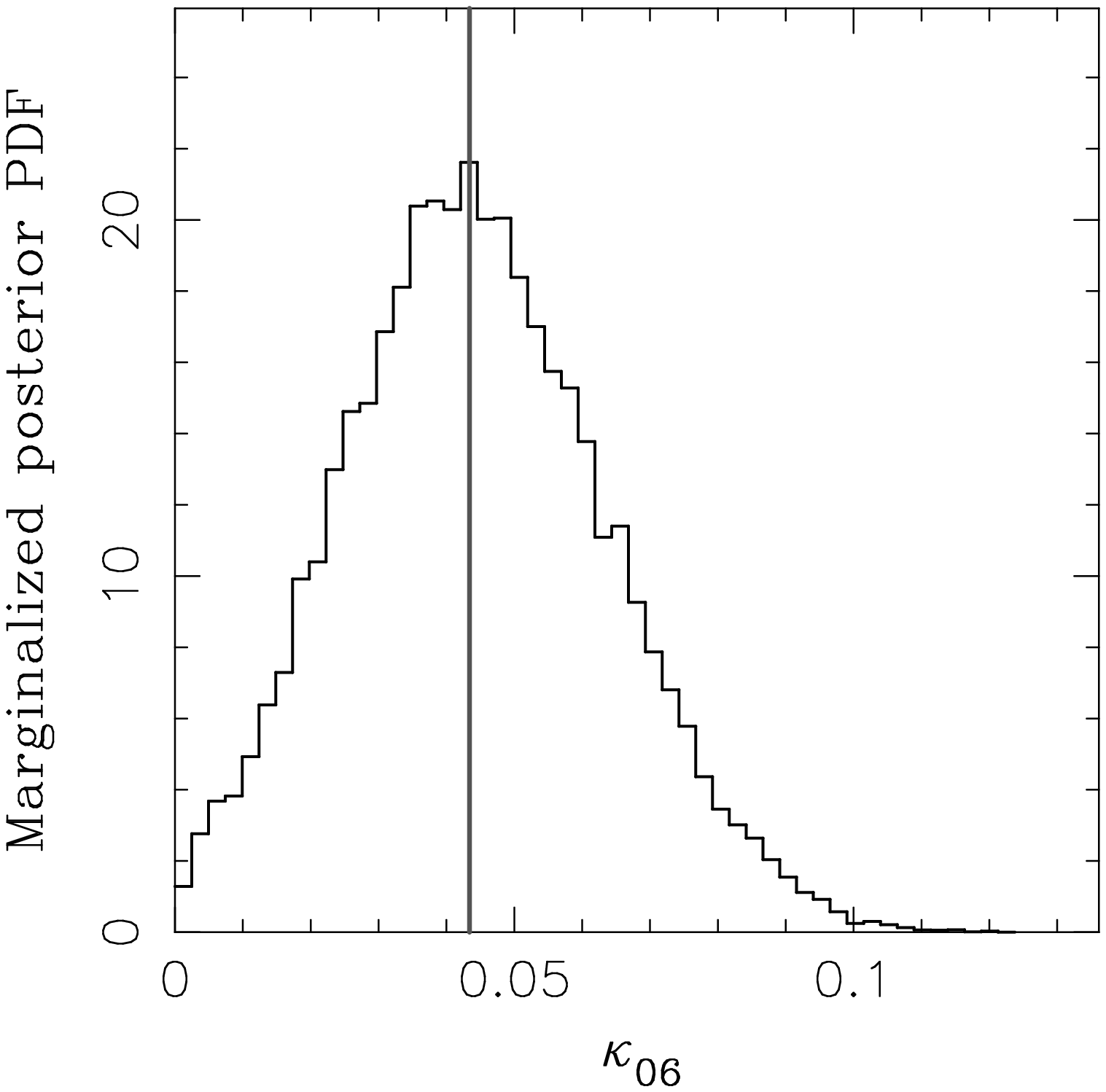} &
 \includegraphics[width=20mm,angle=270]{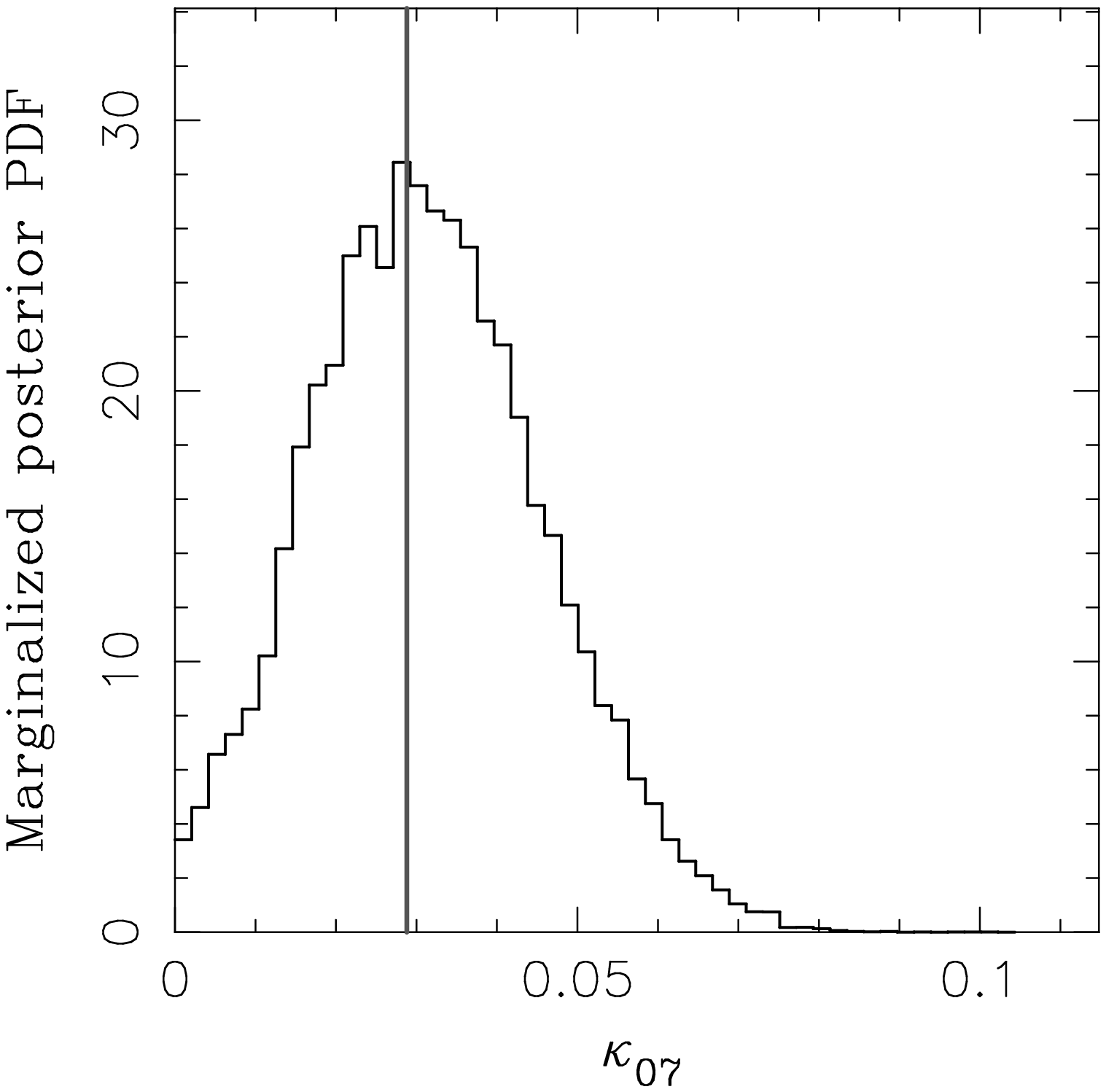} &
 \includegraphics[width=20mm,angle=270]{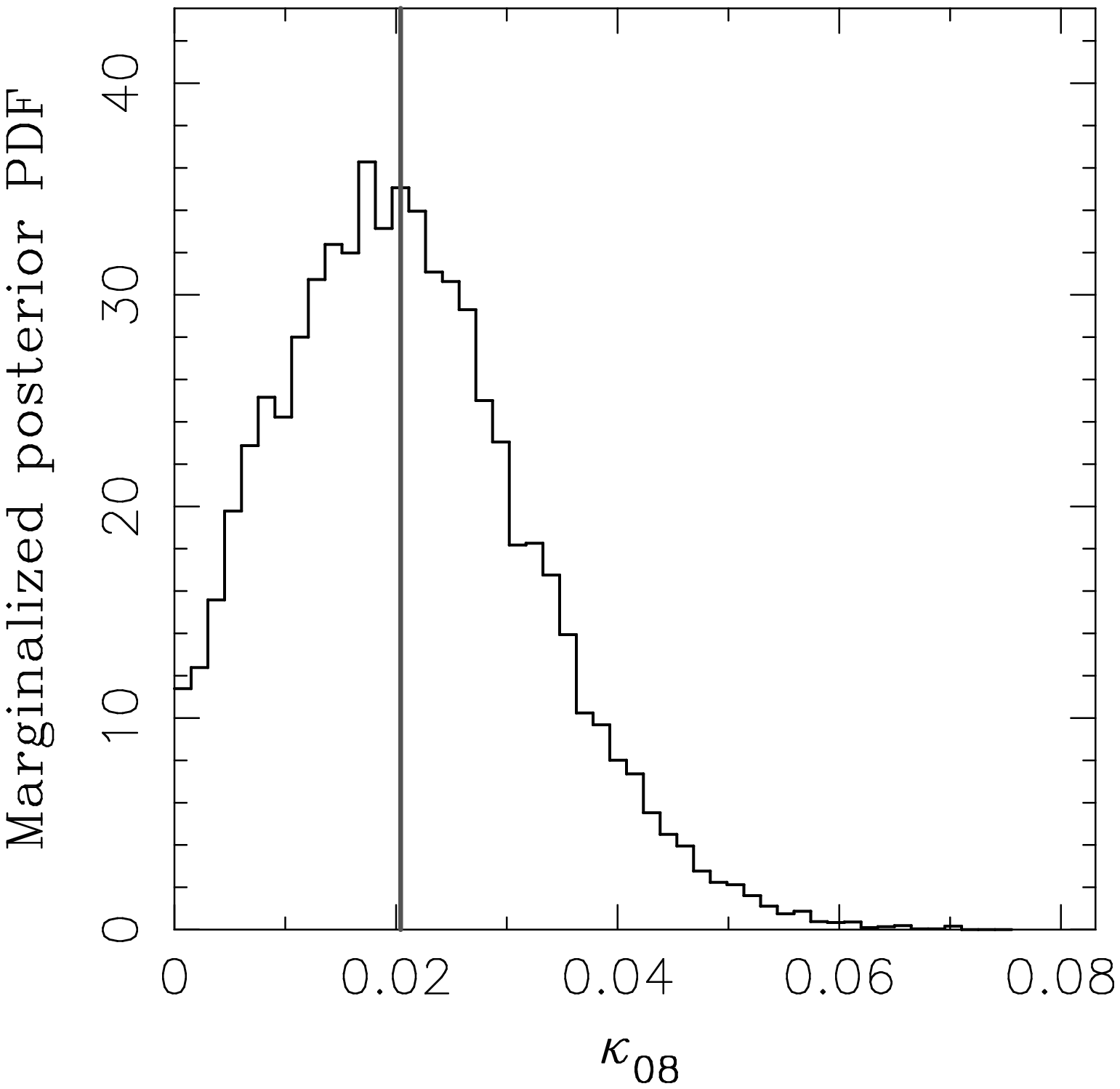} &
 \includegraphics[width=20mm,angle=270]{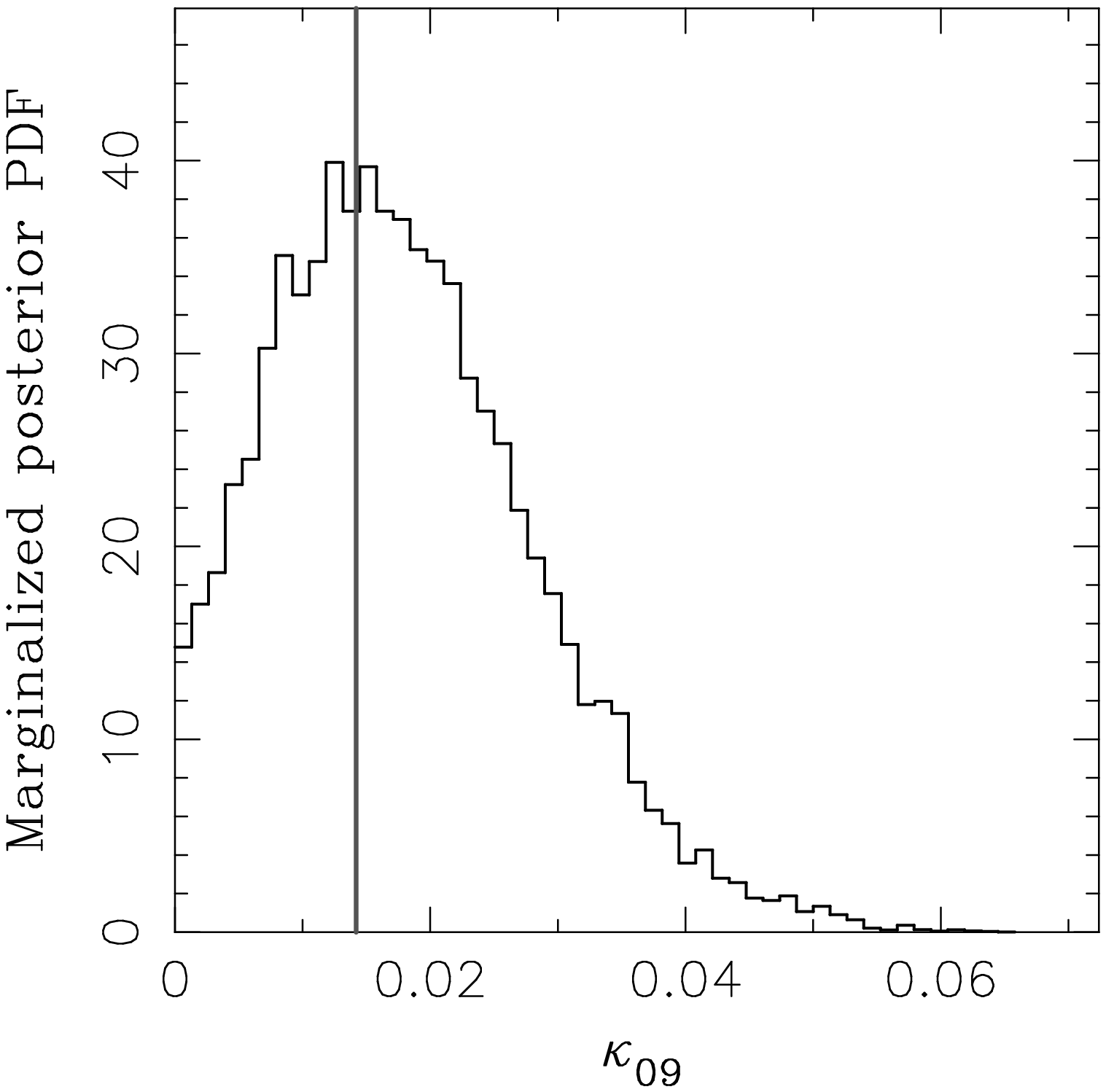} &
 \includegraphics[width=20mm,angle=270]{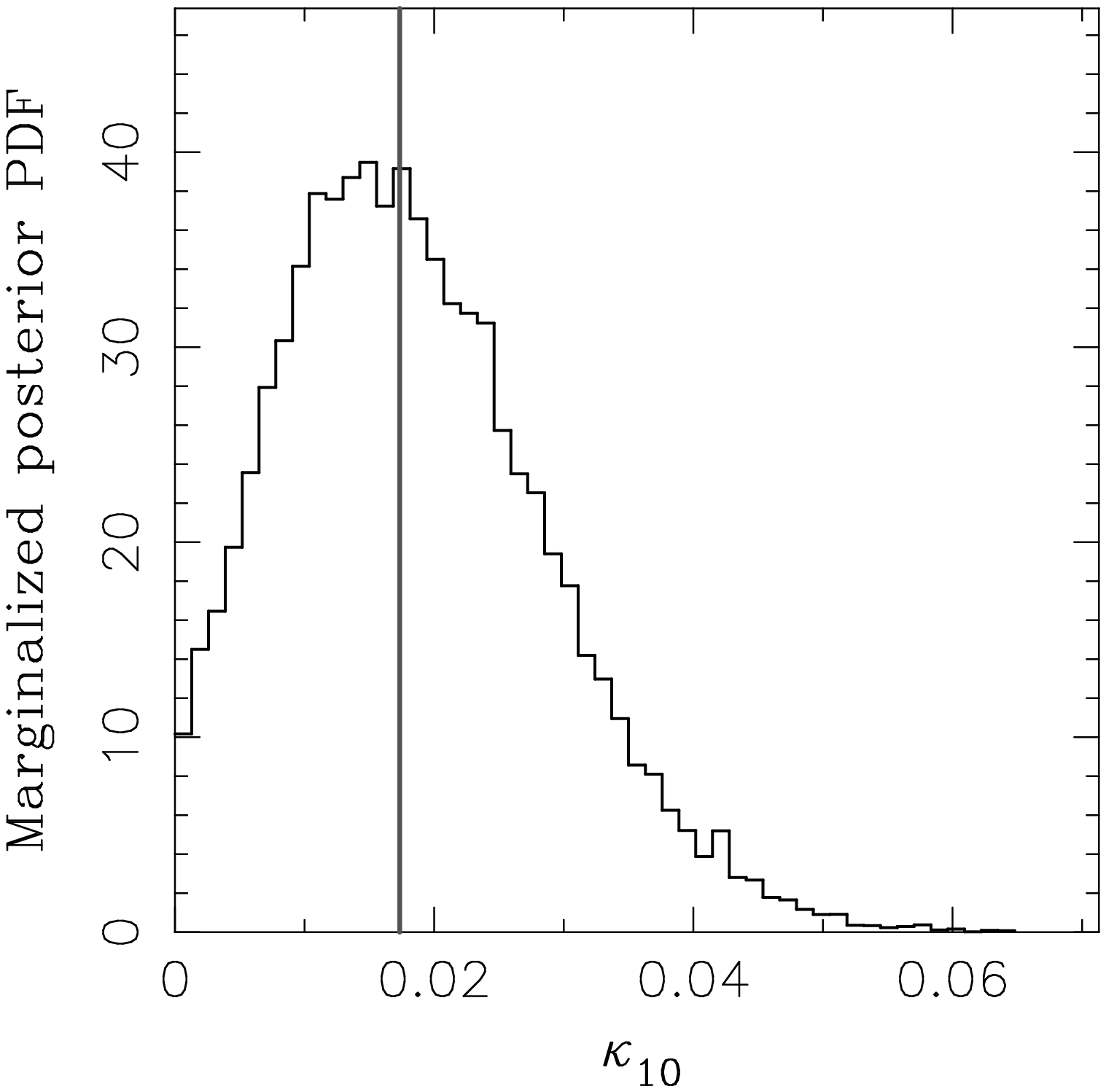} &
 \includegraphics[width=20mm,angle=270]{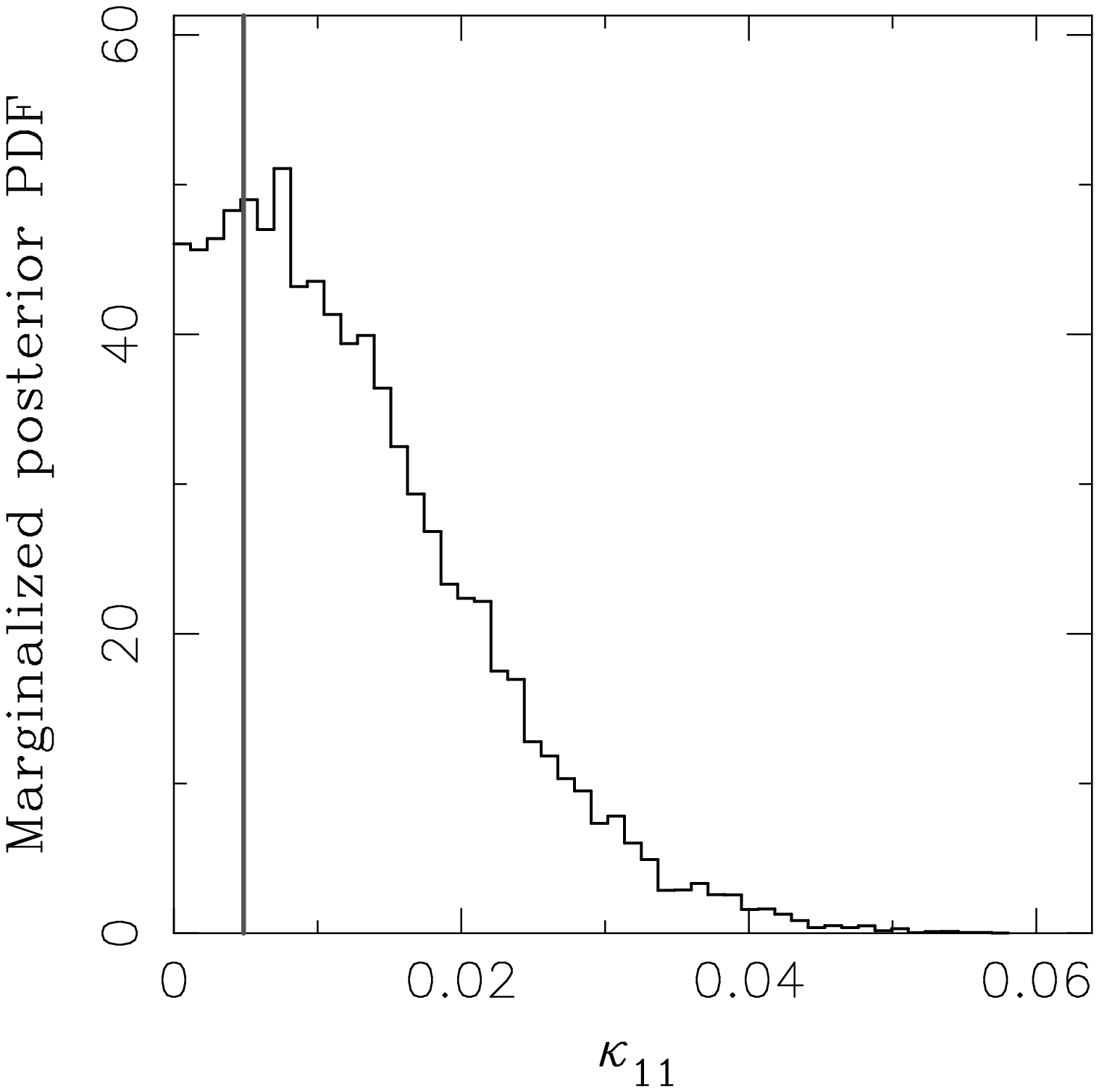}\\
\end{array}
$
 \end{center}
\caption{
One-dimensional marginalized posterior probability density functions (PDFs)
 for the discrete mass profile $\bs=(\overline{\kappa}_{\rm
 min},\kappa_1,\kappa_2,...,\kappa_N)$, shown for
 A1689 (see Figure \ref{fig:kplot}). The solid vertical lines
 show the peak locations (mode values) 
of the a posteriori marginalized distribution for each parameter.
The results are marginalized over all other parameters, including the
 observational parameters $(n_0,s,\omega)$.
The resulting posterior distributions are all single-peaked, and
 nearly Gaussian for most of
 the model parameters.
The mass-sheet degeneracy is broken thanks to the
 inclusion of magnification data.
\label{fig:post}
} 
\end{figure*}


\begin{figure*}[!htb] 
 \begin{center}
$
\begin{array}{c@{\hspace{.1in}}c@{\hspace{.1in}}c@{\hspace{.1in}}c}
 \includegraphics[width=40mm,angle=270]{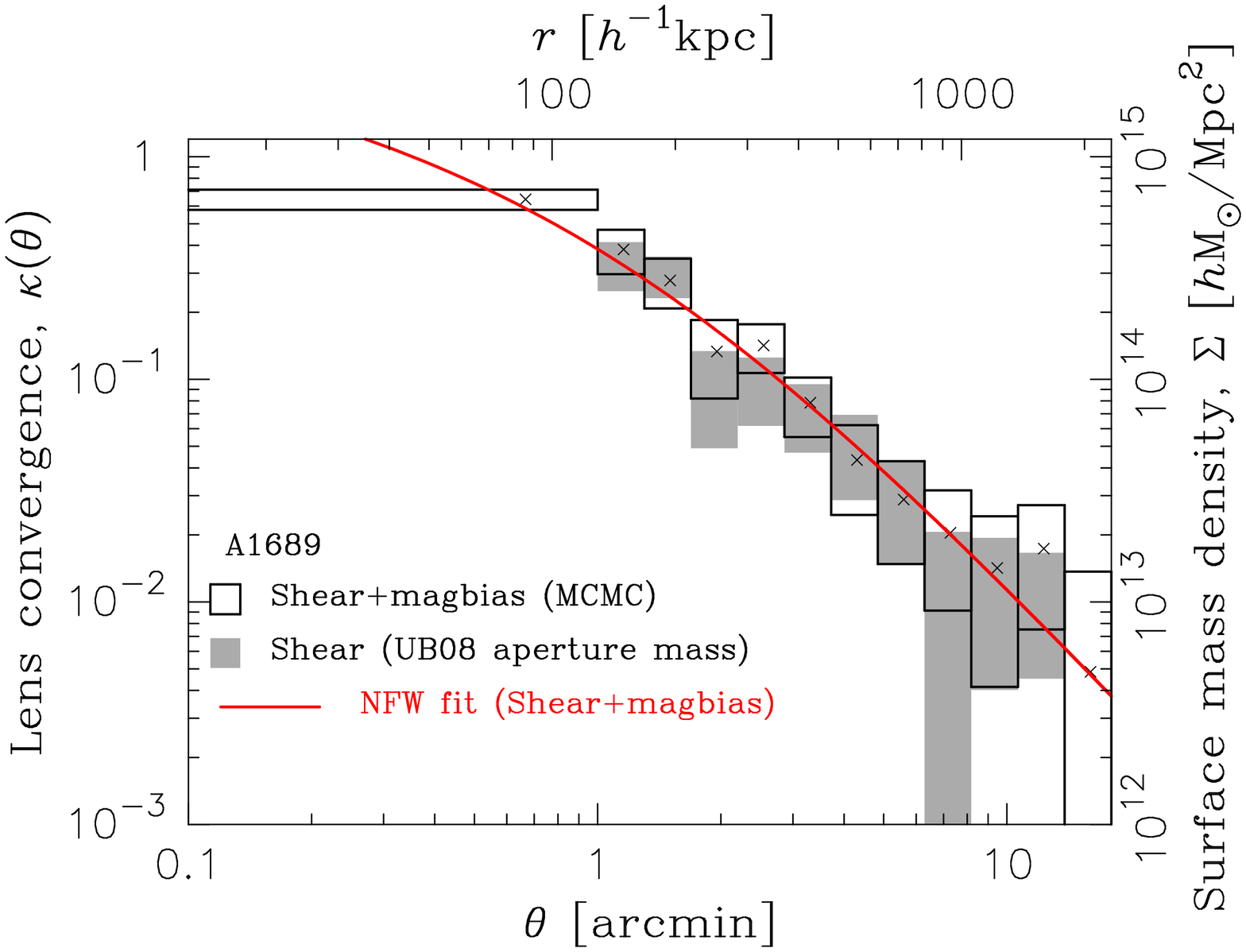} &
 \includegraphics[width=40mm,angle=270]{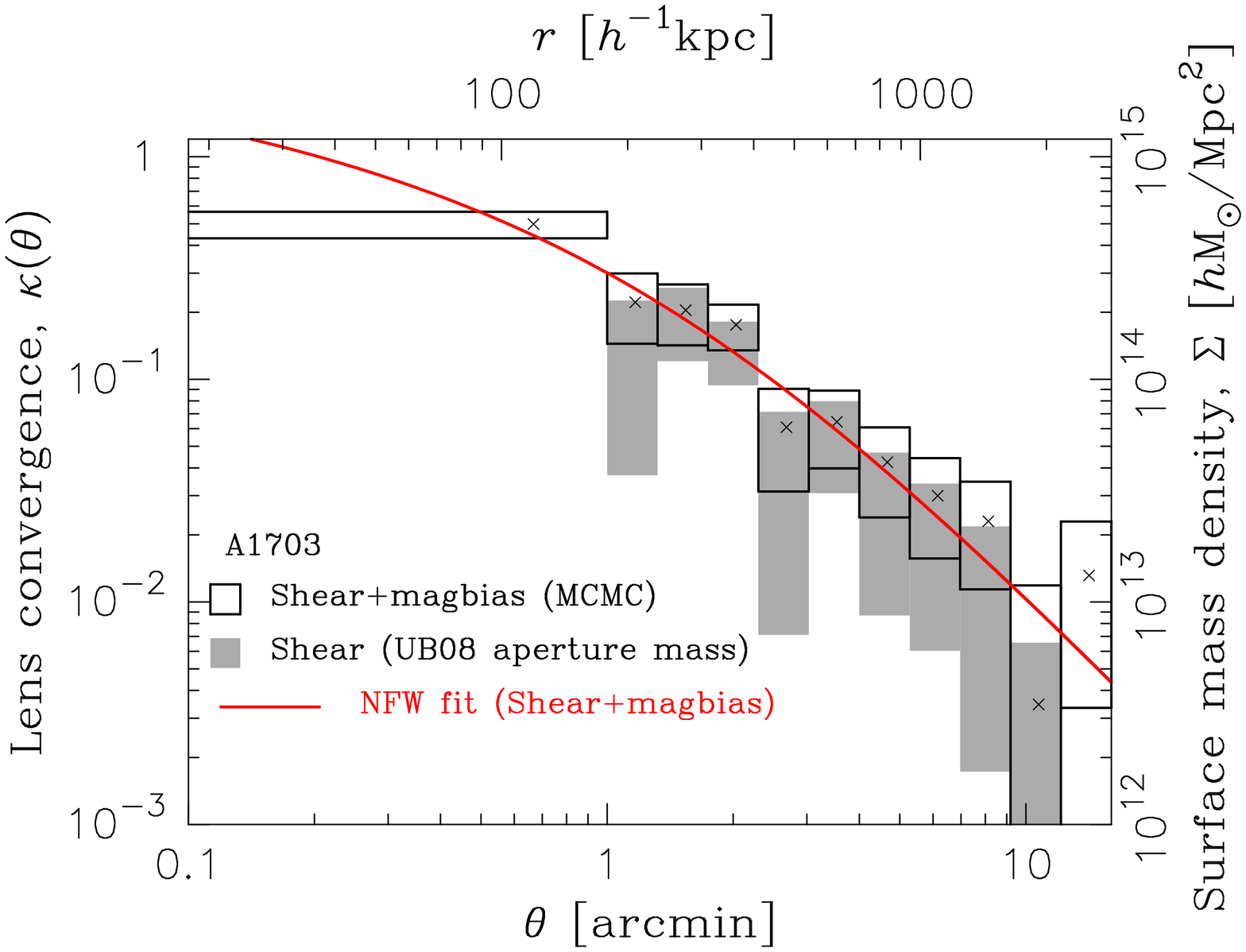} &
 \includegraphics[width=40mm,angle=270]{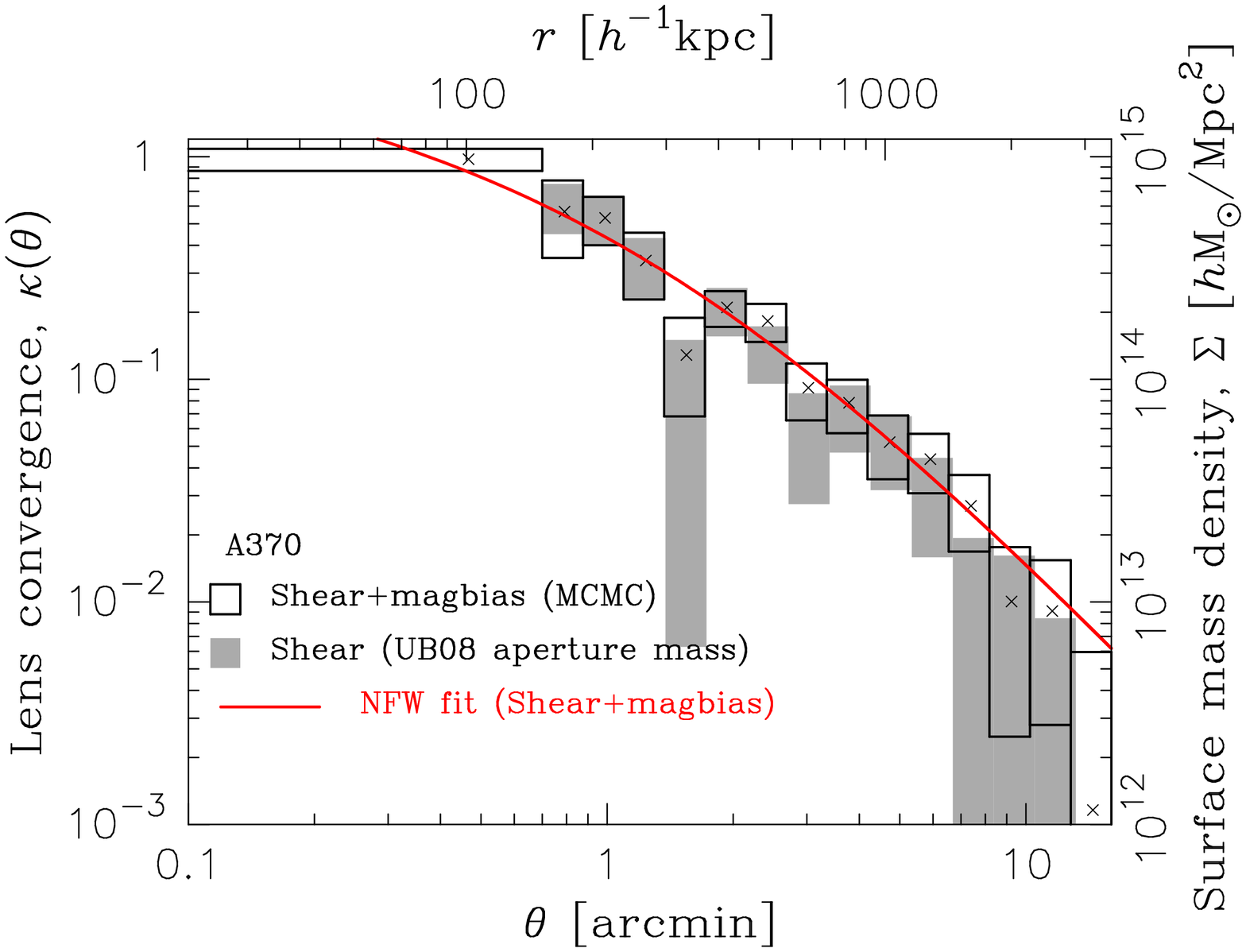} \\
\end{array}
$
$
\begin{array}{c@{\hspace{.1in}}c@{\hspace{.1in}}c}
 \includegraphics[width=40mm,angle=270]{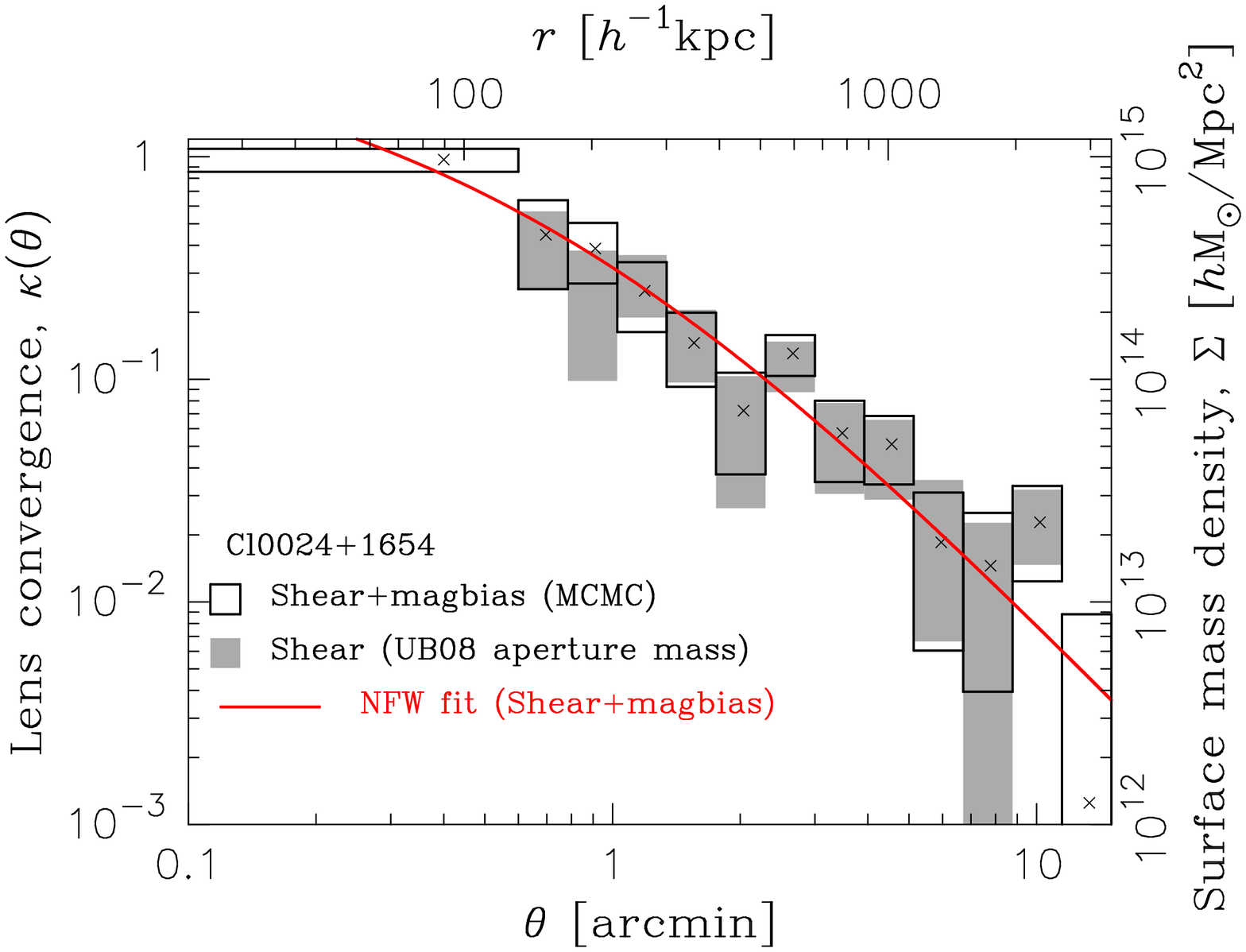} &
 \includegraphics[width=40mm,angle=270]{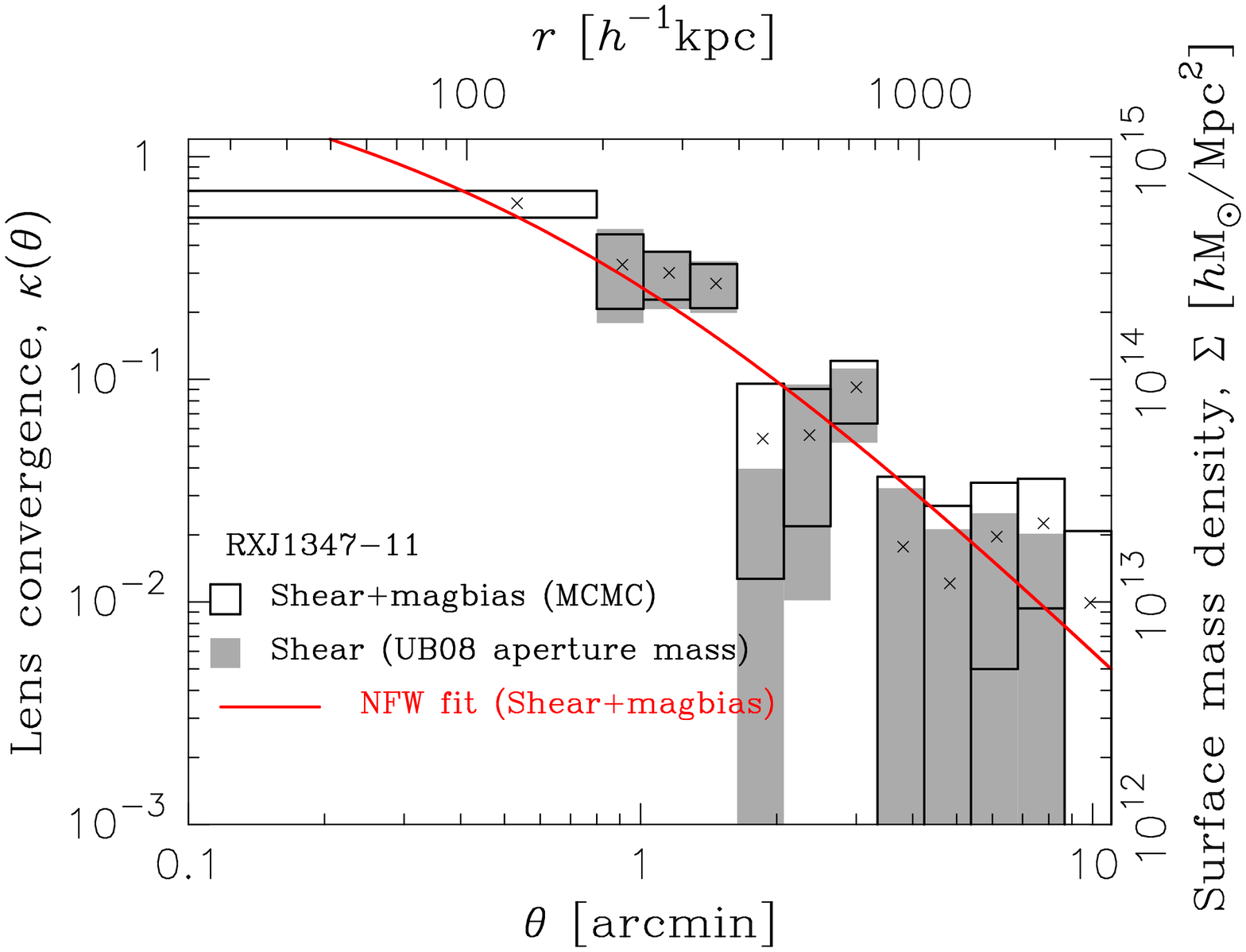}\\
\end{array}
$
 \end{center}
\caption{
Reconstructed projected mass profiles
 $\kappa(\theta)=\Sigma(\theta)/\Sigma_{\rm crit}$ for the five
 clusters, A1689 (top left),  A1703 (top middle), A370 (top right),
 Cl0024+17 (bottom left), and RXJ1347-11 (bottom right).
The open boxes show the results obtained using the Bayesian mass
 reconstruction 
 method from the combined Subaru tangential distortion and magnification 
 bias measurements of background galaxies (Figure
 \ref{fig:data}).  The innermost (first) box represents the average
 convergence $\overline{\kappa}(<\theta_{\rm min})$ interior to the minimum
 radius $\theta_{\rm min}$ of the weak lensing data.
Also shown for each cluster is the best-fit NFW profile (solid
 curve) for the combined Subaru distortion and magnification data.
The gray boxes represent the mass profile reconstructed using the
 nonlinear extension of aperture mass densitometry by \citet{UB2008}
 based on the same tangential distortion data (but without the
 magnification data 
combined), which employs an outer boundary
 condition on the mean convergence in the outermost radial bin. 
The horizontal extent of 
 each box represents the range of the radial band, and its vertical
 extent shows the statistical $1\sigma$ uncertainty in $\kappa$.  The
 cross in each 
 open box indicates the band-averaged convergence, being
 marked at the area-weighted center of the radial band.
The errors are correlated in both reconstructions.
\label{fig:kplot}
} 
\end{figure*}

\subsection{Cluster Sample and Observations}
\label{subsec:sample}

In this section we apply our Bayesian mass reconstruction method to a
sample of five well-studied high-mass clusters ($M\simgt 10^{15}M_\odot$)
at intermediate redshifts,
A1689 ($z=0.183$), A1703 
($z=0.281$), A370 ($z=0.375$), Cl0024+17 ($z=0.395$), and RXJ1347-11
($z=0.451$), observed with the wide-field camera \citep[$34\arcmin\times
27\arcmin$; see][]{2002PASJ...54..833M} on the 8.2m Subaru telescope.
Table \ref{tab:sample} gives a summary of cluster observations.
The clusters were observed deeply in several optical passbands,
with exposures in the range 2000--10000\,s per passband, with seeing in
the detection images ranging from $0.6\arcsec$ to $0.8\arcsec$
\citep[see][]{UB2008,BUM+08,Medezinski+2010,Umetsu+2010_CL0024}.

These massive clusters are known as strong lensing clusters, displaying
prominent strong-lensing features and large Einstein radii of
$\theta_{\rm ein}\simgt 30\arcsec$
\citep[e.g., for a fiducial source redshift
$z_s\sim 2$;][]{Broadhurst+Barkana2008,Oguri+Blandford2009}.  
For the clusters, the central mass distributions have been
recovered in detail by our strong-lens modeling 
\citep{2005ApJ...621...53B,Zitrin+2009_CL0024,Zitrin+2010_A1703}.
Here the models were constrained by a number of multiply-lensed
images identified previously in very deep multi-color imaging with {\it
HST}/ACS  
\citep[e.g.,][]{2005ApJ...621...53B,Bradac+2008_RXJ1347,Halkola+2008_RXJ1347,Limousin+2008_A1703,Richard+2009_A1703,Richard+2010_A370,Zitrin+2009_CL0024,Zitrin+2010_A1703}. 
Table \ref{tab:rein} gives a summary of the Einstein radii of the five
clusters as constrained from our detailed strong lens modeling with the
ACS observations.


\subsection{Background Galaxy Samples}
\label{subsec:back}

Here we will add to our high-quality spin-2 shape measurements 
\citep{UB2008,Medezinski+2010,Umetsu+2010_CL0024}
the independent magnification information based on deep multi-band
imaging with Subaru, 
in order to achieve
the maximum possible lensing precision.
Full details  of the Subaru observations and weak-lensing shape analysis
of these clusters were presented in a series of our papers \citep[][see
also Table
\ref{tab:sample}]{UB2008,BUM+08,Umetsu+2009,Medezinski+2010,Umetsu+2010_CL0024,Zitrin+2010_A1703}. 
The level of shear calibration bias with our implementation of the KSB+
method \citep{1995ApJ...449..460K} has been assessed by
\cite{Umetsu+2010_CL0024} using simulated Subaru Suprime-Cam images
\citep[M. Oguri 2010, in private communication;][]{2007MNRAS.376...13M}. 
We find, typically, $|m|\simlt 5\%$ of the shear calibration bias,
and $c\sim 10^{-3}$ of the residual shear offset which is about 1
order of magnitude smaller than the typical distortion signal in
cluster outskirts. This level of calibration bias
is subdominant compared to the statistical uncertainty ($\Delta M/M\sim 15\%$)
due to the intrinsic scatter in galaxy shapes, and to the dilution effect 
which can lead to an underestimation 
of the true signal for $R\simlt 400\,{\rm kpc}\,h^{-1}$ by a factor of
2--5 \citep[see Figure 1 of][]{BTU+05}.

A careful background selection
is critical for a weak-lensing analysis so that unlensed cluster
members and foreground galaxies do not dilute the true lensing
signal of the background
\citep{BTU+05,Medezinski+07,Medezinski+2010}.
We use undiluted samples of background galaxies derived in our previous
lensing work, as summarized 
in Table \ref{tab:back}. 
When deep multi-color photometry is available in our cluster fields, 
we use the background selection method of \citet[][see also Umetsu et
al. 2010]{Medezinski+2010} to
define blue and red background samples (A1703, A370, Cl0024+17,
RXJ1347-11),  
which relies on empirical correlations for galaxies in color-color space
derived from the deep Subaru photometry, by reference to the
deep photometric-redshift survey in the COSMOS field
\citep{Ilbert+2009_COSMOS}. 
Otherwise, we use the color-magnitude selection method
\citep{BTU+05,Medezinski+07,UB2008} to define 
a sample of red galaxies (A1689) whose colors are redder than the red
sequence of cluster E/S0 galaxies. These red galaxies are expected to
lie in the background by virtue of $K$-corrections which are greater
than the red cluster sequence galaxies, as  convincingly
demonstrated spectroscopically by \citet{Rines+Geller2008}.
Apparent magnitude cuts are applied in the reddest band available for
each cluster to avoid incompleteness near the detection limit.

A flux-limited sample of red background
galaxies is used for the magnification analysis (see
\S~\ref{subsec:magbias}). 
For the distortion analysis, we use a full composite sample of red and blue
(if available) background galaxies, 
where the galaxies used are well resolved to make reliable shape measurements
\citep[see, e.g.,][]{Umetsu+2010_CL0024}. 
Table \ref{tab:back} lists for respective color samples the
mean surface number density $\overline{n}$ of background galaxies, 
the effective source redshift, $\overline{z}_{s,\beta}$,\footnote{The
effective single-lens plane redshift $\overline{z}_{s,D}$, corresponding
to the mean depth $\langle\beta\rangle$, is defined as
$\langle\beta\rangle=\beta(\overline{z}_{s,\beta})$. For details, see \S~3.4
of \cite{Umetsu+2010_CL0024}.}
and the mean distance ratio $\langle\beta\rangle$ averaged over the
source redshift distribution.
We also quote in Table \ref{tab:back} the values of the relative 
mean depth, $\omega=\langle\beta({\rm
red})\rangle/\langle\beta({\rm full})\rangle$,  between the background
samples used for the magnification and distortion measurements (see
\S~\ref{subsubsec:prior} and Appendix \ref{appendix:lprof}).  
We adopt a conservative uncertainty of $5\%$ in the relative mean depth
$\omega$ for all of the clusters based on our previous work
\citep{UB2008,Umetsu+2009,Medezinski+2010,Umetsu+2010_CL0024}.  

The conversion from the observed
counts into magnification
depends on the normalization $n_0$ and the 
slope parameter $s$ of the unlensed number counts, which can be reliably
estimated thanks to the wide field of view of
Subaru/Suprime-Cam (see \S~\ref{subsec:massrec}).
Table \ref{tab:magbias} lists the magnitude cuts ($m_{\rm cut}$) and
unlensed count parameters ($n_0,s$) as measured from the red background
counts in outermost annular regions in cluster outskirts 
($\simgt 10\arcmin$). 


\subsection{Subaru Weak Lensing Profiles}
\label{subsec:lprof}

Following the methodology outlined in \S~\ref{subsec:gt} and
\S~\ref{subsec:magbias} (see also Appendix \ref{appendix:mask}), we
derive lens distortion and magnification profiles of five massive
clusters from Subaru observations.
In order to obtain meaningful radial profiles, one must carefully define
the center of the cluster. 
It is often assumed that the cluster mass centroid coincides with the
position of the brightest cluster galaxy (BCG), whereas the BCGs can be
offset from the mass centroids of the corresponding dark matter halos
\citep{2007arXiv0709.1159J,Oguri+2010_LoCuSS,Oguri+Takada2011}. 
Here we utilize our detailed mass maps in the cluster cores recovered
from strong-lens modeling of ACS observations (\S \ref{subsec:sample}),
providing an independent mass-centroid determination.
We find that for these five clusters there is only a small offset of
typically $\simlt 5\arcsec$
($20\,$kpc$\,h^{-1}$ at the highest cluster redshift of our study,
$z_d=0.451$) 
between the BCG and the dark-matter center of mass
\citep[see also \S~4.2 of][]{Umetsu+2010_CL0024}, 
often implied 
by other massive bright galaxies in the vicinity of the
BCG.  This level of cluster centering offset is substantially small as
compared to the typical inner radial boundary of weak lensing
observations, $\theta_{\rm min}\sim 1\arcmin$.
In the present work, we therefore simply assume that the cluster mass
centroid coincides with the location of the BCG, which is adopted as the
cluster center in our one-dimensional profile analysis.


The lensing profiles were calculated in $N$ discrete radial bins 
over the range of radii $\theta=[\theta_{\rm min},\theta_{\rm max}]$, 
with a constant logarithmic
 radial spacing of $\Delta\ln\theta =\ln(\theta_{\rm max}/\theta_{\rm
 min})/N$, where the inner radial boundary 
 $\theta_{\rm min}$ is taken such that $\theta_{\rm
 min}>\theta_{\rm ein}$ (see Table \ref{tab:rein}).  
The typical inner boundary is $\theta_{\rm min}\sim 1\arcmin
 (>\theta_{\rm ein})$ for
 cluster weak lensing.  The outer radial boundary $\theta_{\rm max}$ was
 chosen to be sufficiently larger than the typical virial radius of high
 mass  clusters with $M_{\rm vir}=(1-2)\times 
 10^{15}M_\odot\,h^{-1}$, $R_{\rm max}=D_d\theta_{\rm max}\simgt
 3\,$Mpc, but sufficiently small  ($\theta_{\rm max}\simlt 18\arcmin$)
 with respect to the size of the
 Suprime-Cam's field-of-view so  as to ensure accurate PSF (point spread
 function) anisotropy correction. 
The number of radial bins $N$ was determined for each cluster
such that the per-pixel detection signal-to-noise ratio (${\rm
S/N}$)
is of the order of unity,\footnote{We quantify the significance of a
detection for a given lensing profile in analogy to equation (38) of
\citet{UB2008}.} 
which is optimal for an inversion problem.
The radial binning scheme is summarized
 in Table \ref{tab:data}.

 
\begin{deluxetable}{cccccc} 
\tablecolumns{8}
\tablecaption{ 
 \label{tab:data}
Subaru weak lensing data 
} 
\tablewidth{0pt} 
\tablehead{ 
 \multicolumn{1}{c}{Cluster} &
 \multicolumn{1}{c}{$\theta_{\rm min},\theta_{\rm max}$} &
 \multicolumn{1}{c}{$N$} &
 \multicolumn{3}{c}{Detection ${\rm S/N}$} 
\\
 \multicolumn{1}{c}{} &
 \multicolumn{1}{c}{($\arcmin$)} &
 \multicolumn{1}{c}{} &
 \multicolumn{1}{c}{$g_+$} &
 \multicolumn{1}{c}{$n_\mu$} &
 \multicolumn{1}{c}{Combined ($\kappa$)} 
}
\startdata 
 A1689 & 
 $1,18$ & 
 11 &
13.8 &
8.8 &
17.8
\\
 A1703  & 
 $1,16$ &
 10 &
9.9 &
7.1 &
12.7
\\
 A370  & 
$0.7,16$ &
14 &
17.3 &
11.0 &
23.8
\\
 Cl0024+17  & 
$0.6,15$ &
12 &
13.0 &
8.5 &
18.5
\\
 RXJ1347-11  & 
$0.8,11$ &
11 &
9.7 &
5.7 & 
12.6
\enddata
\tablecomments{
The lensing profiles are calculated in $N$ discrete radial bins over the
 radial range of $\theta=[\theta_{\rm min},\theta_{\rm max}]$,
with a logarithmic radial spacing of $\Delta\ln\theta =\ln(\theta_{\rm
 max}/\theta_{\rm  min})/N$.  
For each cluster, $\theta_{\rm min}$ is taken such that $\theta_{\rm
 min}>\theta_{\rm ein}$ (see Table \ref{tab:rein}).
}
\end{deluxetable}

In Figure \ref{fig:data} we show the resulting distortion and 
magnification profiles for our sample of five massive lensing clusters.  
In all the clusters, a strong depletion of the red galaxy counts is seen
in the central, high-density region of the cluster, 
and clearly detected
out to several arcminutes from the cluster center. 
The statistical
significance of the detection of the depletion signal is in the range
$6\sigma$--$11\sigma$ 
(see Table \ref{tab:data}).
The detection significance of the tangential distortion derived from the 
full background sample ranges from $10\sigma$ to $17\sigma$, and is
better than the red counts \citep[see, e.g.,][]{1999A&A...345...17B}.
The $\times$-component is
consistent with a null signal in most of radial bins, indicating the
reliability of our weak-lensing analysis.
The magnification measurements with and without the masking correction
are roughly consistent with each other. 
Typically, the masking area is negligible (a few \%) at large
radii, and increases up to $(10-20)\%$ of the sky close to the cluster
center (see Appendix \ref{appendix:mask}).
To test the consistency
between our distortion and depletion measurements, we calculate
the depletion of the counts expected for
the best-fitting NFW profile derived from the distortion measurements
(Figure \ref{fig:data}), normalized to the observed density $n_0$ (Table
\ref{tab:magbias}). 
This comparison shows clear consistency
between two independent lensing observables with different systematics
(see \S~5.5 of Umetsu \& Broadhurst 2008),
which strongly supports the reliability and accuracy of
our weak-lensing analysis (see also supplemental material presented 
in Appendix \ref{appendix:supplement}).

\subsection{Cluster Mass Profile Reconstruction}
\label{subsec:massrec}

 
\begin{deluxetable*}{c|cccc|ccccc} 
\tabletypesize{\tiny}
\tablecolumns{20}
\tablecaption{ 
 \label{tab:nfw}
Best-fit NFW model parameters
} 
\tablewidth{0pt} 
\tablehead{ 
 \multicolumn{1}{c|}{Cluster} &
 \multicolumn{4}{c|}{NFW (weak lensing)} &
 \multicolumn{5}{c}{gNFW (weak+strong lensing)}
\\
 \multicolumn{1}{c|}{} &
 \multicolumn{1}{c}{$M_{\rm vir}$} &
 \multicolumn{1}{c}{$c_{\rm vir}$} &
 \multicolumn{1}{c}{$\chi^2/{\rm dof}$} &
 \multicolumn{1}{c}{$\theta_{\rm ein}$\tablenotemark{a}} &
 \multicolumn{1}{|c}{$M_{\rm vir}$} &
 \multicolumn{1}{c}{$c_{-2}$\tablenotemark{b}} &
 \multicolumn{1}{c}{$\alpha$} &
 \multicolumn{1}{c}{$\chi^2/{\rm dof}$} &
 \multicolumn{1}{c}{$\theta_{\rm ein}$\tablenotemark{a}}
\\
 \colhead{} &
 \multicolumn{1}{|c}{($10^{15}M_\odot h^{-1}$)} &
 \multicolumn{1}{c}{} &
 \multicolumn{1}{c}{} &
 \multicolumn{1}{c}{($\arcsec$)} &
 \multicolumn{1}{|c}{($10^{15}M_\odot h^{-1}$)} &
 \multicolumn{1}{c}{} &
 \multicolumn{1}{c}{} &
 \multicolumn{1}{c}{} &
 \multicolumn{1}{c}{($\arcsec$)}
}
\startdata 
 A1689  & 
 $1.282^{+0.217}_{-0.176}$ &
 $12.80^{+3.09}_{-2.41}$ &
 $4.2/10$ &
 $48.7^{+15.6}_{-13.8}$ &
$1.301^{+0.193}_{-0.156}$  &
$13.71^{+1.19}_{-1.22}$  &
$0.268^{+0.415}_{-0.268}$ &
$4.4/20$  &
$49.7^{+12.9}_{-9.2}$
\\
 A1703  & 
$1.232^{+0.244}_{-0.204}$ &
$7.02^{+2.36}_{-1.70}$ &
$6.1/9$ &
$28.1^{+16.3}_{-13.2}$ &
$1.272^{+0.234}_{-0.191}$ &
$7.07^{+1.08}_{-1.06}$ &
$0.934^{+0.192}_{-0.253}$ &
$7.2/21$ &
$27.9^{+14.7}_{-14.0}$
\\
 A370  & 
$2.451^{+0.309}_{-0.262}$ &
$7.00^{+1.09}_{-0.92}$ &
$10.2/13$ &
$50.4^{+11.0}_{-10.0}$ &
$2.276^{+0.255}_{-0.220}$ &
$5.68^{+0.48}_{-0.48}$ &
$0.392^{+0.158}_{-0.190}$ &
$15.2/26$ &
$30.5^{+9.4}_{-10.7}$
\\
 Cl0024+17  & 
$1.376^{+0.232}_{-0.201}$ &
$8.82^{+2.25}_{-1.68}$ &
$11.7/11$ &
$36.9^{+11.3}_{-10.1}$ &
$1.339^{+0.247}_{-0.203}$ &
$8.05^{+1.24}_{-1.28}$ &
$0.814^{+0.396}_{-0.717}$ &
$11.1/23$ &
$31.3^{+13.4}_{-9.2}$
\\
 RXJ1347-11  & 
$1.488^{+0.277}_{-0.244}$ &
$9.08^{+3.14}_{-2.17}$&
$10.4/10$ &
$43.2^{+13.6}_{-12.6}$ &
$1.435^{+0.114}_{-0.100}$ &
$7.16^{+0.43}_{-0.43}$&
$0.046^{+0.205}_{-0.046}$ &
$58.2/28$ &
$32.4^{+5.1}_{-5.2}$
\enddata
\tablenotetext{a}{Einstein radius 
predicted by the best-fit NFW (gNFW) model, evaluated at the arc
 redshift given  in Table \ref{tab:rein}.}
\tablenotetext{b}{Effective concentration parameter for gNFW,
 $c_{-2}\equiv r_{\rm vir}/r_{-2}=c_{\rm vir}/(2-\alpha)$.}
\end{deluxetable*}

We use a Markov chain Monte Carlo (MCMC) approach
with Metropolis-Hastings sampling to reconstruct the discrete cluster
mass profile $\bs=\{\overline{\kappa}_{\rm min},\kappa_i\}_{i=1}^{N}$
within a  Bayesian statistical framework 
(\S~\ref{subsec:bayesian}). 
We largely follow the sampling procedure outlined in
\citet{Dunkley+2005}, but employ the Gelman-Rubin $R$ statistic 
\citep{Gelman+Rubin1992}
as a simple but reasonable 
convergence criterion of generated chains.
Once convergence to a stationary distribution is achieved, 
we run a long final chain of $300,000$ steps, which
adequately samples 
the underlying posterior probability distribution.  
For all of the parameters, the number of iterations
required for convergence is much less than our final chain length.
Note, only the final
chain is used for our parameter estimations and error analysis.
We estimate the location of the maximum a posteriori probability for
each model parameter using the bisection method in  conjunction with
bootstrap techniques.  The covariance matrix $C_{ij}$ 
($i,j=1,2,...,N+1$) for the discrete mass profile $\bs$ is estimated from
the MCMC samples.   As an example, we show in 
Figure \ref{fig:post} one-dimensional marginalized posterior
PDFs for the mass profile $\bs$ of A1689 $(N=11)$.
The results are marginalized over all other parameters, including the
 observational parameters $(n_0,s,\omega)$.

The resulting posterior distributions are all clearly single-peaked, and
approximately  Gaussian for most of the parameters.
It is clearly evident that the mass-sheet degeneracy is 
broken thanks to the inclusion of magnification information on the local
area distortion.
Excluding magnification data, on the other hand, strongly
modifies the Gaussian shape of the marginalized posterior PDFs, producing
long non-Gaussian tails and broadening the distribution function,
resulting in large errors for the reconstructed mass profile. 
The improvement here from adding the magnification measurements is
significant, $\sim 30\%$ in terms of cluster mass profile measurements
(see Table \ref{tab:data}).

Figure \ref{fig:kplot} shows the lensing convergence profiles
$\bs=\{\overline{\kappa}_{\rm min},\kappa_1,\kappa_2,...,\kappa_N\}$, for
the five clusters reconstructed using our Bayesian method from combined
Subaru distortion and magnification data.
Also shown for comparison are independent reconstructions from the same
tangential distortion data (but without the magnification data combined)
using the one-dimensional method of \citet[][see also Umetsu et
al. 2010]{UB2008} 
based on the nonlinear extension of aperture mass densitometry,
which employs an outer boundary 
condition on the mean convergence in the outermost radial 
bin, $\overline{\kappa}_{\rm
max}\equiv \overline{\kappa}(\theta_{N-1},\theta_{\rm max})$. 
Here $\overline{\kappa}_{\rm max}$ for an isolated NFW halo can be
negligibly  small if  the outermost radii are taken as large as the
cluster virial radius \citep[see][]{Umetsu+2010_CL0024}.
This method has been applied successfully to Subaru weak lensing
observations of massive clusters
including A1689 \citep{UB2008,Umetsu+2009}, 
A1703 \citep{Oguri+2009_Subaru,Zitrin+2010_A1703}, 
and Cl0024+17 \citep{Umetsu+2010_CL0024}.    
Our results with different 
combinations of lensing measurements and boundary conditions, having
different systematics, are in agreement with each other.
This consistency clearly demonstrates
that our results are robust and insensitive to the choice of boundary
condition as well as to systematic errors
in the lensing measurements, such as the shear calibration error, as
found by \citet[][]{UB2008}.

Unlike the distortion effect, the magnification bias due to the
local area distortion falls off sharply with increasing distance from
the cluster center. 
We find from the reconstructed mass profiles  that the lens
convergence at  large
radii of $\theta=[10\arcmin,15\arcmin]$ is of the order $\kappa
=5\times 10^{-3}-0.01$. 
The expected level of the depletion signal in the weak-lensing limit is
$\delta n_\mu/n_0\approx -2\kappa$ for a {\it maximally}-depleted sample with
$s=0$, indicating a depletion signal of $\simlt (1-2)\%$ in the cluster
outskirts where we have estimated the unlensed
background counts, $n_0$.  This level of signal is smaller than
the fractional uncertainties in estimated unlensed counts $n_0$ of
$(2-4)\%$, thus consistent with the assumption.
Note that 
the calibration uncertainties in our observational parameters
$(n_0,s,\omega)$ have been properly taken into account and marginalized
over in our Bayesian analysis.

In the presence of magnification, one probes the number
counts at an effectively fainter limiting magnitude: 
$m_{\rm cut}+2.5\log_{10}\mu(\theta)$.   
The level of magnification is on average small in the weak-lensing
regime but for the innermost bin reaches a factor of 2 to 4 depending on
the cluster.  Here we use the count slope at the fainter effective limit 
($m_{\rm lim}$) when making the magnification estimate, to be
self-consistent.    
In our analysis we have implicitly assumed that the power-law behavior
(equation [\ref{eq:magbias}]) persists down to $\sim 1$ mag fainter than
$m_{\rm cut}$ where the count slope may be shallower.  For a given level
of count depletion, an overestimation of the count slope could lead to
an overestimation of the magnification, thus biasing the resulting mass
profile.  However, the number count slope for our data flattens only
slowly with depth varying from $s\sim 0.1$ to $s\sim 0.05$ from a limit
of $m=25.5$ to $m=26.5$, so  that this introduces a small correction of
only typically 8\%--11\% for the most magnified bins ($\mu=2-4$). 
In fact, we have found a good consistency between the
purely shear-based results and the results based on the combined
distortion and magnification data (see Figure \ref{fig:kplot}).  


To quantify and characterize the cluster mass distribution,
we compare the reconstructed $\kappa$ profile with 
the physically and observationally motivated NFW model.
Here we consider a generalized parametrization of the
NFW model of the following form \citep{Zhao1996,Jing+Suto2000}:
\begin{equation}
\label{eq:gnfw}
\rho(r)=\frac{\rho_s}{(r/r_s)^\alpha(1+r/r_s)^{3-\alpha}},
\end{equation} 
which has an arbitrary power-law shaped central cusp, 
$\gamma_{\rm 3D}=-\alpha$, and an asymptotic outer slope of 
$\gamma_{\rm 3D}=-3$.  This reduces to the NFW model for $\alpha=1$.
We refer to the profile
given by equation (\ref{eq:gnfw}) as the generalized NFW (gNFW, hereafter)
profile. 
It is useful to introduce the radius $r_{-2}$ at which the logarithmic
slope of the density is isothermal, i.e., $\gamma_{\rm 3D}=-2$. For the
gNFW profile, $r_{-2}=(2-\alpha)r_s$, and thus the corresponding
concentration parameter reduces to $c_{-2}\equiv r_{\rm
vir}/r_{-2}=c_{\rm vir}/(2-\alpha)$.
We specify the gNFW model with the central cusp slope, $\alpha$, 
the halo virial mass, $M_{\rm vir}$, and the concentration,
$c_{-2}=c_{\rm vir}/(2-\alpha)$.
We employ the radial dependence of the gNFW lensing profiles given by
\citet{Keeton2001_mass}.

We first fix the central cusp slope to $\alpha=1$ (NFW),
and constrain $(M_{\rm vir},c_{\rm vir})$  from $\chi^2$ fitting to the
discrete cluster mass 
profile $\bs=\{\overline{\kappa}_{\rm min},\kappa_i\}_{i=1}^{N}$
reconstructed from the combined weak-lensing distortion and magnification
measurements. 
The $\chi^2$ function for weak lensing is defined by
\begin{equation}
\chi^2
=\sum_{i=1}^{N+1} \sum_{j=1}^{N+1}
\big[ 
  s_i-\hat{s}_i(M_{\rm vir},c_{\rm vir})
\big]
 C^{-1}_{ij}
\big[ 
  s_j-\hat{s}_j(M_{\rm vir},c_{\rm vir})
\big],
\end{equation}
where $\hat{\bs}(M_{\rm vir},c_{\rm vir})$ is the NFW model prediction
for the discrete mass profile $\bs$.
The resulting constraints on the NFW model parameters
and the predicted Einstein radius $\theta_{\rm ein}$
are shown in Table \ref{tab:nfw}.  
For all the cases, the best-fit NFW model from weak lensing properly
reproduces the observed location of the Einstein radius, consistent with
the independent strong-lensing observations (see Table \ref{tab:rein}).  
In Table \ref{tab:data} we quote
the values of the total detection ${\rm S/N}$ in the reconstructed mass
profile $\bs$ based on the combined distortion and magnification data.

\subsection{Cluster Mass Estimates}
\label{subsec:rvir}


\begin{figure*}[!htb]
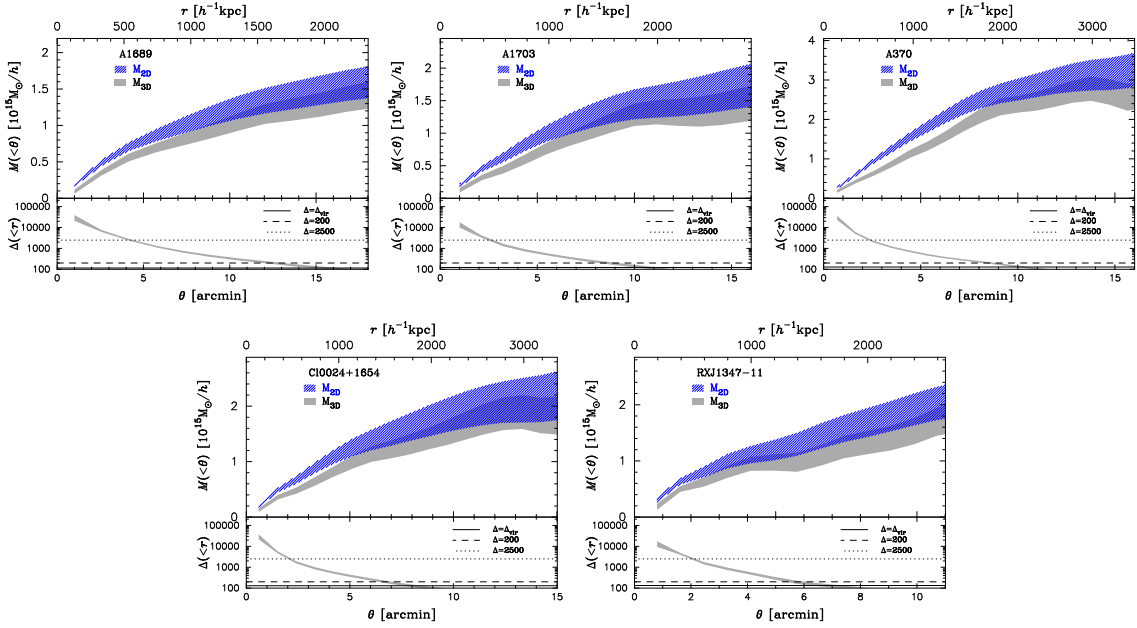
 
 \begin{center}
$
\begin{array}{c@{\hspace{.1in}}c@{\hspace{.1in}}c@{\hspace{.1in}}c}
 \includegraphics[width=40mm,angle=270]{f4a.eps} &
 \includegraphics[width=40mm,angle=270]{f4b.eps} &
 \includegraphics[width=40mm,angle=270]{f4c.eps} \\
\end{array}
$
$
\begin{array}{c@{\hspace{.1in}}c@{\hspace{.1in}}c}
 \includegraphics[width=40mm,angle=270]{f4d.eps} &
 \includegraphics[width=40mm,angle=270]{f4e.eps}\\
\end{array}
$
 \end{center}
\caption{
Non-parametric cumulative projected (blue-hatched) and
 spherical (gray-shaded) mass profiles,
 $M_{\rm 2D}(<\theta)$ and  $M_{\rm  3D}(<\theta)$, respectively, shown
 separately for the five clusters.
For each cluster, the blue-hatched area in the upper panel
shows the $68.3\%$ confidence interval for $M_{\rm
 2D}(<\theta)$
at each
 radius estimated from a 
 Monte-Carlo error analysis taking into account the error covariance
 matrix of the projected mass profile $\kappa(\theta)$ (see Figure
 \ref{fig:kplot}).   Similarly, the gray-shaded area shows the results
 for the deprojected mass profile $M_{\rm 3D}(<\theta)$ assuming
 spherical symmetry.
Shown in the lower panel is the mean spherical overdensity
 $\overline{\rho}(<r)=M_{\rm 3D}(<r)/(4\pi 
 r^3/3)$ interior to radius $r$ 
 in units of the critical density of the universe $\rho_{\rm crit}(z)$,
$\Delta(<r)\equiv
 \overline{\rho}(<r)/\rho_{\rm crit}(z)$, reconstructed from our
 comprehensive weak lensing analysis.  The gray-shaded area shows
 the $68.3\%$ confidence interval at each radius estimated from a
 Monte-Carlo error analysis taking into account the error covariance
 matrix of the projected mass profile $\kappa(\theta)$ (see Figure
 \ref{fig:kplot}).   The horizontal solid, dashed, and dotted lines
 show three representative values for the fractional overdensity,
 $\Delta=\Delta_{\rm vir}, 200$, and $2500$, respectively, for each
 cluster.  See also Table \ref{tab:m3d}.
\label{fig:m23}
} 
\end{figure*}

 
\begin{deluxetable*}{c|cc|cc|cc|cc} 
\tabletypesize{\tiny}
\tablecolumns{8}
\tablecaption{ 
 \label{tab:m3d}
Three-dimensional Cluster Mass from a Non-parametric Deprojection Analysis.
} 
\tablewidth{0pt} 
\tablehead{ 
 \multicolumn{1}{c|}{Cluster} &
 \multicolumn{2}{c|}{$\Delta=2500$} &
 \multicolumn{2}{c|}{$\Delta=500$} & 
 \multicolumn{2}{c|}{$\Delta=200$} &
 \multicolumn{2}{c}{$\Delta=\Delta_{\rm vir}$}  
\\ 
 \colhead{} &
 \multicolumn{1}{|c}{$M_{2500}$} &
 \multicolumn{1}{c|}{$r_{2500}$} &
\multicolumn{1}{|c}{$M_{500}$} &
 \multicolumn{1}{c|}{$r_{500}$} &
\multicolumn{1}{|c}{$M_{200}$} &
 \multicolumn{1}{c|}{$r_{200}$} &
\multicolumn{1}{|c}{$M_{\rm vir}$} &
 \multicolumn{1}{c}{$r_{\rm vir}$} 
\\
 \colhead{} &
 \multicolumn{1}{|c}{($10^{15}M_\odot h^{-1}$)} &
 \multicolumn{1}{c|}{(${\rm Mpc}\,h^{-1}$)} &
 \multicolumn{1}{|c}{($10^{15}M_\odot h^{-1}$)} &
 \multicolumn{1}{c|}{(${\rm Mpc}\,h^{-1}$)} &
 \multicolumn{1}{|c}{($10^{15}M_\odot h^{-1}$)} &
 \multicolumn{1}{c|}{(${\rm Mpc}\,h^{-1}$)} &
 \multicolumn{1}{|c}{($10^{15}M_\odot h^{-1}$)} &
 \multicolumn{1}{c}{(${\rm Mpc}\,h^{-1}$)} 
}
\startdata 
 A1689 &
 $0.569 \pm 0.073$ &
 $0.547 \pm 0.025$ &
 $0.883 \pm 0.117$ &
 $1.083 \pm 0.050$ &
 $1.179 \pm 0.170$ &
 $1.618 \pm 0.085$ &
 $1.300 \pm 0.205$ &
 $2.011 \pm 0.113$ 
\\
 A1703  & 
$0.404 \pm 0.069$ &
$0.471 \pm 0.028$ &
$0.787 \pm 0.126$ &
$1.006 \pm 0.058$ &
$1.143 \pm 0.196$ &
$1.546 \pm 0.097$ &
$1.325 \pm 0.221$ &
$1.915 \pm 0.148$
\\
 A370 &
$0.529 \pm 0.134$ &
$0.496 \pm 0.040$ &
$1.315 \pm 0.161$ &
$1.152 \pm 0.046$ &
$2.213 \pm 0.270$ &
$1.860 \pm 0.079$ &
$2.399 \pm 0.249$ &
$2.215 \pm 0.079$
\\
 Cl0024+17  & 
$0.462 \pm 0.063$ &
$0.472 \pm 0.023$ &
$0.863 \pm 0.136$ &
$0.993 \pm 0.055$ &
$1.194 \pm 0.188$ &
$1.502 \pm 0.081$ &
$1.329 \pm 0.224$ &
$1.799 \pm 0.105$
\\
 RXJ1347-11  & 
$0.587 \pm 0.060$ &
$0.500 \pm 0.017$ &
$0.949 \pm 0.145$ &
$1.003 \pm 0.051$ &
$0.972 \pm 0.208$ &
$1.373 \pm 0.089$ &
$1.150 \pm 0.250$ &
$1.663 \pm 0.115$
\enddata
\tablecomments{
$M_\Delta\equiv M_{\rm 3D}(<r_{\Delta})$
is the three-dimensional mass within a sphere of
a fixed mean interior overdensity $\Delta$ with respect to the critical
density of the universe at the cluster redshift $z_d$.
}
\end{deluxetable*}

Our comprehensive Bayesian 
analysis of the weak lensing distortion and magnification of background
galaxies allows us to recover the mass normalization,  given as the mean
convergence $\overline{\kappa}_{\rm min}$
within the innermost measurement radius $\theta_{\rm min}$
($>\theta_{\rm ein}$),
without employing inner strong lensing information.

We use the non-parametric deprojection method of
\citet{Broadhurst+Barkana2008} 
to derive for each cluster
three-dimensional virial quantities ($r_{\rm vir}, M_{\rm vir}$)
and values of $M_{\Delta}=M_{\rm 3D}(<r_\Delta)$ 
within a sphere of a fixed mean interior overdensity $\Delta$
with 
respect to the critical density $\rho_{\rm crit}(z_d)$ of the universe at
the cluster redshift $z_d$. 
We first deproject the two-dimensional mass profiles obtained in
\S~\ref{subsec:massrec} and 
derive non-parametric three-dimensional mass profiles $M_{\rm 3D}(<r)$
simply assuming spherical symmetry, following the method introduced by
\citet{Broadhurst+Barkana2008}.
This method is based on the fact that
the surface-mass density $\Sigma(R)$ is related to the
three-dimensional mass density $\rho(r)$ by an Abel integral transform;
or equivalently, one finds that
the three-dimensional mass $M_{\rm 3D}(<r)$
out to spherical radius $r$ is written in terms of $\Sigma(R)$ as
\begin{eqnarray}
\label{eq:m3d}
M_{\rm 3D}(<r) &=& 
M_{\rm 2D}(<R_{\rm min}) \\\nonumber
&+&2\pi\int_{R_{\rm min}}^r\! dRR\Sigma(R)
-4\int_r^\infty\!dRR f
\left(
\frac{R}{r}
\right)
\Sigma(R),
\end{eqnarray}
where 
$f(x)=(x^2-1)^{-1/2}-\tan^{-1}(x^2-1)^{-1/2}$
\citep{Broadhurst+Barkana2008,Umetsu+2010_CL0024},
\footnote{
This integral transformation has 
an integrable singularity at the lower limit of the
second integrand ($R=r$), which can be readily avoided by a suitable
coordinate transformation. 
}
and the first term of the right-hand side can be obtained as 
$M_{\rm 2D}(<R_{\rm min})=\pi R_{\rm min}^2\Sigma_{\rm
crit}\overline{\kappa}_{\rm min}$ with $R_{\rm min}\equiv D_d\theta_{\rm
min}$.
The errors are estimated from Monte Carlo
simulations based on the full covariance matrix of the lensing
convergence profile \citep[for details, see][]{Umetsu+2010_CL0024}.
In Figure \ref{fig:m23} we show the non-parametric cumulative projected
(blue-hatched) and spherical (gray-shaded) mass profiles,  $M_{\rm
2D}(<\theta)$ and $M_{\rm 3D}(<\theta)$, separately for the five
clusters.

Table \ref{tab:m3d} gives a summary of the
spherical mass estimates $M_{\rm 3D}(<r_\Delta)$
corresponding to $\Delta=\Delta_{\rm vir}, 200$, and 2500
from our non-parametric deprojection analysis, where $\Delta_{\rm
vir}\simeq 110-130$ is the virial overdensity of the spherical collapse
model  evaluated at the cluster redshift $z_d$.
Overall, we find a good agreement between the virial mass
estimates from the parametric and non-parametric deprojection
approaches (see \S~\ref{subsec:wl+sl} for the results of gNFW fits
to the combined weak and strong lensing data). 

Our virial mass estimate of A1689 is obtained as
$M_{\rm vir}=(1.30\pm 0.21)\times 10^{15}M_\odot\,h^{-1}$
from the combined distortion and 
magnification profiles, consistent with the results
of \citet{UB2008}, who combined strong lensing,
weak lensing distortion and magnification data in a full two-dimensional
analysis, and derived $M_{\rm vir}=1.5^{+0.6}_{-0.3}\times
10^{15}M_\odot \,h^{-1}$, where this $1\sigma$ error 
includes both statistical and systematic uncertainties
\citep[see
also][]{Umetsu+2009,Kawaharada+2010,Molnar+2010_ApJL}.\footnote{Note, 
without the inner strong lensing information combined, \citet{UB2008}
found $M_{\rm vir}=(1.38\pm 0.14)\times
10^{15}M_\odot \,h^{-1}$ (the errors represent only the statistical one)
from an entropy-regularized maximum-likelihood 
combination of
Subaru distortion and magnification data sets, in excellent agreement
with the one dimensional results in this work.}
This is also in good agreement with the results from the recent
high-resolution lensing observations by \citet{Coe+2010},  
$M_{\rm vir}=(1.4^{+0.4}_{-0.2})\times 10^{15}M_\odot\,h^{-1}$.
These lensing results are consistent with careful dynamical
work by \citet{Lemze+2009}, who obtained a virial mass estimate of
$M_{\rm vir}=(1.3\pm 0.4)\times 10^{15}M_\odot\,h^{-1}$ for A1689.
As found early by \citet{BUM+08}, our comprehensive weak-lensing
analysis  implies A370 is the most massive cluster now known,
$M_{\rm vir}=(2.40\pm 0.25)\times 10^{15}M_\odot \,h^{-1}$.  
Our virial mass estimate is slightly higher than that derived in our
earlier weak-lensing work combined with the inner Einstein radius
information, $M_{\rm vir}=(2.1\pm 0.2)\times 10^{15}M_\odot\,h^{-1}$
\citep{BUM+08}, where the difference is primarily due to our improved
background selection and depth estimate presented in
\citet{Medezinski+2010}.  
For RXJ1347-11, we find its NFW virial mass is slightly overestimated
compared to the non-parametric estimate of
$M_{\rm vir}=(1.15\pm 0.25)\times 10^{15}M_\odot \,h^{-1}$
due to the projection of subclumps associated with the large scale
structure around the cluster \citep[see][]{Lu+2010_RXJ1347}.
Our non-parametric mass estimates are in good agreement with independent
X-ray, dynamical and lensing analyses
\citep{Kling+2005_RXJ1347,Miranda+2008_RXJ1347,Lu+2010_RXJ1347}. 
For A1703 and Cl0024+17, our new mass estimates are fully consistent
with our recent weak-lensing results derived from the Subaru distortion
data alone \citep[][see also
\S~\ref{subsec:massrec}]{Zitrin+2010_A1703,Umetsu+2010_CL0024}.


\begin{figure*}[!htb] 
 \begin{center}
 \includegraphics[width=0.35\textwidth,angle=270,clip]{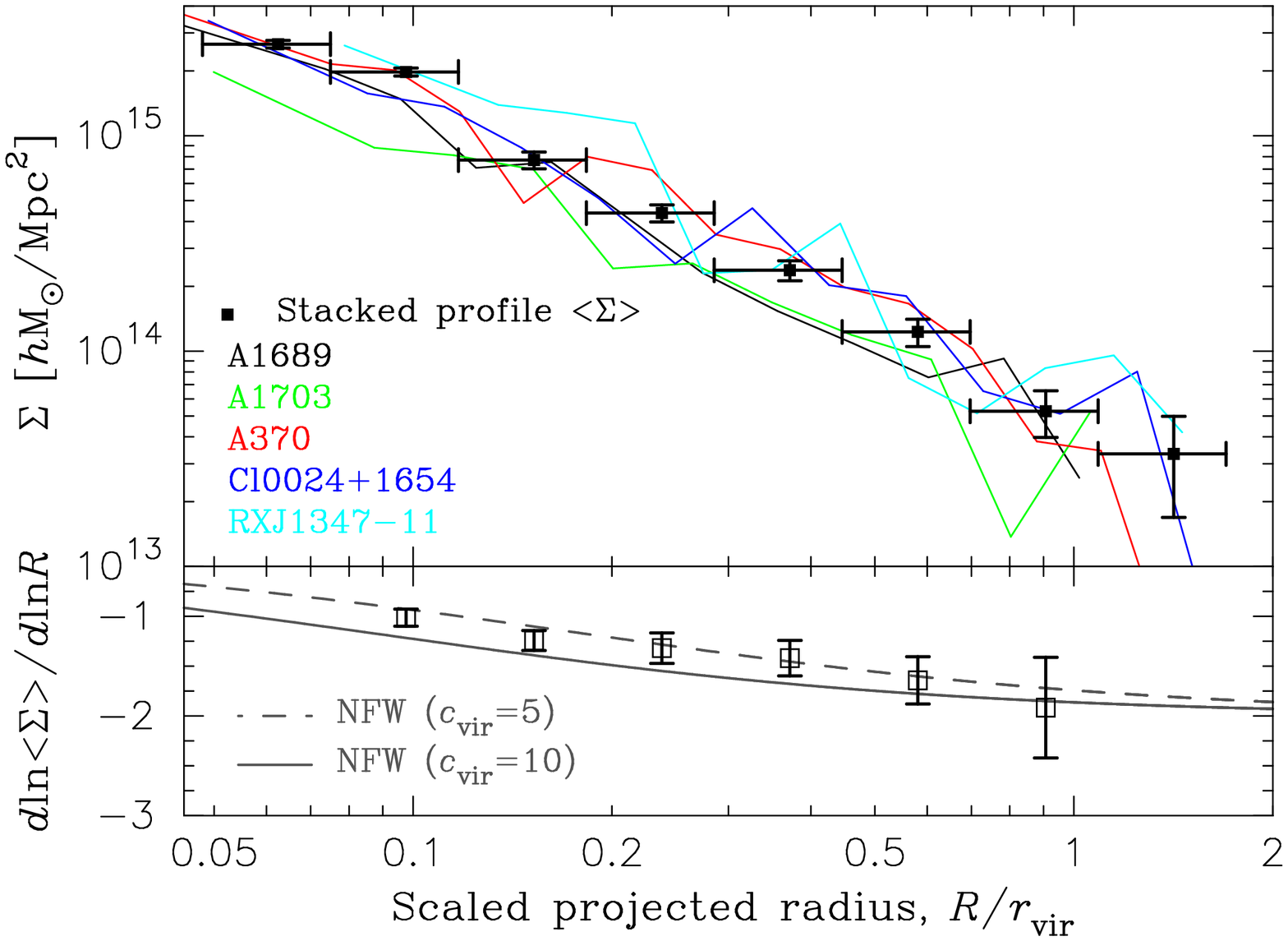} 
 \end{center}
\caption{
({\it Top}):
The model-independent average mass profile 
$\langle \Sigma\rangle(R)$
(filled squares) with its
 statistical $1\sigma$ uncertainty as a 
 function of the projected radius $R$ scaled with $r_{\rm vir}$,
which is obtained by stacking the lensing convergence profiles for the
 five clusters, A1689, A1703, A370, Cl0024+17, and RXJ1347-11, shown as
 solid lines.
({\it Bottom}): The logarithmic slope of the stacked mass profile (open
 squares with error bars),
 $d\ln\langle\Sigma\rangle/d\ln{R}$, is shown as a function of the
 scaled projected radius $R/r_{\rm vir}$, along with NFW model
 predictions with $c_{\rm vir}=5$ and $c_{\rm vir}=10$ for comparison.
 The projected logarithmic slope steepens from
$\gamma_{\rm 2D}=-1.01\pm 0.09$ at $R\approx 0.1r_{\rm vir}$ to 
$\gamma_{\rm 2D}=-1.92\pm 0.51$ at $R\approx 0.9r_{\rm vir}$.
\label{fig:stack}
} 
\end{figure*}

\subsection{Stacking Analysis}
\label{subsec:stack}

The statistical precision of lensing constraints
can be further improved by stacking the signal
from an ensemble of clusters with respect to their centers, providing
average properties of cluster 
mass profiles.  As discussed by \citet{Okabe+2010_WL}, this stacking
analysis has several important advantages. A notable advantage of the
stacking analysis is that the resulting average profile is insensitive
to the inherent asphericity and substructure (in projection) of
individual cluster mass distributions, as well as to uncorrelated
large-scale structure projected along the same line of
sight. Consequently, the statistical precision can be boosted by
stacking together a number of clusters, especially on small angular
scales \citep[see][]{Okabe+2010_WL}.

To do this,
we first define a new set of radial bands in which the
mass profiles of individual clusters are re-evaluated for a stacking
analysis.  
Here we scale each cluster mass profile 
according to the cluster virial radius $r_{\rm vir}$ obtained by our
non-parametric 
method (see \S~\ref{subsec:rvir}).
For each cluster, we construct an $M\times N$ {\it projection} matrix
${\cal P}_{ji}$ 
that projects the mass profile $\kappa_i$ ($i=1,2,...,N$) of the cluster 
onto the new radial bands scaled in units of $r_{\rm vir}$ ($j=1,2,...,M$).  
Assuming a constant density in each radial band,
the projection matrix ${\cal P}_{ji}$
is uniquely specified by the conservation of mass.  With this projection
matrix, the mass profile in the new basis is written as
\begin{equation}
\tilde{\bkappa}={\cal P}\bkappa.
\end{equation}
Accordingly, the error covariance matrix in the new basis is
\begin{equation}
\tilde{C} = {\cal P}C{\cal P}^{t}.
\end{equation}
%

With the mass profiles of individual clusters on a common basis,
we can stack the clusters to 
produce an averaged mass profile.
Here we re-evaluate the mass profiles of the individual clusters in
$8$ logarithmically-spaced radial bins over the range of radii
$R=[0.05,1.7]r_{\rm vir}$.
Since the noise in different clusters is uncorrelated,
the mass profiles of individual clusters can be co-added according to 
\citep[e.g.,][]{Sayers+2009}
\begin{equation}
\label{eq:stack}
\langle \bSigma\rangle = 
\left(\displaystyle\sum_n \tilde{C}_n^{-1} w_n^2\right)^{-1}
 \,
\left(
\displaystyle\sum_n{ \tilde{C}^{-1}_n w_n \tilde{\bkappa}_{n}}
\right), 
\end{equation}
where the index $n$ runs over all of the clusters, and $w_n$ is the
inverse critical surface mass density for the $n$th cluster,
$w_n=\left(\Sigma_{\rm crit}^{-1}\right)_n$.
 The error
covariance matrix for the stacked mass profile
$\langle\Sigma\rangle$ is obtained as 
\begin{equation}
\label{eq:covar_stack}
{\cal C} =\left(\displaystyle\sum_n \tilde{C}^{-1}_n w_n^2\right)^{-1}, 
\end{equation}
where the index $n$ runs over all of the clusters.

We show in the top panel of Figure \ref{fig:stack} the resulting
model-independent average mass profile $\langle \Sigma\rangle$
with its statistical $1\sigma$ uncertainty as a function of the scaled
projected radius $R/r_{\rm vir}$,
obtained by stacking the five clusters using equations (\ref{eq:stack})
and (\ref{eq:covar_stack})
\citep[see also][]{Mandelbaum+2006,2007arXiv0709.1159J,Sheldon+2009a}. 
We note that the effect of different cluster redshifts has been taken
into account by proper error propagation  in terms of the lensing
efficiency functions ($w_n$) of individual clusters to average over.
The stacked mass profile exhibits a fairly smooth radial trend, and is
detected at a high significance level of 
$37\sigma$ out to $R=1.7r_{\rm vir}$.

In the bottom panel of Figure
\ref{fig:stack}, we plot the logarithmic density slope
$\gamma_{\rm 2D}(R)\equiv d\ln{\langle\Sigma\rangle}/d\ln{R}$ of the
stacked mass profile as a function of the scaled projected radius
along with NFW model predictions with $c_{\rm vir}=5$ and $c_{\rm
vir}=10$.  The logarithmic gradient of the average profile shows a
slight steepening trend with increasing radius in projection, consistent
with NFW profiles with $c_{\rm vir}=5-10$.  
Finally, we quote model-independent constraints on the 
average logarithmic density slope to be
$\gamma_{\rm 2D}=-1.01\pm 0.09$ at $R\approx 0.1 r_{\rm vir}$ and
$\gamma_{\rm 2D}=-1.92\pm 0.51$ at $R\approx 0.9 r_{\rm vir}$ for the five
clusters.

\subsection{Combining Weak and Strong Lensing}
\label{subsec:wl+sl}


\begin{figure*}[!htb] 
 \begin{center}
 \includegraphics[width=0.5\textwidth,angle=270,clip]{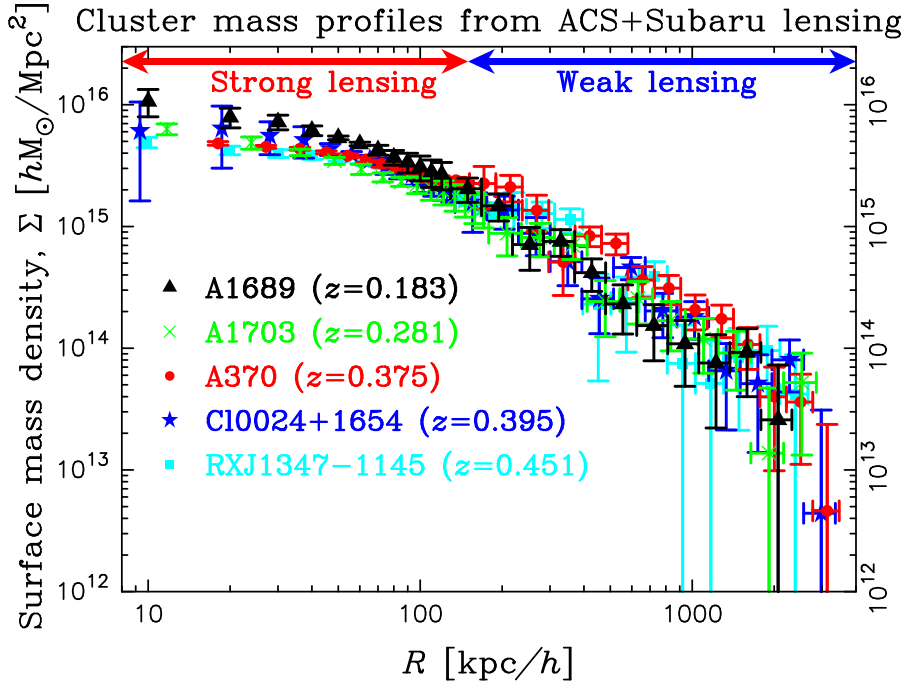} 
 \end{center}
\caption{
Full mass profiles for A1689 (triangles), A1703 (crosses), A370
 (circles), Cl0024+17 (stars), and RXJ1347-11 (squares) 
over a wide range of radius, from 10\,kpc$\,h^{-1}$ to
 $3000$\,kpc$\,h^{-1}$, 
 derived from ACS strong-lensing ($R\simlt 150\,$kpc\,$h^{-1}$)
and Subaru weak-lensing ($R\simgt 150\,$kpc\,$h^{-1}$) measurements,
 showing a continuously steepening radial trend out to and beyond the
 cluster virial radius ($r_{\rm vir}\sim 2\,$Mpc\,$h^{-1}$). 
\label{fig:fullmass}
} 
\end{figure*}

The Subaru data allow the weak lensing profiles of individual clusters
to be accurately measured in several independent radial bins in the
subcritical regime ($\theta> \theta_{\rm ein}$).
The projected mass profile can be unambiguously recovered on an
individual cluster basis from combined weak-lensing shape distortion and
magnification bias 
measurements.   

Here we combine our weak-lensing profiles with detailed
strong-lensing information for the inner $\simlt 200\,$kpc region of
these clusters, for which we have 
identified many new sets of multiple images from deep {\it HST}/ACS
observations
\citep{2005ApJ...621...53B,Zitrin+2009_CL0024,Zitrin+2010_A1703},  
for a full determination of the entire mass profiles of the five
well-studied clusters. 
Figure \ref{fig:fullmass} shows a sample of joint mass profiles for our 
five clusters
recovered over two decades of radius ranging from 10\,kpc$\,h^{-1}$ to
$3000$\,kpc$\,h^{-1}$.
Note in this comparison we have excluded the central weak-lensing bin
$\overline{\kappa}(<\theta_{\rm min})$ and 
the strong-lensing data points at radii overlapping with the Subaru
data.  In each case, the weak and strong 
lensing are in excellent agreement where the data overlap (typically
around $R\sim 150\,$kpc\,$h^{-1}$), 
and the joint mass profiles
form well-defined radial profiles 
with a continuously-steepening radial trend
from the central region to beyond the virial radius ($\simlt 1.7r_{\rm
vir}$).

Our high-quality lensing data, covering the entire cluster,
allow us to place useful constraints on the gNFW structure
parameters (\S~\ref{subsec:lprof}), namely, the central cusp slope
$\alpha$ as well as the NFW virial mass and concentration parameters.
Using our full lensing constraints, we obtain for each cluster the
best-fit gNFW model as summarized in Table \ref{tab:nfw}. 
For the halo mass and concentration parameters ($M_{\rm vir},c_{-2}$),
we find good agreement between the results with and without the inner
strong-lensing profile combined \cite[for Cl0024+17,
see also][]{Umetsu+2010_CL0024}. 
Our joint mass profiles for the entire cluster region are consistent
with a generalized form of the NFW density profile with modest
variations in the central cusp slope ($\alpha\simlt 0.9$), 
except for the ongoing merger RXJ1347-11 \citep[see][]{Mason+2010_RXJ1347}, 
for which we find an unacceptable fit with a reduced 
$\chi^2$ of $58$ for $28$ degrees of freedom,
due to various local deviations from the model, especially in the inner
mass profile which is tightly constrained by strong lensing.
The best-fit value of $\alpha$ derived for the relaxed cluster A1703 is
$\alpha=0.93^{+0.19}_{-0.25}$, 
being consistent with NFW ($\alpha=1$),
which is in excellent agreement with independent lensing results by
\cite{Oguri+2009_Subaru},
$\alpha=0.9^{+0.2}_{-0.4}$,
and 
\cite{Richard+2009_A1703},
$\alpha=0.92^{+0.05}_{-0.04}$.


\section{Summary and Discussion}
\label{sec:discussion}

We have developed a new method for a direct reconstruction of the projected
 mass  profile of galaxy clusters from combined weak-lensing distortion
 and  magnification measurements  (\S~\ref{subsec:gt} and
 \S~\ref{subsec:magbias}) within a Bayesian statistical framework,
which allows for a full parameter-space extraction of the underlying
 signal.    
This method applies to the full range of radius outside the Einstein
 radius, where nonlinearity between the surface mass density and
the observables extends to a radius of a few arcminutes, and
recovers the absolute mass normalization.
A proper Bayesian statistical analysis is essential to
explore the entire parameter 
space and investigate the parameter degeneracies
(\S~\ref{subsec:bayesian}), arising from the 
mass-sheet degeneracy (\S~\ref{sec:basis}).
This method  can be readily generalized for a statistical  
analysis using stacked lensing profiles of a sample of clusters.

We have applied our comprehensive lensing method to a
sample of five high-mass clusters 
for which detailed strong-lensing information is readily available from
{\it HST}/ACS observations.
The deep Subaru multi-band photometry, in conjunction with
our background-selection techniques
\citep{Medezinski+2010,Umetsu+2010_CL0024}, 
allows for a secure selection 
of uncontaminated blue and red background populations. 
In each cluster, a strong depletion of the red galaxy counts has been
detected at a significance level of $6\sigma$--$11\sigma$ (Table
\ref{tab:data}).
A comparison  shows clear consistency between two independent lensing
observables with different systematics, ensuring the reliability of our
weak-lensing analysis (Figure \ref{fig:data}). 
The combination of independent Subaru distortion and magnification data
breaks the mass-sheet degeneracy, as examined by our Bayesian
statistical analysis (Figure \ref{fig:post}).
Excluding magnification data, on the other hand, strongly modifies the
Gaussian shape of the marginalized posterior PDFs, producing long
non-Gaussian tails and broadening the distribution function, resulting
in large reconstruction errors. 
The improvement here from adding the magnification measurements is
significant, $\sim 30\%$ in terms of cluster mass profile measurements
(Table \ref{tab:data}).

We have formed a model-independent mass profile from stacking the
clusters, which is detected at 
$37\sigma$ out to beyond the virial radius, 
$R\approx 1.7r_{\rm vir}$. 
We found that the projected logarithmic slope, 
$\gamma_{\rm 2D}(R)= d\ln{\Sigma}(R)/d\ln{R}$, steepens from
$\gamma_{\rm 2D}=-1.01\pm 0.09$ at $R\approx 0.1r_{\rm vir}$ to 
$\gamma_{\rm 2D}=-1.92\pm 0.51$ at $R\approx 0.9r_{\rm vir}$,
consistent with NFW profiles with $c_{\rm vir}=5-10$. 
We also obtained for each cluster inner strong-lensing based mass 
profiles from deep {\it HST}/ACS observations, which we have shown
overlap well with the outer Subaru-based profiles and together are
well described by a generalized Navarro-Frenk-White profile, except for
the ongoing merger RXJ1347-11 \citep{Mason+2010_RXJ1347}, with modest
variations in the central cusp slope ($\alpha\simlt 0.9$), perhaps
related to the dynamical state of the cluster.

These high-mass lensing clusters with large Einstein radii
appear to be centrally concentrated in projection, as found 
in several other well studied massive clusters from careful lensing work
\citep{2003A&A...403...11G,2003ApJ...598..804K,BTU+05,2007ApJ...668..643L,BUM+08,Broadhurst+Barkana2008,Oguri+2009_Subaru,Umetsu+2010_CL0024,Zitrin+2011_MACS}. 
An accurate characterization of the observed sample is
crucial for any cluster-based cosmological tests.
It has been suggested that clusters selected with giant arcs represent a  
highly biased population in the context of $\Lambda$CDM.
For those clusters with large Einstein radii (say, $\theta_{\rm
ein}>20\arcsec$), a large statistical bias of about $(40-60)\%$ is
derived from $N$-body simulations 
\citep{2007ApJ...654..714H,Oguri+Blandford2009},
representing the most triaxial cases.
The mean level of mass concentration inferred from our weak lensing
analysis is high, 
$\langle c_{\rm vir}\rangle=7.9\pm 0.8$
(simply ignoring the mass and redshift dependencies),  for our sample
with  
$\langle M_{\rm vir}\rangle=(1.48\pm 0.10)\times
10^{15}M_\odot\,h^{-1}$,
as compared to the $\Lambda$CDM predictions,
$\langle c_{\rm vir}\rangle=3.4^{+1.5}_{-1.0}$ 
\citep[the errors quoted represent a $1\sigma$ lognormal scatter of
$\sigma(\log_{10}{c_{\rm vir}})=0.15$; see][] 
{Duffy+2008},
evaluated at the median redshift of our sample, $\overline{z}_d\simeq
0.38$. This represents an overall discrepancy of  $5\sigma$ with
respect to the predictions, without taking into account the effects of
projection bias.
This apparent discrepancy is also evident when the weak-lensing mass
profiles are combined with the inner strong-lensing information from
deep ACS observations (\S~\ref{subsec:wl+sl}). 
Applying a bias correction of $50\%$
\citep[see][]{BUM+08,Oguri+2009_Subaru},  
the discrepancy in $\langle c_{\rm vir} \rangle$ is reduced to the
$3\sigma$ level.

Another possible source of systematic errors in the mass profile
determination is the cluster off-centering effect
(\S~\ref{subsec:lprof}).
The effect is essentially the smoothing of the central lensing signal
\citep[see][]{Oguri+Takada2011},
which flattens the
recovered convergence profile below the offset scale, and therefore
reduces the derived mass concentration ($c_{-2}$) and cusp slope
($\alpha$) parameters.
For our sample of clusters, we found a
cluster centering offset of typically $\Delta\theta \simlt 5\arcsec$
using our 
detailed strong-lensing models.  This level of centering offset
($\Delta R\simlt 20$\,kpc$\,h^{-1}$)  is much smaller
than the estimated values for the inner characteristic radius of our
clusters, $r_{-2} \sim 200\,{\rm kpc}\,h^{-1}$, and hence may not
significantly affect our concentration measurements.  However, this
could potentially lead to an underestimation of the central cusp slope
$\alpha$.

Our results are consistent with previous
lensing work which similarly detected a concentration excess in the 
lensing-based measurements for strong-lensing clusters
\citep{2007MNRAS.379..190C,BUM+08,Oguri+2009_Subaru}.
Our findings could imply, for these clusters, either substantial mass
projected along the line of sight, due in part to intrinsic halo
triaxiality, or a higher-than-expected concentration of dark matter.
Nevertheless,
the overall level of uncertainties may be too
large to robustly test the CDM predictions with the present sample.
The forthcoming space telescope cluster survey, CLASH\footnote{Cluster
Lensing And Supernova Survey with Hubble (P.I.: M. Postman), http://www.stsci.edu/{\large\textasciitilde{}}postman/CLASH/},   
will provide a definitive derivation of mass profiles 
for a larger, lensing-unbiased sample of relaxed X-ray clusters
($>5\,$keV), combining high-resolution panchromatic space imaging with  
deep, wide-field Subaru weak-lensing observations, for a definitive 
determination of the representative mass profile of massive clusters.


\acknowledgments
We thank the anonymous referee for a careful reading of the manuscript 
and and for providing invaluable comments. 
We are very grateful for discussions with Doron Lemze, Sandor Molnar,
Masahiro Takada, Masayuki Tanaka, Nobuhiro Okabe, and Sherry Suyu, whose
comments were very helpful.  
K.U. acknowledges Yuji Chinone for helpful comments on MCMC techniques.
We thank Nick Kaiser for making the IMCAT package publicly available.
The work is partially supported by the National Science Council of Taiwan
under the grant NSC97-2112-M-001-020-MY3. 


\appendix


\begin{figure*}[!htb] 
 \begin{center}
 \includegraphics[width=0.5\textwidth,angle=0,clip]{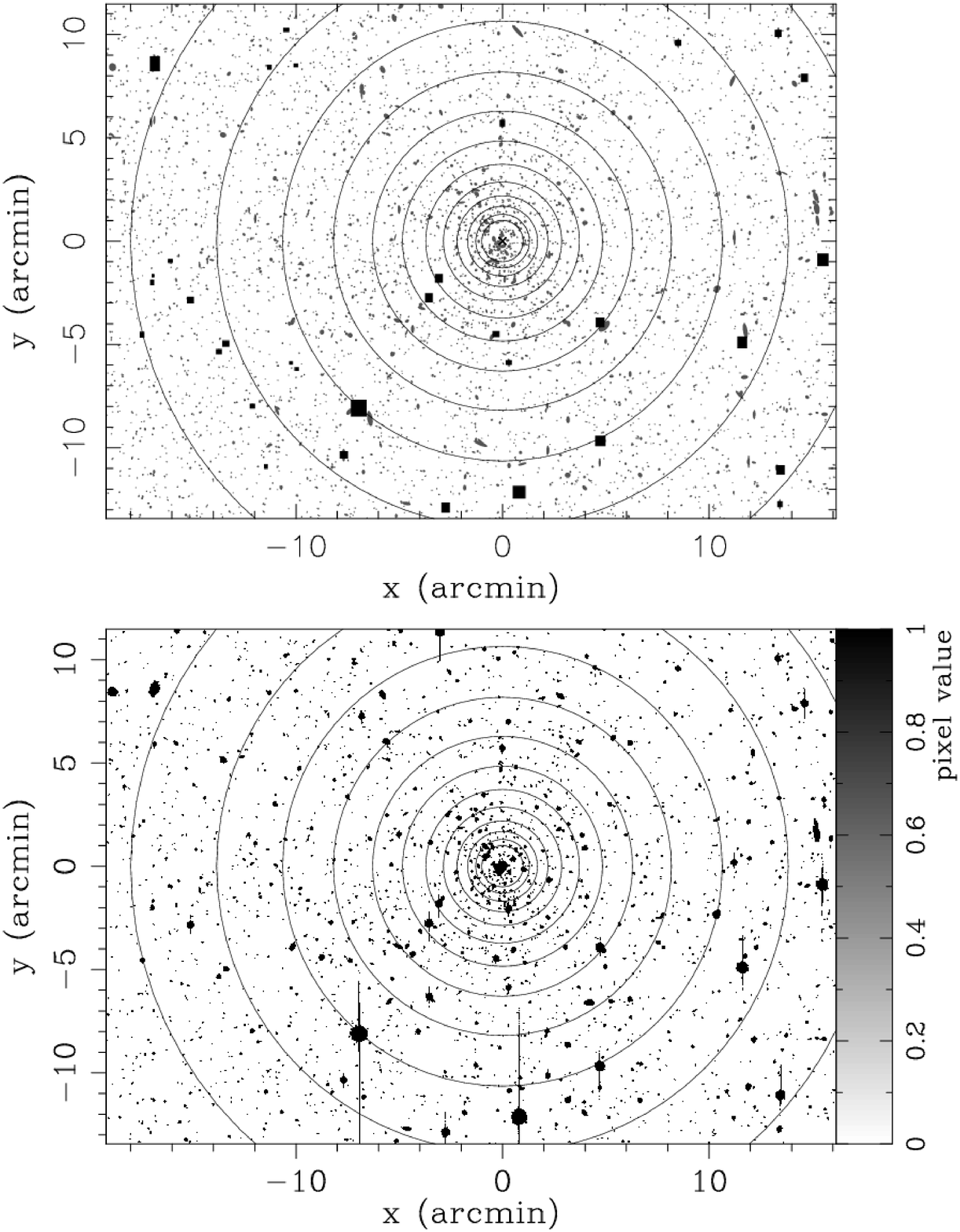}
 \end{center}
\caption{
Distribution of masked regions in A1689 ($z=0.183$).
Superposed are the concentric radial bands used for calculating the
 radial profiles centered on the BCG.
{\it Top}: Results obtained using Method A (see Appendix
 \ref{appendix:method_A}).  The gray area shows bright masking objects
 ($i'<22$ AB mag), each described as an ellipse.  The black rectangular
 regions are masked areas for saturated
stars and bright stellar trails.
{\it Bottom}: Results obtained using 
Method B (see Appendix \ref{appendix:method_B}) based on SExtractor's
 check-image output with ${\tt CHECKIMAGE\_TYPE=OBJECT}$.  The black
 area shows those connected pixels that belong to bright foreground
objects, bright cluster galaxies, saturated stars, and stellar trails.
\label{fig:mask}
}  
\end{figure*}


\begin{figure*}[!htb] 
 \begin{center}
 \includegraphics[width=0.35\textwidth,angle=270,clip]{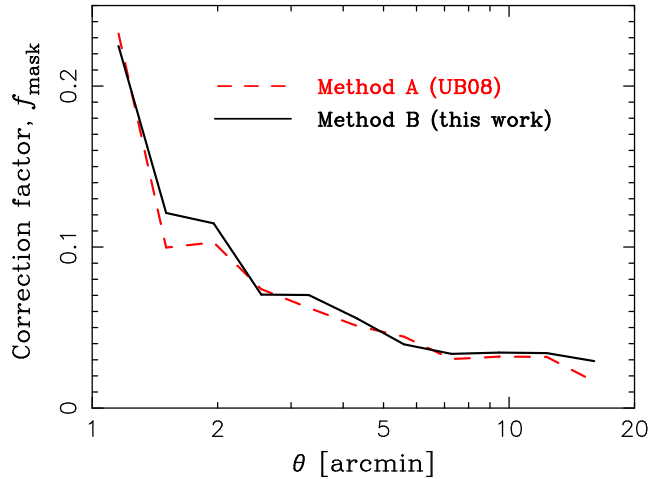} 
 \end{center}
\caption{The azimuthally-averaged correction factor $f_{\rm
 mask}(\theta)$ for the masking effect as 
 a function of  radius from the center of A1689.
The red  dashed curve shows the level of correction required due
 to masking, as estimated by \citet{UB2008} using Method A (Appendix
 \ref{appendix:method_A}). 
The black solid curve shows the results using Method B (Appendix
 \ref{appendix:method_B}). The two methods give consistent results in
 terms of both shape and amplitude of the radial profile of the
 correction factor.  
\label{fig:fmask}
} 
\end{figure*}

\section{Masking Correction}
\label{appendix:mask}

For the cluster magnification analysis,
the masking effect due to bright cluster galaxies, bright
foreground objects, and saturated objects has to be properly taken
into account and corrected for \citep{BTU+05,UB2008,Umetsu+2010_CL0024}.
Here we describe two different approaches for estimating the effect of
masking in the magnification bias measurement. 

\subsection{Method A}
\label{appendix:method_A}

In this method (Method A), we describe each masking object as an ellipse
specified by its structure parameters, such as  the size, axis
ratio, and position angle.  SExtractor provides such useful estimates of
object properties (magnitude, size, axis ratio, position angle, and so
on) for calculating masking areas and the resulting correction.
In practice, we account for the masking of observed
sky by excluding a generous area $\pi a b$ around each masking
object, where $a$ and $b$ are defined as  $\nu_{\rm mask} (=2-4)$ times
the major 
({\tt A\_IMAGE}) and minor axes ({\tt B\_IMAGE}) computed from SExtractor,
corresponding roughly to the isophotal detection limit (see
\citet{UB2008} and \citet{Umetsu+2010_CL0024}).
We then calculate the correction factor for
this masking effect as a function of radius from the cluster center,
and renormalize the number density accordingly.
Note that the actual choice of $\nu_{\rm mask}$ is insensitive to the
resulting magnification measurement $n_\mu/n_0=\mu^{2.5s-1}$
\citep{UB2008,Umetsu+2010_CL0024}, as this effect will cancel out when
taking a ratio of the local surface density $n_\mu$ to the unlensed mean
number density $n_0$ (see \S~5.5.3 of Umetsu \& Broadhurst 2008).
However, this method is not suitable for estimating masking areas by
saturated stars and stellar trails, which cannot be described by
simple ellipses.   For this case, we manually place a rectangular
region mask around each saturated object to exclude a proper area.
This approach has been commonly adopted in previous studies of
cluster magnification bias
\citep{1998ApJ...501..539T,BTU+05,UB2008,Umetsu+2010_CL0024}. 

In the top panel of Figure \ref{fig:mask}, we show as an example 
the masked region in A1689 based on the Subaru $i'$-band
photometry, as obtained using Method A (see also Umetsu \& Broadhurst
2008). 
The gray area shows bright masking objects ($i'<22$ AB mag),
each described as an ellipse.
The black rectangular regions are masked areas for saturated
stars and bright stellar trails.  We show in Figure \ref{fig:fmask} the
level of mask area correction in A1689 as a function of radius from the
cluster center.  The masking area is negligible (a few \%) at large
radii, and increases up to $\sim 20\%$ of the sky close to the cluster
center. Typically, the level of masking correction is found to be
sufficiently small ($\ll 1$), so that the uncertainty in the masking
correction is of the second order, and hence negligible.

\subsection{Method B}
\label{appendix:method_B}

Here we describe an alternative method  (Method B) for the masking
correction which  we have developed in this work.  This method has
practical advantages in comparison to Method A: that is, it is easy to
implement and can be fully automated once the configuration of analysis
parameters is tuned, especially useful for large scale sky surveys such
as the Subaru HSC survey.

In this approach, we use a pixelized flag image to exclude masking
regions, namely those connected pixels that belong to bright foreground
objects, 
bright cluster galaxies, saturated stars, and stellar trails.
This can be done using
SExtractor's check-image output with ${\tt  CHECKIMAGE\_TYPE =
OBJECTS}$. 
We tune SExtractor configuration's detection parameters
in order to optimize the
detection for a particular object size and brightness.  In our analysis of
Subaru images,
the per-pixel detection threshold above the local sky background is set
to $5\sigma$ (${\tt DETECT\_THRESH}=5$), and the minimum number of
connected pixels above threshold is set to $300$ (${\tt
DETECT\_MINAREA=300}$) with $0.202\arcsec\,{\rm pixel}^{-1}$ sampling.
The bottom panel of Figure \ref{fig:mask} shows the distribution of
masked regions in A1689 as obtained using Method B.  
Figure \ref{fig:fmask} compares the masking correction factor in A1689
obtained using Methods A and B, shown as a function of radius from the
cluster center.  It is shown that the two methods give consistent results in terms of both shape and amplitude of the radial profile of the correction factor.

\section{Discretized Estimator for the Averaged Convergence}
\label{appendix:avkappa}

In this Appendix, we aim to derive a discrete expression 
for the mean interior convergence $\overline\kappa(<\theta)$
using the azimuthally-averaged convergence $\kappa(\theta)$.
In the continuous limit,
the mean convergence $\bar\kappa(<\theta)$ interior to radius $\theta$
can be expressed in terms of $\kappa(\theta)$ 
as
\begin{eqnarray}
\bar{\kappa}(\theta)=\frac{2}{\theta^2}\int_0^{\theta}
\!d\ln\theta'\theta'^2\kappa(\theta').
\end{eqnarray}
For a given set of $(N+1)$ annular radii $\theta_l$
$(l=1,2,...,N+1)$,
defining $N$ radial bands in the range $\theta_{\rm min}\equiv\theta_1\le
\theta\le \theta_{N+1}\equiv \theta_{\rm max}$,
a discretized estimator for $\overline{\kappa}(<\theta)$
can be written in the following way:
\begin{equation}
\label{eq:avkappa_d}
\bar{\kappa}(<\theta_l)=
\frac{\theta_{\rm min}^2}{\theta_l^2}\overline{\kappa}(<\theta_{\rm min})+
\frac{2}{\theta_l^2}\sum_{i=1}^{l-1}
\Delta\ln\theta_i
\bar\theta_i^2
\kappa(\bar\theta_i),
\end{equation}
with
$\Delta\ln\theta_i \equiv (\theta_{i+1}-\theta_i)/\bar\theta_i$
and $\bar\theta_i$
being the area-weighted center of the $i$th
annulus defined by $\theta_i$ and $\theta_{i+1}$;
in the continuous limit, we have
\begin{eqnarray}
\label{eq:medianr}
\bar\theta_i
&\equiv& 
2\int_{\theta_i}^{\theta_{i+1}}\!d\theta'\theta'^2/
(\theta_{i+1}^2-\theta_{i}^2)\nonumber\\ 
&=&
\frac{2}{3}
\frac{\theta_{i}^2+\theta_{i+1}^2+\theta_{i}\theta_{i+1}}
{ \theta_{i}+\theta_{i+1} }. 
\end{eqnarray}

\section{Discretized Estimators for the Lensing Profiles}
\label{appendix:lprof}

We derive expressions for the binned tangential
distortion $g_{+}$ and magnification $\mu$ in terms of the binned
convergence $\kappa$, using the following relations:
\begin{eqnarray}
g_+(\overline\theta_i) &=&
\frac{\overline{\kappa}(<\overline\theta_i)-\kappa(\overline\theta_i)}{1-\kappa(\overline\theta_i)},\\
\mu(\overline\theta_i)&=&\frac{1}{\left(1-\kappa(\overline\theta_i)\right)^2
\left(1-g_+^2(\overline\theta_i)\right)},
\end{eqnarray}
where both the quantities depend on the mean convergence
$\overline\kappa$ interior to the radius $\overline\theta_i$, which
is the center of the $i$th radial band 
of $[\theta_i,\theta_{i+1}]$ 
(see Appendix \ref{appendix:avkappa}).
By assuming a constant density in each radial band and by noting that
$\overline\theta_i$ is the {\it median} radius of the $i$th
radial band, $\overline\kappa(<\overline\theta_i)$ can be well
approximated by 
\begin{eqnarray} 
\overline{\kappa}(<\overline\theta_i) = 
\frac{1}{2}\Big(
\overline\kappa(<\theta_i)+\overline\kappa(<\theta_{i+1})
\Big)  
\end{eqnarray} 
where $\overline\kappa(<\theta_i)$ and $\overline\kappa(<\theta_{i+1})$
can be computed using the formulae given in Appendices
\ref{appendix:avkappa} and \ref{appendix:lprof}.

If we are to combine background samples of different populations ($a$
and $b$) and
different lensing depths $\langle\beta\rangle$,
we should take into account population-to-population variations in the
lensing depth,
$\omega_{ba}=\langle\beta\rangle_b/\langle\beta\rangle_a$, which can be
measured from deep multi-band imaging, such as the 30-band COSMOS
database \citep{Ilbert+2009_COSMOS}.   The tangential distortion
$g_+^{(b)}$ for Population ($b$) can be expressed in terms of the
convergence $\kappa^{(a)}$ for Population ($a$) as 
\begin{equation}
g_+^{(b)}=\frac{\omega_{ba}(\overline{\kappa}^{(a)}-\kappa^{(a)})}
{(1-\omega_{ba}\kappa^{(a)})}.
\end{equation}  
Similarly, the magnification $\mu^{(b)}$ for Population ($b$) can be
expressed in terms of $\omega_{ba}\kappa^{(a)}$.

\clearpage

\section{SUPPLEMENTAL MATERIAL}
\label{appendix:supplement}

\begin{figure*}[!htb] 
 \begin{center}
 \includegraphics[width=1.0\textwidth,angle=0,clip]{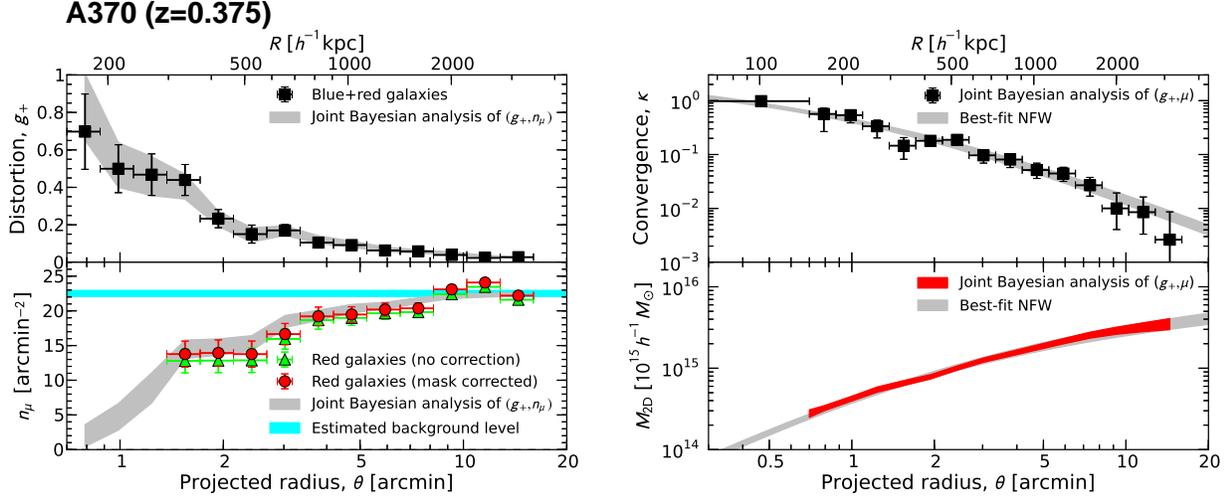} 
 \end{center}
\caption{
{\bf Left}: weak-lensing radial profiles of A370 ($z_d=0.375$)
 as measured from background galaxies registered in Subaru 
$B_{\rm j}R_{\rm c}z'$ images (Table \ref{tab:sample}).
The top panel shows the tangential reduced shear profile $g_+(\theta)$
 (squares) based on Subaru distortion data of the 
full background (red+blue) sample (Table \ref{tab:back}). 
The bottom panel shows the count depletion profiles $n_\mu(\theta)$ due
 to magnification for a flux-limited sample of red background
 galaxies. The circles and  
triangles show the respective results with and without the mask correction due
to bright foreground objects and cluster members. The horizontal bar represents
the constraints on the unlensed count normalization, $n_0$, as estimated from
Subaru data. Also shown in each panel is the joint Bayesian fit (gray
 area, 68\% CL) to both profiles, 
{\bf Right}: the {\it joint mass profile solution} of the shear and
 magnification data sets shown in the left panels, obtained following
 the Bayesian method described in the main text (Section
 \ref{sec:method}), effectively breaking  the mass-sheet degeneracy.
The top panel shows the reconstructed surface mass density profile
 $\kappa=\Sigma/\Sigma_{\rm crit}$  (squares), in good agreement with
 the standard NFW form (gray area, 68\% CL; see Table \ref{tab:nfw}).
The innermost bin represents the average surface mass density
 $\overline{\kappa}_{\rm min}\equiv \overline{\kappa}(<\theta_{\rm min})$ 
interior to the inner radial boundary $\theta_{\rm min}=0.7\arcmin$
of the weak-lensing distortion data.  The bottom panel shows the
 projected cumulative mass profile $M_{\rm 2D}$ (red),
 along with the best-fit NFW model (gray area, 68\% CL; Table \ref{tab:nfw}).
\label{fig:a370}
} 
\end{figure*}

\begin{figure*}[!htb] 
 \begin{center}
 \includegraphics[width=1.0\textwidth,angle=0,clip]{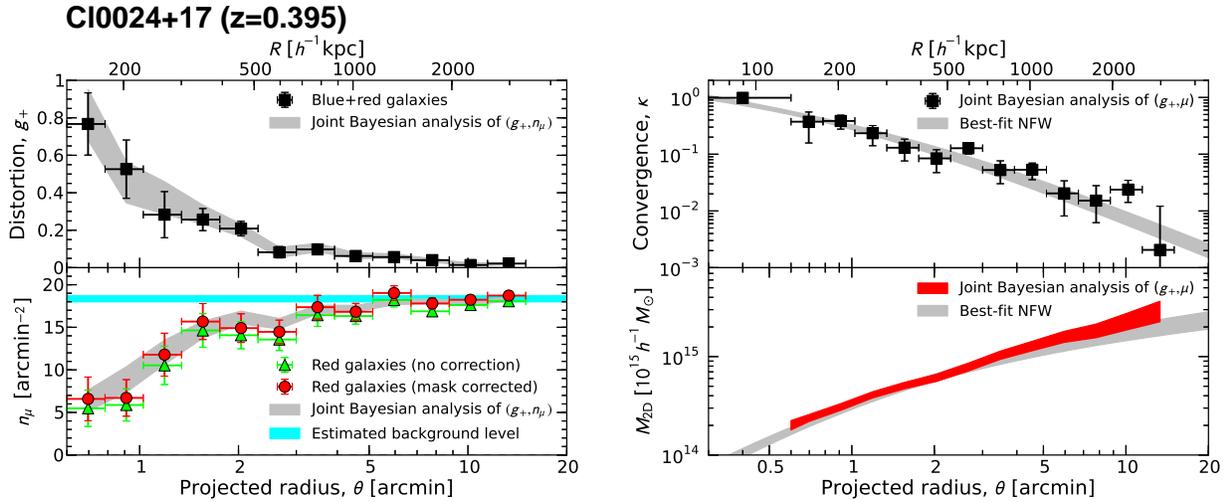} 
 \end{center}
\caption{  
Same as in Figure \ref{fig:a370}, but shown for the cluster Cl0024+17
 ($z_d=0.395$).   A strong depletion of the red galaxy counts is clearly
 detected (bottom left panel) down to the inner radial boundary of the
 weak lensing data, $\theta_{\rm  min}=0.6\arcmin$.  A consistent mass
 profile solution 
 $\bs=\left\{\overline{\kappa}_{\rm min},\kappa_i\right\}_{i=1}^{N}$ 
 (top right panel) is obtained from a joint
 Bayesian fit  to both lensing profiles shown in the left panels.
\label{fig:cl0024}
} 
\end{figure*}

In this supplemental material section, we compare observed 
tangential-distortion (shear) $g_+$ and count-depletion
(magnification) $n_\mu$ radial profiles with their respective
reconstructed profiles, taking A370 at $z_d=0.375$ (Figure \ref{fig:a370}) and
Cl0024+17 at $z_d=0.395$ (Figure \ref{fig:cl0024}) as examples. 
We note that this supplemental material is added for the arXiv version
only, and is not included in the published ApJ version (Umetsu et
al. 2011, ApJ, 729, 127). 

\clearpage

\end{document}